\documentclass[12pt, draftclsnofoot, onecolumn]{IEEEtran}
\usepackage[margin=0.788in]{geometry}
\IEEEoverridecommandlockouts
\usepackage[utf8]{inputenc}
\usepackage{caption} 
\ifCLASSINFOpdf
\usepackage[pdftex]{graphicx}

\usepackage{verbatim}
\else
\fi
\ifCLASSOPTIONcaptionsoff
\usepackage[nomarkers]{endfloat}
\let\MYoriglatexcaption\caption
\renewcommand{\caption}[2][\relax]{\MYoriglatexcaption[#2]{#2}}
\fi
\usepackage{cite}
\usepackage{amsmath,amssymb,amsfonts}
\usepackage{cases}
\usepackage{graphicx}
\usepackage{epstopdf}
\usepackage{textcomp}
\usepackage{xcolor}
\def\BibTeX{{\rm B\kern-.05em{\sc i\kern-.025em b}\kern-.08em
	T\kern-.1667em\lower.7ex\hbox{E}\kern-.125emX}}
\usepackage{amsthm}
\usepackage{amssymb}
\usepackage{amsmath}

\DeclareMathOperator{\diag}{diag}
\usepackage{amsmath,amsfonts,amsthm,bm} 
\usepackage{cite}
\usepackage{booktabs}
\usepackage{float}
\makeatletter

\newcommand{\Rmnum}[1]{\expandafter\@slowromancap\romannumeral #1@}
\makeatother

\DeclareMathOperator{\subto}{subject \hspace{0.125em} to}
\usepackage{color, soul} 
\usepackage{subfigure}
\usepackage{threeparttable}
\usepackage{graphicx} 
\usepackage{graphics} 
\graphicspath{{figure/}}
\usepackage{amsmath} 
\usepackage{verbatim}
\usepackage{mathrsfs}
\usepackage{caption}
\usepackage{multirow}
\usepackage{float}
\usepackage{epstopdf}
\usepackage{mleftright}
\mleftright
\newcommand{\dd}{\mathop{}\!\mathrm{d}}
\usepackage{tikz}

\usepackage[linesnumbered,ruled,vlined]{algorithm2e}
\usepackage{algpseudocode}
\SetKwInOut{Input}{\textbf{Input :}}  
\SetKwInOut{Output}{\textbf{Output :}} 
\usepackage{tikz}

\usepackage{array}
\renewcommand{\Re}{\operatorname{Re}}
\renewcommand{\Im}{\operatorname{Im}}
\usepackage{caption}
\usepackage{subfigure}
\usepackage{color}
\definecolor{gray}{rgb}{0.6, 0.6, 0.6}
\definecolor{blue}{rgb}{0.0, 0.0, 1}
\newcommand{\tabincell}[2]{\begin{tabular}{@{}#1@{}}#2\end{tabular}}
\usepackage{makecell}

\usepackage{enumerate}

\usepackage{centernot}
\usepackage{mathtools}
\usepackage{stmaryrd }
\begin{document}
%
\title{Deep Multi-Task Learning for Cooperative NOMA: System Design and Principles}
	\author{Yuxin Lu, Peng Cheng, Zhuo Chen, Wai Ho Mow, Yonghui Li, and Branka Vucetic
			\thanks{Yuxin Lu and Wai Ho Mow are with the Department of Electronic and Computer Engineering, the Hong Kong University of Science and Technology, Hong Kong S.A.R. (e-mail: ylubg@ust.hk; eewhmow@ust.hk).} \thanks{Yonghui Li and Branka Vucetic  are with the School of Electrical and Information Engineering, the University of Sydney, Australia, (e-mail:  yonghui.li@sydney.edu.au; branka.vucetic@sydney.edu.au).} \thanks{Peng Cheng is with the Department of Computer Science and Information Technology, La Trobe University, Melbourne, VIC 3086, Australia, and also with the School of Electrical and Information Engineering, the University of Sydney, Sydney, NSW 2006, Australia (e-mail: p.cheng@latrobe.edu.au; peng.cheng@sydney.edu.au). } \thanks{ Zhuo Chen is with the CSIRO DATA61, Marsfield, NSW 2122, Australia (e-mail: zhuo.chen@ieee.org). }
	}

\maketitle


%


%
\begin{abstract}

Envisioned as a promising component of the future wireless Internet-of-Things (IoT) networks, the non-orthogonal multiple access (NOMA) technique can support massive connectivity with a significantly increased spectral efficiency. Cooperative NOMA is able to further improve the communication reliability of users under poor channel conditions. However, the conventional system design suffers from several inherent limitations and is not optimized from the bit error rate (BER) perspective. In this paper, we develop a novel deep cooperative NOMA scheme, drawing upon the recent advances in deep learning (DL). We develop a novel hybrid-cascaded deep neural network (DNN) architecture such that the entire system can be optimized in a holistic manner. On this basis, we construct multiple loss functions to quantify the BER performance and propose a novel multi-task oriented two-stage training method to solve the end-to-end training problem in a self-supervised manner. The learning mechanism of each DNN module is then analyzed based on information theory, offering insights into the proposed DNN architecture and its corresponding training method. We also adapt the proposed scheme to handle the power allocation (PA) mismatch between training and inference and incorporate it with channel coding to combat signal deterioration.  Simulation results verify its advantages over orthogonal multiple access (OMA) and the conventional cooperative NOMA scheme in various scenarios.

\end{abstract}

\section{Introduction}	
\label{sec:intro}
 
Massive wireless device connectivity under limited spectrum resources is considered as cornerstone of the wireless Internet-of-Things (IoT) evolution. As a transformative physical-layer technology, non-orthogonal multiple access (NOMA)  \cite{Intro-NOMA-book-vaezi2019multiple,NOMA-liu2020non} leverages superposition coding (SC) and successive interference cancellation (SIC) techniques to support  simultaneous multiple  user  transmission  in  the  same  time-frequency  resource  block. Compared with its conventional orthogonal multiple access (OMA) counterpart, NOMA 
can significantly increase the spectrum efficiency, reduce access latency, and achieve more balanced user fairness \cite{Intro-NOMA-ding2017}. Typically, NOMA functions in either the power domain, by multiplexing different  power levels, or the code domain, by utilizing partially overlapping codes \cite{intro-NOMA-dai2018asur}.

Cooperative NOMA, which integrates cooperative communication techniques into NOMA, can further improve the communication reliability of users under poor channel conditions, and therefore largely extend the radio coverage  \cite{coNOMA-ding2015}. Consider a downlink transmission scenario, where there are two classes of users: 1) near users, which have better channel conditions and are usually located close to the base station (BS); and 2) far users, which have worse channel conditions and are usually located close to the cell edge.  The near users perform SIC or joint maximum-likelihood (JML) detection to detect their own information, thereby obtaining the prior knowledge of the far users' messages. Then, the near users act as relays and forward the prior information to the far users, thereby improving the reception reliability and reducing the outage probability for the far users. Many novel information-theoretic NOMA contributions have  been proposed. It was shown in  \cite{intro-CRS-NOMA-kim2015cap,coopNOMA2017-8108407,coNOMA-AF-eb2019} that a significant improvement in terms of the outage probability can be achieved, compared to the non-cooperative counterpart. The impact of user pairing on the outage probability and throughput was investigated in \cite{copNOMA-zhou2018dynamic}, where both random and distance-based pairing strategies were analyzed. To address the issue that the near users are energy-constrained, the energy harvesting technique was introduced into cooperative NOMA in \cite{copNOMA-liu2016cooperative}, where three user selection schemes were proposed and their performances were analyzed. 

Different  from the information-theoretic approach aforementioned, in this paper, we aim to uplift the performance of cooperative NOMA from the bit error rate (BER) perspective, and provide specific guidance to a practical system design.  Our further investigation indicates that the conventional cooperative NOMA suffers from three main limitations (detailed in Section \ref{sec:lim}). 
First, the conventional composite constellation design at the BS adopts a separate mapping rule. Based on a standard constellation such as quadrature amplitude modulation (QAM), bits are first mapped to user symbols, which in turn are mapped into a composite symbol using SC. This  results in a reduced minimum Euclidean distance. 
Second,  while forwarding the far user's signal, the near user does not dynamically design the corresponding constellation \cite{BER-coNOMA-kara2019,BER-coNOMA-kara2020,NOMA-ber-li2019spatial}, but only reuses the same far user constellation at the BS.  
Last,  the far user treats the near user's interference signal as additive white Gaussian noise (AWGN), which is usually not the case. Besides, it applies maximal-ratio combining (MRC)  for signal detection, which ignores the potential error propagation from the near user \cite{BER-coNOMA-kara2019}.

These limitations motivate us to develop a novel cooperative NOMA design referred to as deep cooperative NOMA. The essence lies in its  holistic approach, taking into account the three limitations simultaneously to perform an end-to-end multi-objective  joint optimization. 
However, this task is quite challenging, because it is intractable to transform the multiple objectives into explicit expressions, not to mention to optimize them simultaneously. To address this challenge, we leverage the interdisciplinary synergy from deep learning (DL)  \cite{book-goodfellow2016deep,review-DL-Pqin2019deep,intro-zhang2019deep,onoffline-he2019model,6G-letaief2019roadmap,ch-est-dong2019deep,DL-MIMO-he2020model,ours-lu2020deep}. We develop a novel  hybrid-cascaded deep neural network (DNN) architecture to represent the entire system, and construct multiple loss functions to quantify the BER performance. The DNN architecture consists of several structure-specific DNN modules, capable of tapping the strong capability of universal function approximation and integrating  the communication domain knowledge with combined analytical and data-driven modelling.  

The remaining task is how to train the proposed DNN architecture through learning the parameters of all the DNN modules in an efficient manner. To handle multiple loss functions, we propose a novel multi-task oriented training method with two stages. In stage I, we minimize the loss functions for the near user, and determine the mapping and demapping between the BS and the near user. In stage II,  by fixing the DNN modules learned in stage I, we minimize the loss function for the entire network, and determine the mapping and demapping for the near and far users, respectively. 
Both stages involve self-supervised training, utilizing the input training data as the class labels and thereby eliminating the need for human labeling effort. Instead of adopting the conventional symbol-wise training methods \cite{ae-o2017introduction,ae-jsac-aoudia2019model,deepNOMA-ye2020},  
we propose a novel bit-wise training method to obtain bit-wise soft probability outputs, facilitating the incorporation of channel coding and soft  decoding to combat signal deterioration. 

Then we examine the specific probability distribution that each DNN module has learned, abandoning the ``black-box of learning in DNN" \cite{DL-inte-9061001} and offering insights into the mechanism and the rationale behind the proposed DNN architecture and its corresponding training method. Besides, we  propose a solution to handle the power allocation (PA) mismatch between the training and inference processes to enhance the model adaptation. Our simulation results demonstrate that the proposed deep  cooperative NOMA significantly outperforms both OMA and the conventional cooperative NOMA in terms of the BER performance. Besides, the proposed scheme features a low computational complexity in both uncoded and coded cases.  

%
The main contributions can be summarized as follows.
\begin{itemize}
	\item We propose a novel deep cooperative NOMA scheme with  bit-wise soft probability outputs, where the entire system is re-designed by a hybrid-cascaded DNN architecture, such that it can be optimized in a holistic manner.
	\item By constructing multiple loss functions to quantify the BER performance, we propose a novel multi-task oriented two-stage training method to solve the end-to-end training problem in a self-supervised manner. 
	\item We carry out theoretical analysis based on information theory to reveal the learning mechanism of each DNN module. We also adapt the proposed scheme to handle the PA mismatch between training and inference, and incorporate it with channel coding.
	\item  Our simulation results demonstrate the superiority of the proposed scheme over OMA and the conventional cooperative NOMA  in various channel scenarios. 
\end{itemize}

The rest of this paper is organized as follows. In Section \Rmnum{2}, we introduce the cooperative NOMA system model and the limitations of the conventional scheme. In Section \Rmnum{3}, our deep cooperative NOMA and the multi-task learning problem is introduced, and the two-stage training method is presented, followed by the analysis of the bit-wise loss function. Section \Rmnum{4} provides the  theoretical perspective of the design principles. Section \Rmnum{5} discusses
the adaptation of the proposed scheme. Simulation results are shown in Section \Rmnum{6}. Finally, the conclusion is presented in Section \Rmnum{7}.

\textit{Notation}:  Bold lower case letters denote vectors.  $(\cdot)^T$ and  $(\cdot)^*$  denote the transpose and conjugate operations, respectively. $\diag(\bm{a})$ denotes a diagonal matrix whose diagonal entries starting in the upper left corner are $a_1, \dots, a_n$. $\mathbb{C}$ represents the set of complex numbers.   $\mathbb{E}[\cdot]$ denotes the expected value. $\bm{x}(r)$ denotes the $r$-th element of $\bm{x}$. Random variables are denoted by capital font, e.g., $X$ with the realization $x$. Multivariate random variables are represented by capital bold   font, e.g., ${\bf Y} = [Y_1, Y_2 ]^T$, ${\bf X}(r)$, with realizations $\bm{y} = [y_1, y_2 ]^T$, $\bm{x}(r)$, respectively.  $p(x,y)$, $p(y|x)$, and $I(X; Y )$  represent the  joint probability distribution, conditional probability distribution, and mutual information of the two random variables $X$ and $Y$. 
The cross-entropy of two discrete distributions $p(x)$ and $q(x)$ is denoted by $H( p(x), q(x)) = -\sum_{x} p(x) \log q(x)  $.

\section{Cooperative NOMA Communication System}
\subsection{System Model}
\label{sec:sys-model}
\begin{figure}[!t]
	\centering
	\includegraphics[width=0.9\textwidth]{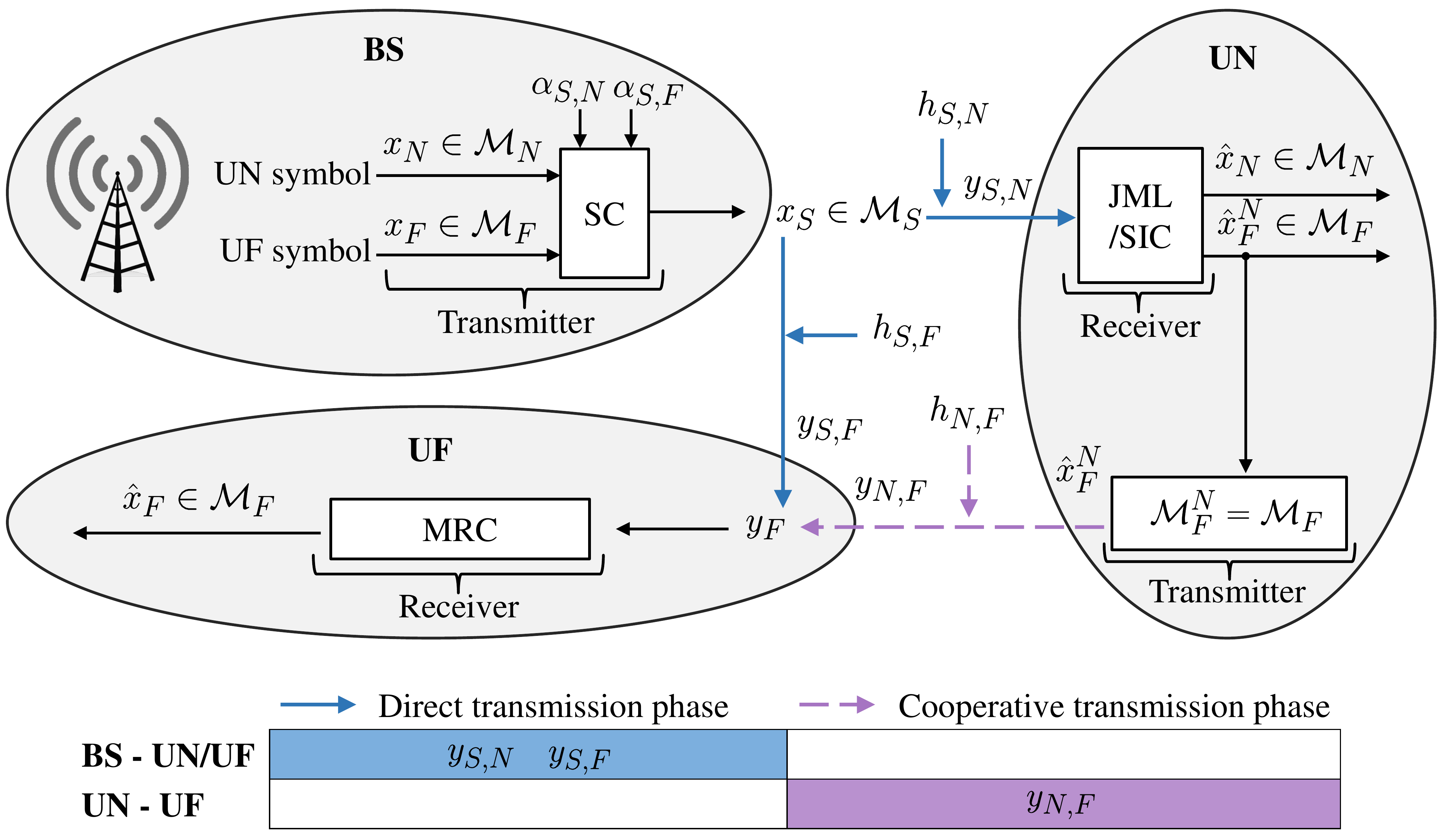}
	\caption{System model of the cooperative NOMA. UN can adopt the JML or SIC detector. }
	\label{fig:sys-coopeNOMA} 
\end{figure}

We consider a downlink cooperative NOMA system with a BS and two users (near
user UN and far user UF), as shown in Fig.~\ref{fig:sys-coopeNOMA}. The BS and users are assumed to be equipped with a single antenna. It is considered that only the statistical channel state information (CSI), such as the average channel gains, are available at the BS, the instantaneous CSI of the BS to UN link is available at UN, and the instantaneous CSI of the BS/UN to UF links are available at UF. UN and UF are classified according to their statistical CSI. Typically, they have better and worse channel conditions, respectively. Correspondingly, UN acts as a decode-and-forward (DF) relay and assists the signal transmission to UF. The complete signal transmission consists of two phases, described as follows. In the direct transmission phase, the BS transmits the composite signal to both users. In the cooperative transmission phase, UN performs joint detection, and then forwards the re-modulated UF signal to UF. 

Let $\bm{s}_{N} \in \{ 0,1 \}^{k_{N}}$ and $ \bm{s}_{F} \in  \{ 0,1 \}^{k_{F}}$ denote the transmitted bit blocks for UN and UF, with lengths $k_{N}$ and $k_{F}$, respectively. $\bm{s}_{N}$ and $\bm{s}_{F}$ are mapped to user symbols $x_{N}$ and $x_{F}$, taking from $ M_{N}$- and $M_{F}$-ary unit-power  constellations $\mathcal{M}_{N} \subset \mathbb{C}$ and $\mathcal{M}_{F} \subset \mathbb{C}$, respectively, where $2^{k_{N}} = M_{N}$ and $2^{k_{F}} = M_{F}$. The detailed transmission process is as follows. 

In the direct transmission phase, the BS uses SC to obtain a composite symbol 
\begin{align}
x_{S} = \sqrt{{\alpha}_{S,N}} x_{N} + \sqrt{{\alpha}_{S,F}} x_{F}, \  x_{S} \in \mathcal{M}_{S} \subset \mathbb{C}  \label{eq:xS-comp}
\end{align}
and transmits $x_{S}$ to the two users, where ${\alpha}_{S,N}$ and  ${\alpha}_{S,F}$  are the PA coefficients with ${\alpha}_{S,N} < {\alpha}_{S,F}$ and ${\alpha}_{S,N} + {\alpha}_{S,F}=1$. $\mathcal{M}_{S}$ is called the composite constellation, and can be written as the sumset $\mathcal{M}_{S} = \sqrt{{\alpha}_{S,N} }\mathcal{M}_{N} + \sqrt{{\alpha}_{S,F}} \mathcal{M}_{F} \triangleq \{ \sqrt{{\alpha}_{S,N} } t_{N} + \sqrt{{\alpha}_{S,F} } t_{F} : t_{N} \in \mathcal{M}_{N},  t_{F} \in \mathcal{M}_{F} \}$. The received signal at the users can be expressed as  
\begin{align}
y_{S,J} =  \sqrt{P_{S}}   h_{S,J} (\sqrt{{\alpha}_{S,N}} x_{N} + \sqrt{{\alpha}_{S,F}} x_{F} ) + n_{S,J},  \ J \in \{ N,F \},  \label{eq:ySFN}
\end{align}
where $P_{S}$ is the transmit power of the BS, $n_{S,J} \sim \mathcal{CN}(0,2 \sigma^2_{S,J} )$ denotes the i.i.d complex AWGN, and $h_{S,J}$ denotes the fading channel coefficient. We define the transmit signal-to-noise ratio as SNR$= \frac{P_S}{2\sigma_{S,F}^2} $.  After receiving $y_{S,N}$, UN performs JML detection\footnote{Here we introduce JML as an example. Note that SIC can also be used.} given by 
\begin{align}
\label{eq:N-JML}
(	\hat{x}_{N}, \hat{x}_{F}^{N}) = \arg\min_{(x_{N},  x_{F}) \in  \{ \mathcal{M}_{N} \times \mathcal{M}_{F} \} }  & \ \Big\vert y_{S,N} -\sqrt{P_{S}}    h_{S,N} (\sqrt{{\alpha}_{S,N}}  x_{N} +   \sqrt{{\alpha}_{S,F}} x_{F}) \Big\vert^2,
\end{align}
where $\hat{x}_{N}$ denotes the estimate of $x_{N}$ and $\hat{x}_{F}^{N}$ denotes the estimate of $x_{F}$ at UN. The  corresponding estimated user bits $(\hat{\bm{s}}_{N}, \hat{\bm{s}}_{F}^{N}) \in (\{ 0,1 \}^{k_{N}}, \{ 0,1 \}^{k_{F}})$ can be demapped from $(	\hat{x}_{N}, \hat{x}_{F}^{N})$.

In the cooperative transmission phase, UN transmits the re-modulated signal $\hat{x}_{F}^{N}$ to UF with $\hat{x}_{F}^{N} \in \mathcal{M}_{F}^{N} = \mathcal{M}_{F}$. The received signal at UF can be written as  
\begin{align}
y_{N,F} =  \sqrt{P_{N}}   h_{N,F} \hat{x}_{F}^{N} + n_{N,F}, \label{eq:yNF}
\end{align}
where $P_{N}$ is the transmit power of UN, $n_{N,F} \sim \mathcal{CN}(0,2 \sigma^2_{N,F} )$ denotes the AWGN, and $h_{N,F}$ denotes the channel fading coefficient.

The entire transmission for UF can be considered as a cooperative transmission with a DF relay, i.e., UN.  As UF has the knowledge of $h_{S,F}$ and $h_{N,F}$, by treating  the interference term $\sqrt{{\alpha}_{S,N}}x_{N}$ in $y_{S,F}$  as AWGN and leveraging the widely used MRC \cite{NOMA-coop-xu2016novel,coopNOMA2017-8108407,NOMA-ber-li2019spatial}, UF first combines $y_{S,F}$ and  $y_{N,F}$ as 
\begin{equation}
y_{F}  = \beta_{S,F}  y_{S,F} + \beta_{N,F}  y_{N,F},
\end{equation}
where $\beta_{S,F}  = \frac{\sqrt{P_{S}{\alpha}_{S,F}} h_{S,F}^{*} }{P_{S}{\alpha}_{S,N} |h_{S,F} |^2 + 2\sigma_{S,F}^2 }  $ and $\beta_{N,F}  = \frac{\sqrt{P_{N}}  h_{N,F}^{*} }{2\sigma_{N,F}^2}$ \cite{NOMA-ber-li2019spatial}.
Then, UF detects its own symbol $x_{F}$ from $y_{F}$ as
\begin{align}
\label{eq:MRC-F}
\hat{x}_{F} = \arg\min_{x_{F} \in \mathcal{M}_{F}} & \ \Big| y_{F}  - \big( \beta_{S,F} \sqrt{P_{S}{\alpha}_{S,F}} h_{S,F} + \beta_{N,F} \sqrt{P_{N}} h_{N,F}
 \big) x_{F}  \Big|^2.
\end{align}
The corresponding estimated bits $\hat{\bm{s}}_{F} \in \{ 0,1 \}^{k_{F}}$ can be demapped from $\hat{x}_{F}$. 

Hereafter, for convenience, we denote the bit to composite symbol mappings at the BS and UN as $f_{S}$ and $f_{N}$, respectively, and denote the demappings at UN and UF as $g_{N}$ and $g_{F}$, respectively. They are defined as 
\begin{align}
f_{S}: & \ (\{ 0,1 \}^{k_{N}}, \{ 0,1 \}^{k_{F}}) \to \mathcal{M}_{S} \subset \mathbb{C}, \label{eq:fS} \\
f_{N}: & \ \hat{\bm{s}}_{F}^{N} \to  \mathcal{M}_{F}^{N} \subset \mathbb{C}, \label{eq:fN}
\end{align}
and
\begin{align}
g_{N}: & \ y_{S,N} \to (\hat{\bm{s}}_{N}, \hat{\bm{s}}_{F}^{N}) \in (\{ 0,1 \}^{k_{N}}, \{ 0,1 \}^{k_{F}}), \label{eq:DetN} \\
g_{F}: & \  (y_{S,F}, y_{N,F}) \to \hat{\bm{s}}_{F} \in \{ 0,1 \}^{k_{F}} . \label{eq:DetF} 
\end{align}

The average symbol error rate (SER) and BER are respectively  denoted as $\mathcal{P}_{N,e_s}$ and $\mathcal{P}_{N,e_b}$ for UN to detect the UN signal, as $\mathcal{P}_{F,e_s}^{N}$ and $\mathcal{P}_{F,e_b}^{N}$ for UN to detect the UF signal, and as $\mathcal{P}_{F,e_s}$ and $\mathcal{P}_{F,e_b}$ at UF. They are defined as $\mathcal{P}_{N,e_s} = \mathbb{E}_{ {x}_{N} } \big[  \Pr \{ {x}_{N} \neq  \hat{x}_{N} \} \big]$, $\mathcal{P}_{N,e_b} =  \mathbb{E}_{ \bm{s}_{N} } \big[  \Pr \{ \bm{s}_{N} \neq  \hat{\bm{s}}_{N} \} \big]$,   $\mathcal{P}_{F,e_s}^{N} =  \mathbb{E}_{ {x}_{F}  } \big[  \Pr \{ {x}_{F} \neq  \hat{x}_{F}^{N} \} \big]$, $\mathcal{P}_{F,e_b}^{N} =  \mathbb{E}_{ \bm{s}_{F} } \big[  \Pr \{ \bm{s}_{F} \neq  \hat{\bm{s}}_{F}^{N} \} \big]$, $\mathcal{P}_{F,e_s} =  \mathbb{E}_{ {x}_{F}  } \big[  \Pr \{ {x}_{F} \neq  \hat{x}_{F} \} \big]$, and $\mathcal{P}_{F,e_b} =  \mathbb{E}_{ \bm{s}_{F} } \big[  \Pr \{ \bm{s}_{F} \neq  \hat{\bm{s}}_{F} \} \big]$. Note that SER and BER are functions of the constellation mappings (i.e., $f_{S}$ and $f_{N}$) and demappings (i.e., $g_{N}$ and $g_{F}$).  For a given design problem, the parameters $ \{  k_{N}, k_{F}, {\alpha}_{S,N},  {\alpha}_{S,F} \} $ are fixed and we let $P_S=P_N=1$.

\subsection{Limitation}
\label{sec:lim}
The system design above has been widely adopted in the literature \cite{intro-CRS-NOMA-kim2015cap,NOMA-coop-xu2016novel,coopNOMA2017-8108407, NOMA-ber-li2019spatial}.  In the following, we specify its three main limitations ({\bf L1})-({\bf L3}), which serve as the underlying motivation for a new system design in Section \ref{sec:proposed-CNOMA}. 

\begin{figure} [!t] 
	\centering
	\subfigure[For $x_{N} \in \mathcal{M}_{N}$,  $x_{F} \in \mathcal{M}_{F}$, and  $\hat{x}_{F}^{N} \in \mathcal{M}_{F}^{N}$ (all QPSK)]{
		\label{fig:cons-xN}
		\includegraphics[width=0.46\textwidth]{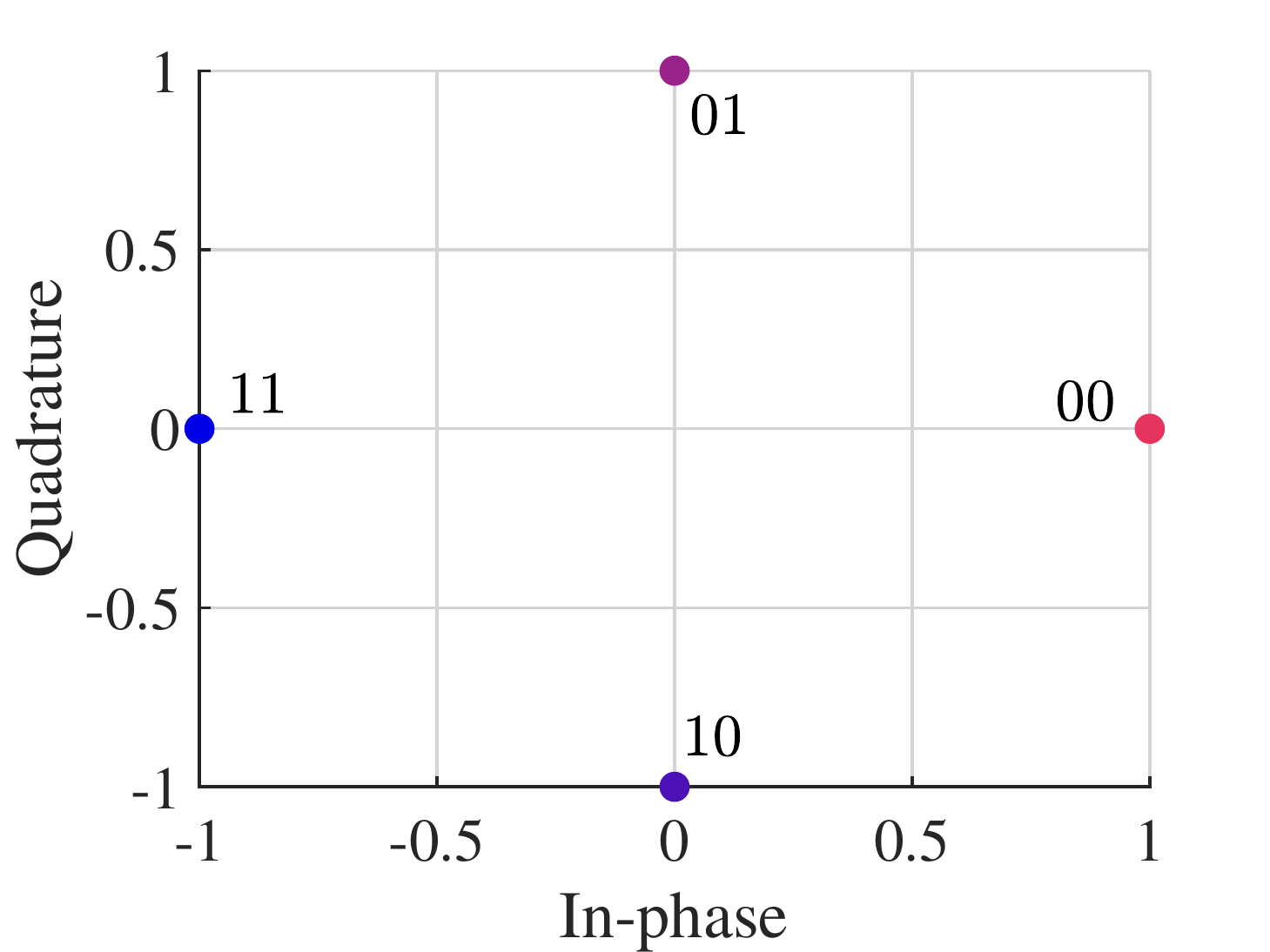}}
	\subfigure[For $x_{S} \in \mathcal{M}_{S}$ (composite constellation)]{
		\label{fig:cons-xS}
		\includegraphics[width=0.46\textwidth]{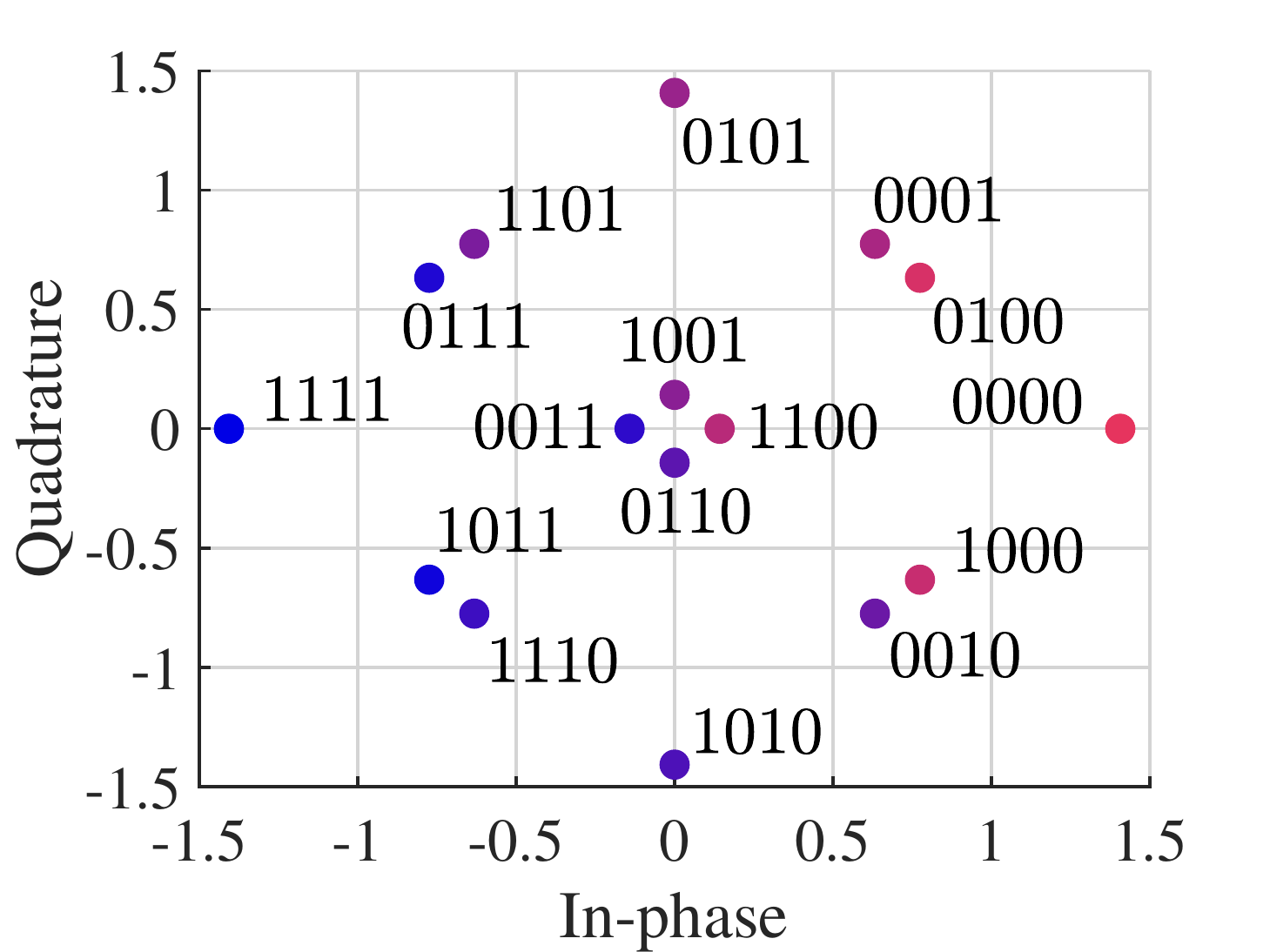}}	
	\caption{Conventional constellations for $M_{N}=M_{F}=4$ and $( {\alpha}_{S,N}, {\alpha}_{S,F} ) = ( 0.4, 0.6 )$.} 
	\label{fig:cons-conven-M4-S1}
\end{figure}

({\bf L1}) \textbf{Bit Mapping at the BS:}
From the signal detection perspective, the conventional mapping from bit to composite symbol (c.f. \eqref{eq:fS}) uses a separate mapping: first $ (\{ 0,1 \}^{k_{N}}, \{ 0,1 \}^{k_{F}}) \to ( 	\mathcal{M}_{N},\mathcal{M}_{F} ) $, and then $ ( 	\mathcal{M}_{N},\mathcal{M}_{F} ) \to \mathcal{M}_{S}$. Typically,  we can adopt Gray mapping for $ \{ 0,1 \}^{k_{N}} \to \mathcal{M}_{N}$ and $ \{ 0,1 \}^{k_{F}} \to \mathcal{M}_{F} $, while  $\mathcal{M}_{N}$ and $\mathcal{M}_{F}$ are chosen from the standard constellations, e.g.,  QAM. Then, for designing $f_{S} $ in \eqref{eq:fS}, only $ ( 	\mathcal{M}_{N},\mathcal{M}_{F} ) \to \mathcal{M}_{S}$ needs to be optimized  as follows 
\begin{IEEEeqnarray}{rCl} 
	& \min_{ \substack{( \mathcal{M}_{N}, \ \mathcal{M}_{F} ) \to \mathcal{M}_{S} \subset \mathbb{C} }} & \quad  \Big\{ \mathcal{P}_{N,e_s} (f_{S}, g_{N}), \ \mathcal{P}_{F,e_s}^{N}  (f_{S}, g_{N}) \Big\} \label{eq:L1}
	\\
	&  \subto	& \quad 	 \mbox{predefined condition},  \notag
\end{IEEEeqnarray}
where $ g_{N}$ here is the JML detector in  \eqref{eq:N-JML}, $ \mathcal{P}_{N,e_s} (f_{S}, g_{N})$ and $\mathcal{P}_{F,e_s}^{N}  (f_{S}, g_{N})$ characterize the SERs associated with  \eqref{eq:N-JML}, and for example, the predefined condition can be  the constellation rotation in  \cite{NOMA-cons-ye2017constellation}. 
Clearly, this disjoint design is suboptimal, resulting in a degraded error performance. For example, in Fig.~\ref{fig:cons-xN},  $x_{N}$ and $x_{F}$ are QPSK symbols with Gray mapping. Accordingly, in Fig.~\ref{fig:cons-xS},  $x_{S}$ is the composite symbol for $( {\alpha}_{S,N}, {\alpha}_{S,F} ) = ( 0.4, 0.6 )$. It can be clearly seen that at the symbol level, the composite constellation $\mathcal{M}_{S}$ for $x_{S}$ results in a very small minimum Euclidean distance. Furthermore, a close look at  $\mathcal{M}_{S}$ reveals that, at the bit level, the mapping $ (\{ 0,1 \}^{k_{N}}, \{ 0,1 \}^{k_{F}}) \to \mathcal{M}_{S}$ is not optimized. 

({\bf L2}) \textbf{Constellation at UN:}
In the cooperative NOMA system, UN acts as a DF relay: first detects ${x}_{F}$ (or equivalently, $\bm{s}_{F}$), and then forwards the re-modulated signal $\hat{x}_{F}^{N}$ to UF. 
Here, $\mathcal{M}_{F}^{N}$ is assumed in the literature to be exactly the same as the UF constellation $\mathcal{M}_{F}$ at the BS. 
Clearly, such design for UF may not be optimal because  (1) detection errors may occur at UN; (2) UF receives the signals not only from UN, but also from the BS ($y_{S,F}$ including non-AWGN interference).
In this case, $\mathcal{M}_{F}^{N}$ should be further designed, rather than simply let $\mathcal{M}_{F}^{N}=\mathcal{M}_{F}$ (known as repetition coding \cite{intro-AF-classic}). 

({\bf L3}) \textbf{Detection at UF:}
In practice, MRC is widely adopted as it only needs $h_{S,F}$ and $h_{N,F}$. Its design principle can be written as
\begin{IEEEeqnarray}{rCl} 
	& \min_{  g_{F} } & \quad   \mathcal{P}_{F,e_s} ( f_{S}, f_{N},   g_{N},   g_{F} ) \label{eq:L3}
	\\
	&  \subto	& \quad  \mathcal{M}_{F}^{N} =  \mathcal{M}_{F} ,  \notag \\
	 & & \quad \hat{x}_{F}^{N} =  x_{F}  , \notag 
\end{IEEEeqnarray}
where $ \mathcal{P}_{F,e_s} ( f_{S}, f_{N},   g_{N},   g_{F} )$ characterizes the SER associated with \eqref{eq:MRC-F}, $f_{S}$ and $ g_{N}$ are given, and $\mathcal{M}_{F}^{N} =  \mathcal{M}_{F}$ is for $f_{N}$. However, it is sub-optimal due to the potential signal detection error at UN (i.e., $\hat{x}_{F}^{N}\neq x_{F}$) \cite{BER-coNOMA-kara2019} and the ideal assumption in \eqref{eq:MRC-F} that the interference term $ \sqrt{{\alpha}_{S,N}}x_{N}$ in $y_{S,F}$ is AWGN. 
\section{The Proposed Deep Cooperative NOMA Scheme}
\label{sec:proposed-CNOMA}
\subsection{Motivation}	%
\label{sec:moti}
To overcome ({\bf L1}), a desirable approach is to solve the following problem
\begin{IEEEeqnarray}{rCl} 
	%
	& \min_{ f_{S} }   & \quad  \Big\{ \mathcal{P}_{N,e_b} (f_{S}, g_{N}), \ \mathcal{P}_{F,e_b}^{N} (f_{S}, g_{N}) \Big\} \label{eq:Q1}
\end{IEEEeqnarray}
with given $ g_{N}$. 
That is, we use BER as the performance metric, and directly optimize the mapping $ f_S: (\{ 0,1 \}^{k_{N}}, \{ 0,1 \}^{k_{F}}) \to \mathcal{M}_{S} \subset \mathbb{C}$. 	
To handle ({\bf L2}) and minimize the end-to-end BER $\mathcal{P}_{F,e_b} ( f_{S}, f_{N},   g_{N},   g_{F} )$, the constellation $\mathcal{M}_{F}^{N}$ in $ f_{N}$ should be designed by solving the following problem
\begin{IEEEeqnarray}{rCl} 
	&
	\min_{ \substack{  f_{N} } } & \quad   \mathcal{P}_{F,e_b} ( f_{S}, f_{N},   g_{N},   g_{F} ) \label{eq:Q2} 
\end{IEEEeqnarray}  
with  given $f_{S}$, $ g_{N}$, and $g_{F}$.
To handle ({\bf L3}), the optimization problem can be re-designed as
\begin{align}
%
\min_{ g_{F} }  & \quad   \mathcal{P}_{F,e_b} ( f_{S}, f_{N},   g_{N},   g_{F} )  \label{eq:Q3} 
\end{align}
with  given $f_{S}$, $ f_{N}$, and $g_{N}$, where the ideal assumptions in \eqref{eq:L3}, i.e., $\hat{x}_{F}^{N} =  x_{F}$ and $\sqrt{{\alpha}_{S,N}}x_{N}$ is AWGN, are removed.

However, addressing ({\bf L1})-({\bf L3}) separately is suboptimal due to the disjoint nature of the mapping and demapping design. This motivates us to take a holistic approach,  taking into account ({\bf L1})-({\bf L3}) simultaneously to perform an end-to-end multi-objective optimization as 
\begin{align}
\raisebox{-0.0\normalbaselineskip}[0pt][0pt]{%
	({\bf P1})}  \qquad 
\min_{ \substack{ f_{S}, \ f_{N}, \ g_{N}, \ g_{F} } }  & \quad \Big\{ \mathcal{P}_{N,e_b} ( f_{S},  g_{N} )  , \ \mathcal{P}_{F,e_b}^{N} ( f_{S},  g_{N} ) , \   \mathcal{P}_{F,e_b} ( f_{S}, f_{N},   g_{N},   g_{F} )  \Big\} . \notag 
\end{align}
Clearly, ({\bf P1}) represents a joint  $\big \{ f_{S}, f_{N}, g_{N},  g_{F} \big\}$ design  for all objectives in \eqref{eq:Q1}-\eqref{eq:Q3}. 
\begin{center}
	\fbox{%
		\parbox{0.98\textwidth}{%
			\textbf{Challenge 1:} It is very challenging to find the solutions for  ({\bf P1}), because it is difficult to transform the objective functions $\big\{ \mathcal{P}_{N,e_b} ( f_{S},  g_{N} ) , \mathcal{P}_{F,e_b}^{N} ( f_{S},  g_{N} ) ,   \mathcal{P}_{F,e_b} ( f_{S}, f_{N},   g_{N},   g_{F} ) \big\}$ and optimization variables $\big \{ f_{S}, f_{N}, g_{N},g_{F} \big\}$  into explicit expressions.
		}%
	}
\end{center}
\begin{center}
	\fbox{%
		\parbox{0.98\textwidth}{
			\textbf{Challenge 2:} Moreover, the three objectives correspond to different users' BER and may be  mutually conflicting \cite{deepNOMA-ye2020}. So it is very difficult to minimize them simultaneously \cite{book-deb2014}.
		}%
	}
\end{center}

To overcome these challenges, we propose a novel deep multi-task oriented learning scheme from a combined model- and  data-driven perspective. Specifically, by tapping the strong nonlinear mapping and demapping capability of DNN (universal function approximation), we first express $ \big\{ f_{S}, f_{N}, g_{N},  g_{F} \big\}$ by constructing a hybrid-cascaded DNN architecture and then transfer $\big\{ \mathcal{P}_{N,e_b} ( f_{S},  g_{N} ) , \\ \mathcal{P}_{F,e_b}^{N} ( f_{S},  g_{N} ) ,   \mathcal{P}_{F,e_b} ( f_{S}, f_{N},   g_{N},   g_{F} ) \big\}$ using the bit-level loss functions, so that they can be evaluated empirically. Then, we develop a multi-task oriented two-stage training method to  minimize the  loss functions  through optimizing the DNN parameters in a self-supervised manner. Thereby the input training data also serve as the class labels. 


\subsection{Deep Cooperative NOMA} \label{sec:AE-CoopNOMA}	 
\begin{figure*}[!t]
	\centering
	\includegraphics[width=1\textwidth]{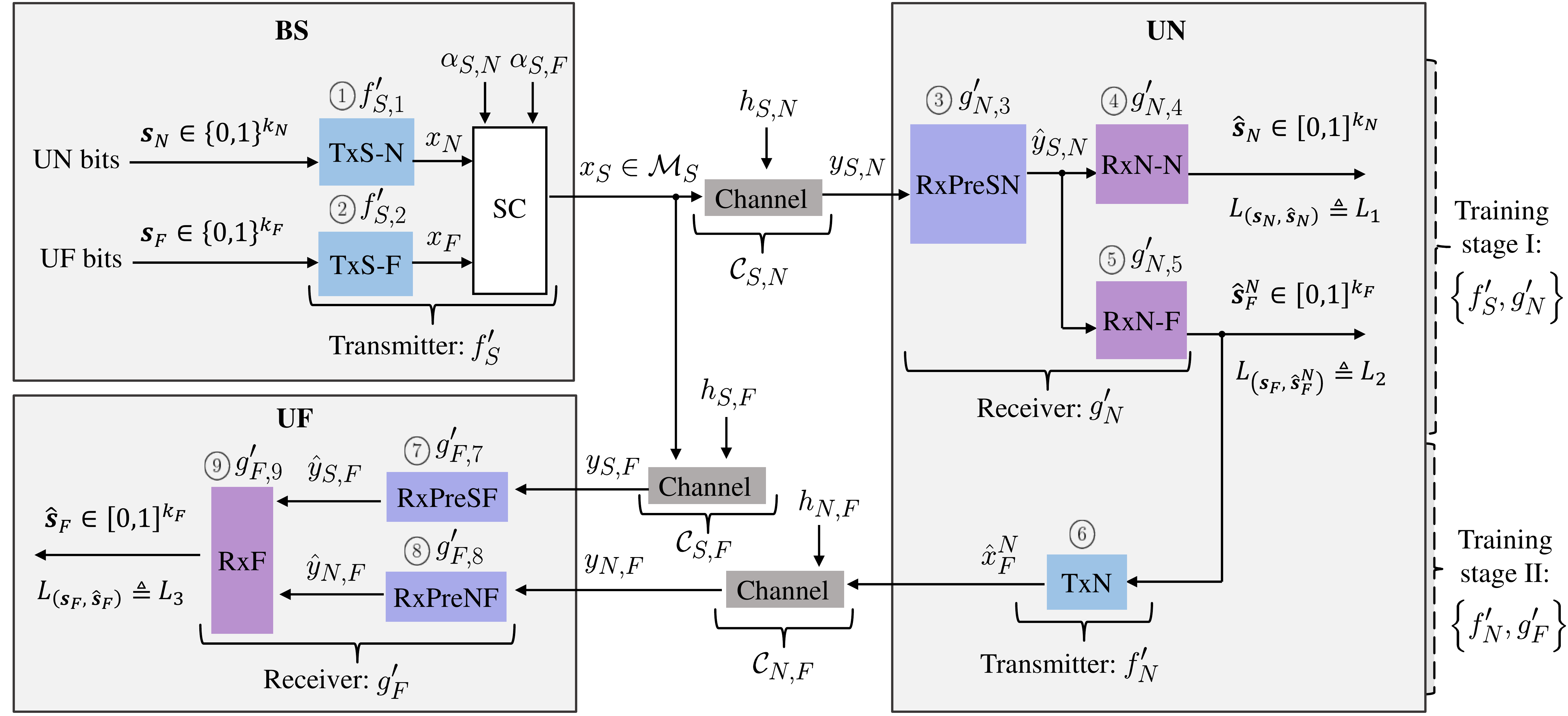}
	\caption{Block diagram of  the proposed deep cooperative NOMA including nine trainable DNN modules $\textcircled{\scriptsize 1}$-$\textcircled{\scriptsize 9}$, where  $\textcircled{\scriptsize 1}$, $\textcircled{\scriptsize 2}$, and $\textcircled{\scriptsize 6}$ are mapping modules, while the remaining are demapping modules.   The inputs $\{ \bm{s}_{N}, \bm{s}_{F} \}$ are bits, and the outputs $\{ \hat{\bm{s} }_{N}, \hat{\bm{s}}_{F}^{N}, \hat{\bm{s}}_{F}\}$ are bit-wise soft probabilities from sigmoid function, e.g., $\hat{\bm{s}}_{F}=[0.96, 0.02]$. The corresponding loss functions are $L_1$ and  $L_2$ for UN,  and $L_3$ for UF.
	}
	\label{fig:AE-NOMA}
\end{figure*}

The block diagram of the proposed deep cooperative NOMA is shown in Fig.~\ref{fig:AE-NOMA},  
where the entire system (c.f. Fig.~1) is re-designed as a novel hybrid-cascaded DNN architecture including nine trainable DNN modules, i.e., three mapping modules and six demapping modules. In essence, the whole DNN architecture learns the mapping between the BS inputs and users outputs to combat the channel fading and noise. Each DNN module consists of multiple hidden layers describing its input-output mapping, including the learnable parameters, i.e., weights and biases. Here, we adopt the offline-training and online-deploying mode in DL. This means that all the DNN modules are deployed without retraining after initial training.

At the BS, we propose to use two parallel DNN mapping modules ($\textcircled{\scriptsize 1}$TxS-N and $\textcircled{\scriptsize 2}$TxS-F) with an SC operation to represent the direct mapping $f_{S}$ in \eqref{eq:fS}, which is hereafter  referred to as  $f_{S}^{\prime} : \{ f_{S,1}^{\prime}, f_{S,2}^{\prime} \}$, denoting the mapping parameterized by the associated DNN parameters. Note that $f_{S,1}^{\prime}$ and $f_{S,2}^{\prime}$ are for $\textcircled{\scriptsize 1}$TxS-N and $\textcircled{\scriptsize 2}$TxS-F, respectively. Their outputs $x_{N}$ and $x_{F}$ are normalized to ensure  $\mathbb{E}\{ | x_{N}|^2\} = 1$ and  $\mathbb{E}\{ | x_{F}|^2\} = 1$.  The composite symbol (c.f. \eqref{eq:xS-comp}) now can be re-expressed by $x_{S} = f_{S}^{\prime}(\bm{s}_{N}, \bm{s}_{F})$. 
In the direct transmission phase, the received signal at the users can be expressed as  
\begin{align}
y_{S,J} =  h_{S,J} f_{S}^{\prime}(\bm{s}_{N}, \bm{s}_{F} ) + n_{S,J}, \  J \in \{ N,F \} . \label{eq:ySFN-DNN}
\end{align}




At UN, we use three DNN demapping modules ($\textcircled{\scriptsize 3}$RxPreSN, $\textcircled{\scriptsize 4}$RxN-N, and $\textcircled{\scriptsize 5}$RxN-F) to represent the demapping in  \eqref{eq:DetN}, referred to as $g_{N}^{\prime} : \{ g_{N,3}^{\prime}, g_{N,4}^{\prime}, g_{N,5}^{\prime} \}$. Note that $g_{N,3}^{\prime}$, $g_{N,4}^{\prime}$, and $g_{N,5}^{\prime}$ are for $\textcircled{\scriptsize 3}$RxPreSN, $\textcircled{\scriptsize 4}$RxN-N, and $\textcircled{\scriptsize 5}$RxN-F, respectively.  The received $y_{S,N}$ is equalized as  $\frac{h_{S,N}^*y_{S,N}}{|h_{S,N}|^2}$, processed by $\textcircled{\scriptsize 3}$RxPreSN, and then demapped by two parallel DNNs ($\textcircled{\scriptsize 4}$RxN-N and $\textcircled{\scriptsize 5}$RxN-F) to obtain the estimates $\hat{\bm{s} }_{N}$ and $\hat{\bm{s}}_{F}^{N}$, respectively.	This process can be expressed as  
\begin{align}
(\hat{\bm{s} }_{N}, \hat{\bm{s}}_{F}^{N})  =  g_{N}^{\prime} (y_{S,N}) \in \big( [0,1 ]^{k_{N}}, [ 0,1 ]^{k_{F}} \big) , \label{gN-DNN}
\end{align}
where $(\hat{\bm{s} }_{N}, \hat{\bm{s}}_{F}^{N}) $ are soft probabilities for each element in the vectors. Integrating \eqref{eq:ySFN-DNN}-\eqref{gN-DNN}, this demapping process at UN can be described as
\begin{align}
\underbrace{(\hat{\bm{s}}_{N}, \hat{\bm{s}}_{F}^{N} ) = g_{N}^{\prime}  }_{\eqref{gN-DNN}} \circ \underbrace{ \mathcal{C}_{S,N} \circ f_{S}^{\prime} (\bm{s}_{N}, \bm{s}_{F})}_{\eqref{eq:ySFN-DNN} \text{ with  } J=N}, \label{eq:det-1}
\end{align}
where $\circ$ is the composition operator and $\mathcal{C}_{S,N} \triangleq \mathcal{C}_{S,N}(y_{S,N}\vert x_{S},h_{S,N})$ denotes the  channel function from the BS to UN. We refer to \eqref{eq:det-1} as the
first demapping phase.

After obtaining $\hat{\bm{s}}_{F}^{N}$, we use the DNN mapping module $\textcircled{\scriptsize 6}$TxN to represent the mapping in 
\eqref{eq:fN}, denoted as $\hat{x}_{F}^{N} = f_{N}^{\prime}(\hat{\bm{s}}_{F}^{N})$, where  $f_{N}^{\prime} = f_{N,6}^{\prime}$. A normalization layer is used at the last layer of $\textcircled{\scriptsize 6}$TxN to ensure  $\mathbb{E}\{ | \hat{x}_{F}^{N}|^2\} = 1$. In the cooperative transmission phase, UF receives
\begin{align}
y_{N,F} =  h_{N,F} f_{N}^{\prime}(\hat{\bm{s}}_{F}^{N} ) + n_{N,F}. \label{eq:yNF-DNN}
\end{align}

Finally at UF, we use three DNN demapping modules ($\textcircled{\scriptsize 7}$RxPreSF, $\textcircled{\scriptsize 8}$RxPreNF, and  $\textcircled{\scriptsize 9}$RxF) to represent the demapping in \eqref{eq:DetF} as $g_{F}^{\prime} :\{ g_{F,7}^{\prime}, g_{F,8}^{\prime}, g_{F,9}^{\prime} \} $. Note that $g_{F,7}^{\prime}$, $g_{F,8}^{\prime}$, and $g_{F,9}^{\prime}$ are for $\textcircled{\scriptsize 7}$RxPreSF, $\textcircled{\scriptsize 8}$RxPreNF, and  $\textcircled{\scriptsize 9}$RxF, respectively. The received $y_{S,F}$ and $y_{N,F}$ are equalized as $\frac{h_{S,F}^*y_{S,F}}{|h_{S,F}|^2}$ and $\frac{h_{N,F}^*y_{N,F}}{|h_{N,F}|^2}$, processed by the parallel  $\textcircled{\scriptsize 7}$RxPreSF and $\textcircled{\scriptsize 8}$RxPreNF, respectively, and then fed into $\textcircled{\scriptsize 9}$RxF to obtain $\hat{\bm{s} }_{F}$.  This process can be described as  
\begin{align}
\hat{\bm{s}}_{F} =  g_{F}^{\prime} (y_{S,F}, y_{N,F}) \in [ 0,1 ]^{k_{F}} . \label{gF-DNN}
\end{align}
Note that the soft probability output $\hat{\bm{s}}_{F}$ can serve as the input of a soft channel decoder, which will be explained in Section \ref{sec:c-coding}. Integrating \eqref{eq:ySFN-DNN}-\eqref{gF-DNN}, the end-to-end  demapping process at UF  can be described as
\begin{align}
\underbrace{\hat{\bm{s}}_{F} =   g_{F}^{\prime}}_{\eqref{gF-DNN}}  \big( & \underbrace{\mathcal{C}_{S,F} \circ f_{S}^{\prime}  (\bm{s}_{N}, \bm{s}_{F})}_{\eqref{eq:ySFN-DNN} \text{ with  } J=F} ,  \underbrace{ \mathcal{C}_{N,F} \circ  f_{N}^{\prime}}_{\eqref{eq:yNF-DNN}} \circ \underbrace{g_{N}^{\prime} \circ \mathcal{C}_{S,N} \circ f_{S}^{\prime}  (\bm{s}_{N}, \bm{s}_{F})}_{\eqref{eq:det-1}}  \big), \label{eq:det-2}
\end{align}
where $\mathcal{C}_{S,F} \triangleq \mathcal{C}_{S,F}(y_{S,F}\vert x_{S},h_{S,F})$ and $\mathcal{C}_{N,F} \triangleq \mathcal{C}_{N,F} (y_{N,F}\vert \hat{x}_{F}^{N},h_{N,F})$ denote the  channel functions from the BS and UN to UF, respectively. We refer to \eqref{eq:det-2} as the second demapping phase. 
\begin{figure} [!t] 
	\centering
	\subfigure[For  $ \text{Tx} \in \{ \textcircled{\scriptsize 1}\text{TxS-N},  \textcircled{\scriptsize 2}\text{TxS-F}, \textcircled{\scriptsize 6}\text{TxN} \}$ and $ \text{Rx} \in \{ \textcircled{\scriptsize 4}\text{RxN-N},  \textcircled{\scriptsize 5}\text{RxN-F}, \textcircled{\scriptsize 9}\text{RxF} \}$]{
		\label{fig:TxRx}
		\includegraphics[width=0.48\textwidth]{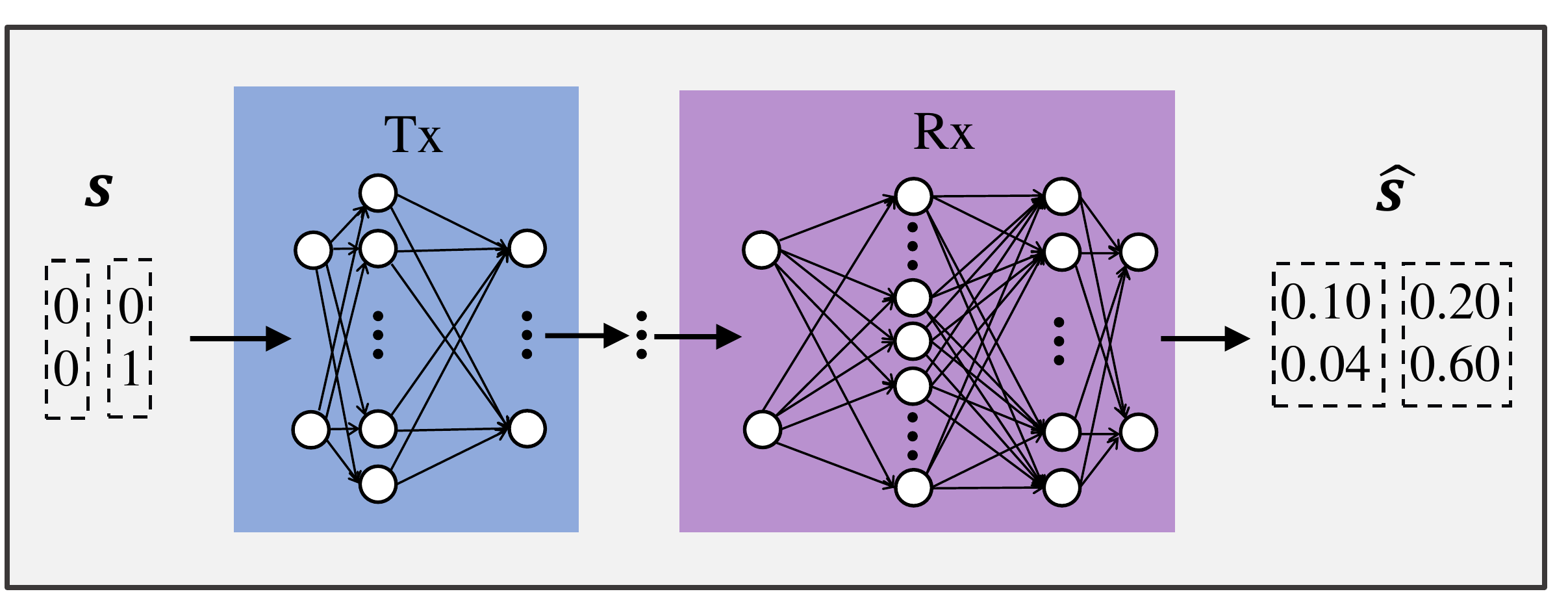}}
	\subfigure[For $ \text{RxPre} \in \{ \textcircled{\scriptsize 3}\text{RxPreSN},  \textcircled{\scriptsize 7}\text{RxPreSF}, \textcircled{\scriptsize 8}\text{RxPreNF} \}$]{
		\label{fig:RxPre}
		\includegraphics[width=0.48\textwidth]{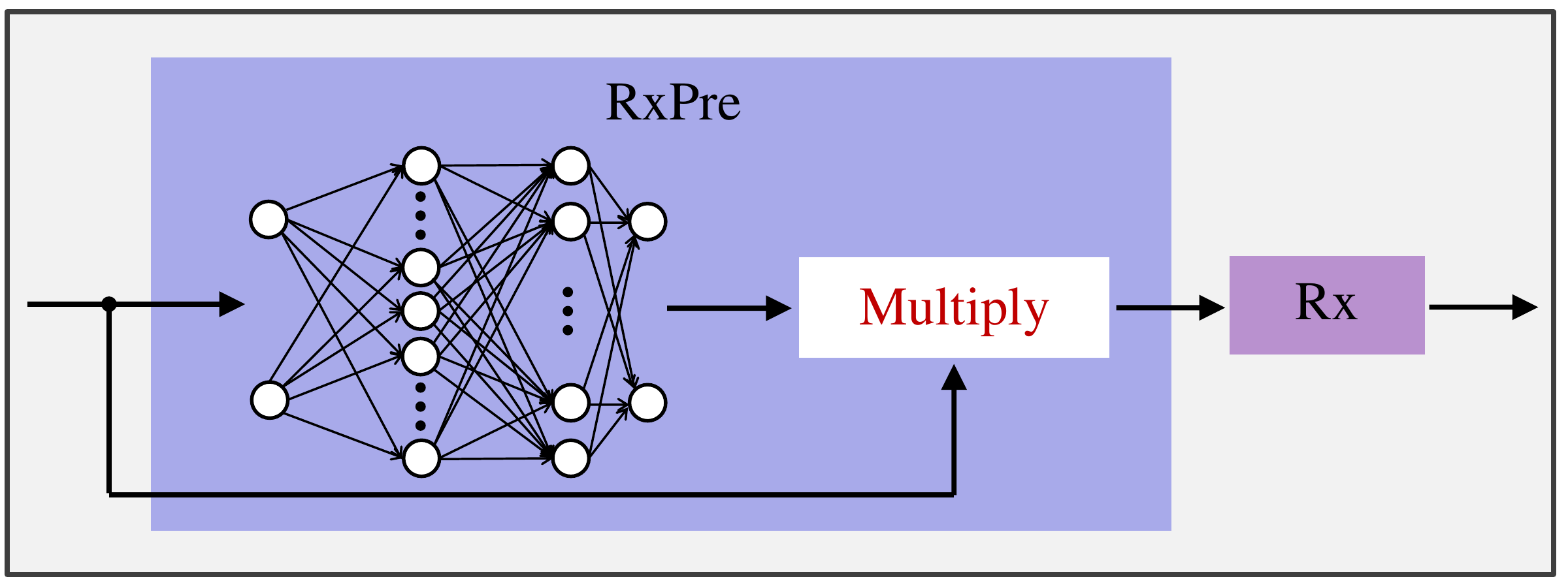}}	
	\caption{Block diagram of the layer structure for the DNN modules.}
	\label{fig:TR}
\end{figure}
 
Having presented the overall picture of the proposed DNN architecture, next we scrutinize the layer structure for each individual module. Fig.~\ref{fig:TxRx} shows the modules $ \text{Tx} \in \{ \textcircled{\scriptsize 1}\text{TxS-N}, \textcircled{\scriptsize 2}\text{TxS-F}, \\ \textcircled{\scriptsize 6}\text{TxN} \}$ and $ \text{Rx} \in \{ \textcircled{\scriptsize 4}\text{RxN-N},  \textcircled{\scriptsize 5}\text{RxN-F}, \textcircled{\scriptsize 9}\text{RxF} \}$, which share a common structure  with multiple cascaded DNN layers. Fig.~\ref{fig:RxPre} shows  the  modules  
$ \text{RxPre} \in \{  \textcircled{\scriptsize 3}\text{RxPreSN},  \textcircled{\scriptsize 7}\text{RxPreSF}, \textcircled{\scriptsize 8}\text{RxPreNF} \}$, which share a common structure with an element-wise  multiplication operation at the output layer. The main purpose of the multiplication operation is to extract the key feature for signal demapping. For example, $\textcircled{\scriptsize 3}$RxPreSN is to learn the feature $ |y_{S,N}-h_{S,N}x_{S}|^2 =  \vert h_{S,N} x_{S} \vert^2 - 2 \Re\{ h_{S,N}^* x_{S}^* y_{S,N} \} + \vert y_{S,N} \vert^2 $ containing $\vert x_{S} \vert^2$, which is key to signal demapping (c.f. \eqref{eq:N-JML}). The input of $\textcircled{\scriptsize 3}$RxPreSN is $\frac{h_{S,N}^*y_{S,N}}{|h_{S,N}|^2}$. After the  multiple cascaded layers learn an estimate of $x_{S}$, e.g., $ax_{S}+b$, the element-wise multiplication operation computes $ \Re \Big\{ \frac{h_{S,N}^*y_{S,N}}{|h_{S,N}|^2}  \Big\} \Re \{ ax_{S}+ b \}  = \Re \Big\{ x_{S}+ \frac{h_{S,N}^*n_{S,N}}{|h_{S,N}|^2} \Big\} \Re \{ ax_{S}+ b \} $ containing $\Re \{ x_{S} \} ^2$ and $\Im  \Big\{ x_{S}+ \frac{h_{S,N}^*n_{S,N}}{|h_{S,N}|^2} \Big\} \Im \{ ax_{S}+ b \} $ containing $\Im \{ x_{S} \} ^2$.

Given the above, the DNN based joint optimization problem for the two demapping phases \eqref{eq:det-1} and  \eqref{eq:det-2} can now be reformulated as  
\begin{align}
\raisebox{-0.0\normalbaselineskip}[0pt][0pt]{%
	({\bf P2})}  \quad 
\min_{ \substack{f_{S}^{\prime}, \ f_{N}^{\prime}, \ g_{N}^{\prime}, \ g_{F}^{\prime} } }   &   \quad 
\Big\{  L_{( \bm{s}_{N},   \hat{\bm{s}}_{N} )}(f_{S}^{\prime}, g_{N}^{\prime}) ,\  L_{( \bm{s}_{F},   \hat{\bm{s}}_{F}^{N} )}(f_{S}^{\prime}, g_{N}^{\prime}) , \  L_{( \bm{s}_{F},   \hat{\bm{s}}_{F} )}(f_{S}^{\prime},  f_{N}^{\prime},  g_{N}^{\prime},  g_{F}^{\prime} )  \Big\},
\notag 
\end{align}
where  $L_{( \bm{s}_{N},   \hat{\bm{s}}_{N} )}(f_{S}^{\prime}, g_{N}^{\prime}) \triangleq L_1$ denotes the loss between the input-output pair  $(\bm{s}_{N}, \hat{\bm{s}}_{N})$ as a function of $\{ f_{S}^{\prime}, g_{N}^{\prime}\}$, and similar definition follows for $ L_{( \bm{s}_{F},   \hat{\bm{s}}_{F}^{N} )}(f_{S}^{\prime}, g_{N}^{\prime}) \triangleq L_2 $ and $  L_{( \bm{s}_{F},   \hat{\bm{s}}_{F} )}(f_{S}^{\prime},  f_{N}^{\prime},  g_{N}^{\prime},  g_{F}^{\prime} ) \triangleq L_3 $. These losses measure the demapping errors for their respective input, and they will be mathematically defined in Section \ref{sec:bit-train}. Note that $\{ L_1, L_2 \}$ are associated with \eqref{eq:det-1}, and $ L_3$ associated with \eqref{eq:det-2} is the end-to-end loss for the entire network.   Clearly, ({\bf P1}) has been translated into ({\bf P2}) in a more tractable form, where highly nonlinear mappings and demappings are learned by training the DNN parameter set  $\big \{ f_{S}^{\prime},  f_{N}^{\prime},  g_{N}^{\prime},  g_{F}^{\prime} \big\}$. This provides a solution to \textbf{Challenge 1}. 

However, we still need to address \textbf{Challenge 2}, as ({\bf P2}) involves three loss functions. Typically, this is a multi-task learning (MTL) problem \cite{MTL-ruder2017overview}, which is more complex than the conventional single-task learning. Moreover, the outputs $\{ \hat{\bm{s} }_{N}, \hat{\bm{s}}_{F}^{N}, \hat{\bm{s}}_{F} \}$ are bit-wise probabilities for each input bit, rather than the widely used symbol-wise probabilities for each input symbol \cite{Sync-dorner2018deep,deepNOMA-ye2020}. Therefore, a bit-wise self-supervised training method needs to be developed and analyzed. We will address the MTL  in Section \ref{sec:MTL}, and the bit-wise self-supervised  training in Sections \ref{sec:self-train} and \ref{sec:bit-train}. 

\subsection{The Proposed Two-Stage Training Method}
\label{sec:2stage}
\subsubsection{Multi-Task Learning} \label{sec:MTL}
In this MTL problem, minimizing $\{ L_1, L_2, L_3 \}$ simultaneously may lead to a poor error performance. For example, we may arrive at a situation where $L_2$ and $L_3$ are sufficiently small but $L_1$ is still very large. To avoid this, we develop a novel two-stage training method by analyzing the relationship among $\{ L_1, L_2 , L_3\}$. 

It is clear that $ L_1$ and $L_2$ are related to $\{ f_{S}^{\prime}, g_{N}^{\prime}  \}$, while $L_3$ is related to $\{ f_{S}^{\prime},  f_{N}^{\prime},  g_{N}^{\prime},  g_{F}^{\prime} \}$. As $ \{ f_{S}^{\prime}, g_{N}^{\prime}  \} \subset  \{ f_{S}^{\prime},  f_{N}^{\prime},  g_{N}^{\prime},  g_{F}^{\prime} \}$, this implies
a causal structure between $\{ L_1, L_2 \}$ and $L_3$. A more rigorous analysis on this relationship is provided in Appendix \ref{app:P3}. On this basis, ({\bf P2}) can be translated into the following problem 
\begin{IEEEeqnarray}{rCl} 
	\raisebox{-1.25\normalbaselineskip}[0pt][0pt]{%
		({\bf P3})
	} \quad 
	\text{Stage I: } \quad & \min_{ f_{S}^{\prime}, \  g_{N}^{\prime} } &  \ 
   \big\{  L_1, \ L_2  \big\}  \notag \\
	\text{Stage II: } \quad &   \min_{ f_{N}^{\prime}, \  g_{F}^{\prime} } & \ L_3  \notag \\
	&  \subto \quad	& \ f_{S}^{\prime}, \  g_{N}^{\prime}. \notag
\end{IEEEeqnarray}
For ({\bf P3}), as shown in Fig.~\ref{fig:AE-NOMA}, in stage I we minimize $L_1$ and $L_2$ through learning $\{ f_{S}^{\prime},  g_{N}^{\prime}\}$ by data training. In stage II, we minimize $L_3$ through learning $\{ f_{N}^{\prime},  g_{F}^{\prime} \}$ by fixing the obtained $\{ f_{S}^{\prime},  g_{N}^{\prime}\}$ in stage I. It is worth noting that stage I is still a MTL problem, but we can  minimize $L_1$ and $L_2$ simultaneously since they share the same $\{ f_{S}^{\prime},  g_{N}^{\prime}\}$.

\subsubsection{Self-Supervised Training} \label{sec:self-train}
For convenience, we express the three loss functions $ L_1$, $L_2$, and $L_3$ in a unified form. On this basis, we elaborate on the self-supervised training method for fading channels. Without loss of generality, we let $k_{N}=k_{F}=k$, and $( k, {\alpha}_{S,N}, {\alpha}_{S,F} )$ are fixed during the training. 

From ({\bf P2}), $ L_1$, $L_2$, and $L_3$ can be written as   
\begin{align}
L_{( \bm{s},   \hat{\bm{s}} )}( {f^{\prime}},   {g^{\prime}} )  \triangleq &  \mathbb{E}_{\bm{s}}  \big[ \mathcal{L}( \bm{s},   \hat{\bm{s}} )  \big],  \  
( \bm{s},   \hat{\bm{s}} ) \in \big\{  ( \bm{s}_{N},   \hat{\bm{s}}_{N} ), ( \bm{s}_{F},   \hat{\bm{s}}_{F} ), ( \bm{s}_{F},   \hat{\bm{s}}_{F}^{N} )  \big\}, \label{eq:opt-loss-multi}
\end{align}
where the input bits $\bm{s}$ also serve as the labels, $\hat{\bm{s}}$ denotes the output soft probabilities, and  $\mathcal{L}( \bm{s},   \hat{\bm{s}} )$ denotes the adopted loss function such as mean squared error and cross-entropy (CE)  \cite[Ch. 5]{book-goodfellow2016deep}.  For  $ {f^{\prime}}$ and  ${g^{\prime}} $ specifically, we have 
\begin{align}
( f^{\prime}, g^{\prime}) = \left\{\begin{array}{lc}
( f_{S}^{\prime},  g_{N}^{\prime}), &  \text{for } ( \bm{s},   \hat{\bm{s}} ) \in \big\{  ( \bm{s}_{N},   \hat{\bm{s}}_{N} ), ( \bm{s}_{F},   \hat{\bm{s}}_{F}^{N} )  \big\}, \\
\big( \{ f_{S}^{\prime}, f_{N}^{\prime} \}, \{ g_{N}^{\prime} , g_{F}^{\prime} \} \big),  &  \text{for } ( \bm{s},   \hat{\bm{s}} ) =  ( \bm{s}_{F},   \hat{\bm{s}}_{F} ).
\end{array} \right.
\end{align}

For a random batch of training examples $\{ ( \bm{s}^{b},   \hat{\bm{s}}^{b})\}_{b=1}^{B}$ of size $B$, the loss in \eqref{eq:opt-loss-multi} can be estimated through sampling as
\begin{align}
    L_{( \bm{s},   \hat{\bm{s}} )}( {f^{\prime}},   {g^{\prime}} )  = \frac{1}{B} \sum_{b=1}^{B}      \mathcal{L}( \bm{s}^{b},   \hat{\bm{s}}^{b}) . \label{eq:L-sample}
\end{align}
We use the stochastic gradient decent (SGD) algorithm to update the DNN parameter set $\{ f^{\prime},   g^{\prime} \}$ through backpropagation \cite[Ch. 6.5]{book-goodfellow2016deep} as
\begin{align} \label{eq:sgd-update}
\{ f^{\prime},   g^{\prime} \}^{(t)} = \{ f^{\prime},   g^{\prime} \}^{(t-1)} - \tau \nabla L_{( \bm{s},   \hat{\bm{s}} )} \big( \{ f^{\prime},   g^{\prime} \}^{(t-1)} \big) , 
\end{align} 
starting with a random initial value $ \{ f^{\prime},   g^{\prime} \}^{(0)}$, where $\tau > 0$, $t$, and $\nabla$ denote the learning rate, iteration index, and  gradient operator, respectively.

For the specific offline training of ({\bf P3}), following the proposed two-stage training method, the DNN parameter set $\{ f_{S}^{\prime}, f_{N}^{\prime},  g_{N}^{\prime},  g_{F}^{\prime} \}$ is first learned under AWGN channels ($\bm{h} = [ h_{S,N}, h_{S,F}, h_{N,F} ]^T = [3,1,3]^T$) to combat the noise. Then, by fixing $\{ f_{S}^{\prime}, f_{N}^{\prime} \} $, only $\{ g_{N}^{\prime},  g_{F}^{\prime} \} $ are fine-tuned under fading channels ($\bm{h} \sim \mathcal{CN} (0, {\bf \Lambda}) $ with ${\bf \Lambda} = \diag \big( [ \lambda_{S,N}, \lambda_{S,F},\lambda_{N,F} ]^T \big)$) to combat signal fluctuation.



Another critical issue is that, in the most literature \cite{ae-o2017introduction,deepNOMA-ye2020}, $L_{( \bm{s},   \hat{\bm{s}} )}( {f^{\prime}},   {g^{\prime}} )  $  only represents the  symbol-level CE loss with softmax activation function \cite{book-goodfellow2016deep}, where $\bm{s}$ is represented by a one-hot vector of length $2^k$, i.e., only one element equals to one and others zero \cite{ae-o2017introduction}. Fundamentally different from \cite{ae-o2017introduction,deepNOMA-ye2020}, $L_{( \bm{s},   \hat{\bm{s}} )}( {f^{\prime}},   {g^{\prime}} )$ here characterizes the bit-level loss, thereby requiring further analysis. 
\subsubsection{Bit-Level Loss} \label{sec:bit-train}

Because the inputs $\{ \bm{s}_{N}, \bm{s}_{F} \}$ are binary bits,  $L_{( \bm{s},   \hat{\bm{s}} )}( {f^{\prime}},   {g^{\prime}} )  $  minimization is a binary classification problem, where we use the binary cross-entropy (BCE) loss to quantify the demapping error. Accordingly, sigmoid activation function, i.e., $\phi(z)=\frac{1}{1+e^{-z}}$, is used at the output layers of $\textcircled{\scriptsize 4}$RxN-N, $\textcircled{\scriptsize 5}$RxN-F, and $\textcircled{\scriptsize 9}$RxF to obtain bit-wise soft probabilities  $\hat{\bm{s}}_{N}$, $\hat{\bm{s}}_{F}^{N}$, and $\hat{\bm{s}}_{F}$, respectively. 
In this case, following \eqref{eq:opt-loss-multi}, the BCE loss function can be written as 
\begin{align}
\mathcal{L}( \bm{s},   \hat{\bm{s}} )
= & \sum_{r=1}^{k}  \mathcal{L}( \bm{s}(r),   \hat{\bm{s}}(r) )  \notag \\
=  & -\sum_{r=1}^{k} \Big(  \bm{s}(r) \log \hat{\bm{s}}(r) + (1-\bm{s}(r) ) \log ( 1-\hat{\bm{s}}(r) )
\Big), \notag \\
& ( \bm{s},   \hat{\bm{s}} ) \in \big\{  ( \bm{s}_{N},   \hat{\bm{s}}_{N} ), ( \bm{s}_{F},   \hat{\bm{s}}_{F}^{N} ), ( \bm{s}_{F},   \hat{\bm{s}}_{F} )  \big\}. \label{eq:BCE}
\end{align}	
In another form, $\mathcal{L}( \bm{s},   \hat{\bm{s}} )$ can be shown as  
\begin{align}
\mathcal{L}( \bm{s},   \hat{\bm{s}} ) =  & H(p_{ {f^{\prime}}}  (  \bm{s} ), \hat{p}_{ {g^{\prime}}} (  \bm{s}  )) \notag \\
= & \sum_{r=1}^{k}  \mathbb{E}_{\bm{s}(r)}  \big[ H( \bm{s}(r),   \hat{\bm{s}}(r) )  \big], \label{eq:H-bit}
\end{align}
where $H(\cdot)$ represents the cross-entropy between the parameterized distributions $p_{ {f^{\prime}}}  (  \bm{s} )$ and $\hat{p}_{ {g^{\prime}}} (  \bm{s}  )$. $p_{ {f^{\prime}}}  (  \bm{s} )$ denotes the true  distribution of $\bm{s}$ for the transmitter with $f^{\prime}$, while $\hat{p}_{ {g^{\prime}}} (  \bm{s}  )$ denotes the estimated distribution of $  \bm{s} $ for the receiver with $g^{\prime}$. We can see from \eqref{eq:H-bit} that the optimization is performed for each individual bits in $\bm{s}$.

Then, during training, $L_{( \bm{s},   \hat{\bm{s}} )}( {f^{\prime}},   {g^{\prime}} )$ can be computed through averaging over all possible channel outputs $\bm{y} =[ y_{S,N}, y_{S,F}, y_{N,F} ]^T$ according to
\begin{align}
L_{( \bm{s},   \hat{\bm{s}} )}( {f^{\prime}},   {g^{\prime}} ) 
= & \sum_{r=1}^{k}  \mathbb{E}_{\bm{s}(r) ,\bm{y}}  \big[ H(  
p_{ {f^{\prime}}} (  \bm{s}(r) \vert \bm{y} ), \hat{p}_{ {g^{\prime}}} (  \bm{s}(r) \vert \bm{y} )
)\big]  \notag \\ 
= & H({\bf  S}) - \sum_{r=1}^{k} I_{{f^{\prime}} }( {\bf  S}(r) ;  {\bf  Y})  +    \sum_{r=1}^{k}  \mathbb{E}_{\bm{y}}  \big[ D_{\text{KL}}(  
p_{ {f^{\prime}}} (  \bm{s}(r) |\bm{y} ) \| \hat{p}_{ {g^{\prime}}} (  \bm{s}(r) |\bm{y} )
)\big], \label{eq:BMI-deri-iid}
\end{align} where $I( \cdot ; \cdot )$ is the mutual information (MI), and $D_{\text{KL}}( p \| \hat{p})$ is the Kullback-Leibler (KL) divergence between distributions $p$ and $\hat{p}$ \cite{MI-book-cover2012elements}.  The first term on the right side of \eqref{eq:BMI-deri-iid} is the entropy of $\bm{s}$, which is a constant. The second term can be viewed as learning $f^{\prime}$  at the transmitter, i.e.,  $(\{ 0,1 \}^{k_{N}}, \{ 0,1 \}^{k_{F}}) \to \mathcal{M}_{S}$ and $\hat{\bm{s}}_{F}^{N} \to  \mathcal{M}_{F}^{N} $. The third term  measures the difference between the true distribution $p_{ {f^{\prime}}} (  \bm{s}(r) |\bm{y} )$ at the transmitter and the learned distribution $\hat{p}_{ {g^{\prime}}} (  \bm{s}(r) |\bm{y} )$ at the receiver, which corresponds to $y_{S,N} \to (\hat{\bm{s}}_{N}, \hat{\bm{s}}_{F}^{N}) \in (\{ 0,1 \}^{k_{N}}, \{ 0,1 \}^{k_{F}})$ and $(y_{S,F}, y_{N,F}) \to \hat{\bm{s}}_{F} \in \{ 0,1 \}^{k_{F}}$.

\section{A Theoretical Perspective of the Design Principles}
\label{sec:theo-prob}
In Section \ref{sec:proposed-CNOMA}, we illustrated the whole picture of the proposed DNN architecture for deep cooperative NOMA. In this section, we further analyze the specific probability distribution  that each DNN module has learned, through studying the loss functions in \eqref{eq:BMI-deri-iid} for each training stage of ({\bf P3}).

\subsection{Training Stage I}

In essence, training stage I is MTL over a multiple access channel with inputs $\{ \bm{s}_{N}, \bm{s}_{F}\}$, transceiver $\{ f_{S}^{\prime},   g_{N}^{\prime}  \}$, channel function $\mathcal{C}_{S,N}$, and outputs $\{ \hat{\bm{s}}_{N}, \hat{\bm{s}}_{F}^{N}\}$. From information theory \cite{MI-book-cover2012elements}, the corresponding  loss functions $L_1$ and $L_2$ for the two tasks can be expressed as 
\begin{align}
L_1 
= & H({\bf  S}_{N}) - \underbrace{\sum_{r=1}^{k} I_{f_{S}^{\prime} }( {\bf  S}_{N}(r) ;  Y_{S,N})}_{\text{Conflicting  MI}}  +  \sum_{r=1}^{k}  \mathbb{E}_{ y_{S,N} }  \bigg[  D_{\text{KL}}(  
p_{ f_{S}^{\prime}} (  \bm{s}_{N}(r) | y_{S,N} )   \| \hat{p}_{ g_{N}^{\prime}} (  \bm{s}_{N}(r) |y_{S,N} )
)\bigg]  \label{eq:LossNN-1} \\
= & H({\bf  S}_{N}) - \underbrace{\sum_{r=1}^{k}  I_{f_{S}^{\prime} }( {\bf  S}_{N}(r) , {\bf  S}_{F}(r) ; Y_{S,N} )}_{\text{Common MI}} + \sum_{r=1}^{k}   I_{f_{S,2}^{\prime} }( {\bf  S}_{F}(r) ; Y_{S,N} | {\bf  S}_{N}(r) ) +    \sum_{r=1}^{k}  \mathbb{E}_{ y_{S,N} }  \bigg[ D_{\text{KL}} \notag \\
&   \Big(  \int_{x_{S} } \underbrace{p_{ f_{S}^{\prime}}  (  \bm{s}_{N}(r) | x_{S} )}_{\text{Individual  distribution}}
\underbrace{p(  x_{S}| y_{S,N} )}_{\text{Common  distribution}}  \dd x_{S}  \Big\| \int_{ \hat{y}_{S,N}}  \underbrace{\hat{p}_{ g_{N,4}^{\prime}} (  \bm{s}_{N}(r) | \hat{y}_{S,N} )}_{\text{ Individual module}} \underbrace{\hat{p}_{g_{N,3}^{\prime}} (  \hat{y}_{S,N} | y_{S,N} )}_{\text{ Common module}}   \dd  \hat{y}_{S,N}
\Big)\bigg]    , \label{eq:LossNN}
\end{align}
where $\hat{y}_{S,N}$ denotes the output signal of $\textcircled{\scriptsize 3}$RxPreSN, and the derivations for \eqref{eq:LossNN-1} and \eqref{eq:LossNN} are given in Appendix \ref{app:A}.  	 

Similarly, we have
\begin{align}
L_2 
= & H({\bf  S}_{F}) - \underbrace{\sum_{r=1}^{k}  I_{f_{S}^{\prime} }  ( {\bf  S}_{N}(r) , {\bf  S}_{F}(r) ; Y_{S,N} )}_{\text{Common MI }} + \underbrace{\sum_{r=1}^{k}   I_{f_{S,1}^{\prime} }( {\bf  S}_{N}(r) ; Y_{S,N} | {\bf  S}_{F}(r) )}_{\text{Conflicting  MI}} +     \sum_{r=1}^{k}  \mathbb{E}_{ y_{S,N} }  \bigg[ D_{\text{KL}} \notag \\
&   \Big(  \int_{x_{S} }   \underbrace{p_{ f_{S}^{\prime}} (  \bm{s}_{F}(r) | x_{S} )}_{\text{Individual distribution}} 
\underbrace{p(  x_{S}| y_{S,N} )}_{\text{Common  distribution}}  \dd x_{S}  \Big\| \int_{ \hat{y}_{S,N}}   \underbrace{\hat{p}_{ g_{N,5}^{\prime}} (  \bm{s}_{F}(r) | \hat{y}_{S,N} )}_{\text{ Individual module}} \underbrace{\hat{p}_{g_{N,3}^{\prime}} (  \hat{y}_{S,N} | y_{S,N} )}_{\text{ Common module}} \dd  \hat{y}_{S,N}
\Big)\bigg]  .    \label{eq:LossNF}  
\end{align}
Now we analyze the components of $L_1$ and $L_2$ in \eqref{eq:LossNN-1}-\eqref{eq:LossNF}. Specifically, on one hand,  \eqref{eq:LossNN} and \eqref{eq:LossNF} share a common MI term  $ \sum_{r=1}^{k} I_{f_{S}^{\prime} }( {\bf  S}_{N}(r) , {\bf  S}_{F}(r) ; Y_{S,N} )$, which corresponds to the learning of $f_{S}^{\prime}$.  On the other hand, \eqref{eq:LossNN-1} and \eqref{eq:LossNF} have conflicting MI terms. That is, minimizing  \eqref{eq:LossNN-1} leads to maximizing the second term 
$ \sum_{r=1}^{k}   I_{f_{S}^{\prime} }( {\bf  S}_{N}(r) ; Y_{S,N}) $, while  minimizing \eqref{eq:LossNF} results in minimizing the third term 
$ \sum_{r=1}^{k}   I_{f_{S,1}^{\prime} }( {\bf  S}_{N}(r) ; Y_{S,N} | {\bf  S}_{F}(r) ) $ with $f_{S,1}^{\prime} \subset f_{S}^{\prime}$. Clearly, these two objectives are contradictory for learning $f_{S}^{\prime}$.


%

Next, let us observe the KL divergence terms in \eqref{eq:LossNN}-\eqref{eq:LossNF} at the receiver side. The true distributions in \eqref{eq:LossNN} and \eqref{eq:LossNF} share a common distribution term $p(x_{S} | y_{S,N})$, and individual (but related) distribution terms $p_{ f_{S}^{\prime}} (  \bm{s}_{J}(r) | x_{S} )$, $J \in \{N,F\}$.
By exploiting this relationship, we use a common demapping module  $\textcircled{\scriptsize 3}$RxPreSN to learn the common distribution  $\hat{p}_{g_{N,3}^{\prime}} (  \hat{y}_{S,N} | y_{S,N} ) $ for $p(x_{S} | y_{S,N})$, such that $\hat{y}_{S,N}$ learns to estimate $x_{S}$. Then, two individual demapping modules $\textcircled{\scriptsize 4}$RxN-N and $\textcircled{\scriptsize 5}$RxN-F are used to learn $\hat{p}_{ g_{N,4}^{\prime}} (  \bm{s}_{N}(r) | \hat{y}_{S,N} )$ and $\hat{p}_{ g_{N,5}^{\prime}} (  \bm{s}_{F}(r) | \hat{y}_{S,N} )$ for estimating  $p_{ f_{S}^{\prime}} (  \bm{s}_{N}(r) | x_{S} )$ and $p_{ f_{S}^{\prime}} (  \bm{s}_{F}(r) | x_{S} )$, respectively. 

%
\subsection{Training Stage II}

Training stage II is end-to-end training with fixed $\{ f_{S}^{\prime}, g_{N}^{\prime}\}$ learned from stage I. As such, $L_3$ can be expressed as (c.f. \eqref{eq:BMI-deri-iid})
\begin{align}
L_3
= & H({\bf  S}_{F}) - \sum_{r=1}^{k} I_{f_{N}^{\prime} }( {\bf  S}_{F}(r) ;  Y_{S,F}, Y_{N,F} )  +   \sum_{r=1}^{k} \mathbb{E}_{ y_{S,F},y_{N,F} }  \bigg[ D_{\text{KL}}(  
p_{ f_{N}^{\prime} } (  \bm{s}_{F}(r) | y_{S,F},y_{N,F} ) \| \notag \\
& \hat{p}_{ g_{F}^{\prime}} (  \bm{s}_{F}(r) | y_{S,F},y_{N,F} )
)\bigg]   .  \label{eq:LossFF}
\end{align}
Minimizing $L_3$ results in maximizing the second term $\sum_{r=1}^{k} I_{f_{N}^{\prime} }( {\bf  S}_{F}(r) ;  Y_{S,F}, Y_{N,F} )$, corresponding to optimizing $f_{N}^{\prime}$. By probability factorization, the true distribution in the third term in \eqref{eq:LossFF} can be expressed  as
\begin{align}
 p_{ f_{N}^{\prime} } (  \bm{s}_{F}(r) | y_{S,F},y_{N,F} )
= &  \int_{x_{S}} \int_{\hat{\bm{s}}_{F}^{N}}   p( \bm{s}_{F}(r)  |  x_{S}, \hat{\bm{s}}_{F}^{N}, y_{S,F},y_{N,F}) p (  x_{S} | y_{S,F} )   p_{ f_{N}^{\prime} } (  \hat{\bm{s}}_{F}^{N} | y_{N,F} )  \dd \hat{\bm{s}}_{F}^{N}  \dd x_{S} \notag \\
= &  \int_{x_{S}} \int_{\hat{\bm{s}}_{F}^{N}}   \underbrace{p( \bm{s}_{F}(r)  |  x_{S}, \hat{\bm{s}}_{F}^{N})}_{\text{Learned by {\textcircled{\tiny 9}}}} \underbrace{p (  x_{S} | y_{S,F} )}_{\text{Learned by {\textcircled{\tiny 7}}}}  \underbrace{p_{ f_{N}^{\prime} } (  \hat{\bm{s}}_{F}^{N} | y_{N,F} )}_{\text{Learned by {\textcircled{\tiny 8}}}}   \dd \hat{\bm{s}}_{F}^{N}  \dd x_{S}  \label{eq:pTx-st2}  ,
\end{align}
where $p( \bm{s}_{F}(r)  |  x_{S}, \hat{\bm{s}}_{F}^{N}) p (  x_{S} | y_{S,F} ) $  is determined through the stage I training. 
To exploit such factorization, we introduce auxiliary variables $\hat{y}_{S,F}$ and $\hat{y}_{N,F}$ to estimate  $ x_{S}$ and $ \hat{\bm{s}}_{F}^{N} $, respectively, and express the distribution $\hat{p}_{ {g^{\prime}}} (  \bm{s}_{F}(r) | y_{S,F},y_{N,F} )$ in \eqref{eq:LossFF} as
\begin{align}
\hat{p}_{ {g^{\prime}}} (  \bm{s}_{F}(r) | y_{S,F},y_{N,F} )  = & \int_{ \hat{y}_{S,F} }  \int_{ \hat{y}_{N,F} }   \hat{p}_{g_{F,9}^{\prime}} (  \bm{s}_{F}(r) | \hat{y}_{S,F},\hat{y}_{N,F} ) \hat{p}_{g_{F,7}^{\prime}} (  \hat{y}_{S,F}| y_{S,F} ) \hat{p}_{g_{F,8}^{\prime}} (  \hat{y}_{N,F}| y_{N,F} ) \notag \\
&  \dd \hat{y}_{N,F} \dd \hat{y}_{S,F},  \label{eq:pRx-st2} 
\end{align}
where $\hat{y}_{S,F}$ and $\hat{y}_{N,F}$ denote the outputs of demapping modules $\textcircled{\scriptsize 7}$RxPreSF and $\textcircled{\scriptsize 8}$RxPreNF, respectively. Correspondingly, $\hat{p}_{g_{F,7}^{\prime}} (  \hat{y}_{S,F}| y_{S,F} )$ and $ \hat{p}_{g_{F,8}^{\prime}} (  \hat{y}_{N,F}| y_{N,F} )$ describe the learned  distributions for these two modules. 
It can be observed that  $\hat{p}_{g_{F,7}^{\prime}} (  \hat{y}_{S,F}| y_{S,F} )$ and $ \hat{p}_{g_{F,8}^{\prime}} (  \hat{y}_{N,F}| y_{N,F} )$ can estimate the true distributions $p (  x_{S} | y_{S,F} )  $ and $p_{ f_{N}^{\prime} } (  \hat{\bm{s}}_{F}^{N} | y_{N,F} ) $,  respectively. 
\begin{table}[!t] 
	\centering
	\captionsetup{font={small}}
	\caption{Learned distributions by the DNN demapping modules and the corresponding true ones
	} 
	\label{table:function-NN}
	\centering
	\scalebox{1}{
		\begin{tabular}{lll}
			\toprule
			Demapping Module & Learned Distribution & True Distribution \\
	    	\midrule
			$\textcircled{\scriptsize 3}$RxPreSN & $\hat{p}_{g_{N,3}^{\prime}} (  \hat{y}_{S,N} | y_{S,N} ) $	 & 
			$p(  x_{S}| y_{S,N} )$ \\
			$\textcircled{\scriptsize 4}$RxN-N & $\hat{p}_{ g_{N,4}^{\prime}} (  \bm{s}_{N} | \hat{y}_{S,N} )$ &  $p_{ f_{S}^{\prime}}  (  \bm{s}_{N} | x_{S} )$ \\
			
			$\textcircled{\scriptsize 5}$RxN-F & $\hat{p}_{ g_{N,5}^{\prime}} (  \bm{s}_{F} | \hat{y}_{S,N} )$ & $p_{ f_{S}^{\prime}}  (  \bm{s}_{F} | x_{S} )$ \\
			$\textcircled{\scriptsize 7}$RxPreSF & $\hat{p}_{g_{F,7}^{\prime}} (  \hat{y}_{S,F}| y_{S,F} )$	 & 
			$p(  x_{S}| y_{S,F} )$ \\
			$\textcircled{\scriptsize 8}$RxPreNF & $\hat{p}_{g_{F,8}^{\prime}} (  \hat{y}_{N,F}| y_{N,F} )$	 & 
			$p_{ f_{N}^{\prime} } (  \hat{\bm{s}}_{F}^{N} | y_{N,F} )$ \\
			$\textcircled{\scriptsize 9}$RxF & $\hat{p}_{g_{F,9}^{\prime}} (  \bm{s}_{F} | \hat{y}_{S,F},\hat{y}_{N,F} )$ & $p( \bm{s}_{F}  |  x_{S}, \hat{\bm{s}}_{F}^{N})$ \\
			\bottomrule
	\end{tabular} 
	}
\end{table}  
Table \ref{table:function-NN} summarizes the  distributions that the DNN demapping modules have learned. In Section \ref{sec:simu}, we will show that the learned distribution is consistent with the true one. 
\section{Model Adaptation}

In this section, we adapt the proposed DNN scheme to suit more practical scenarios. We first address the PA mismatch between training and inference. Then, we investigate the incorporation of the widely adopted channel coding into our proposed scheme. 
In both scenarios, our adaptation enjoys the benefit of reusing the original trained DNN modules without carrying out a new training process.



\subsection{Adaptation to Power Allocation}
\label{sec:pa}
In Section \ref{sec:proposed-CNOMA}, the PA coefficients $( {\alpha}_{S,N}, {\alpha}_{S,F} ) $ at the BS are fixed during the training process. However, their values might change during the inference process due to the nonlinear behaviors of the power amplifier in different power regions  \cite{PA-popovic2017amping,PA-sun2019behavioral}, resulting in the mismatch between the two processes.  Denote the new PA coefficient for inference as $\hat{\alpha}_{S,N}$ for UN, and $\hat{\alpha}_{S,F}$ for UF.

As a solution,  we propose to scale the received signals for $g_N^{\prime}$ and $g_F^{\prime}$. 
The goal is to ensure that their input signal-to-interference-plus-noise ratios (SINRs) are equal to those during the inference process, i.e., 
$\frac{ \hat{\alpha}_{S,N} |h_{S,N}|^2 }{  \hat{\alpha}_{S,F} |h_{S,N}|^2 + 2 \sigma_{S,N}^2 }$ for  $\bm{s}_{N}$ demapping by $g_N^{\prime}$,  $\frac{ \hat{\alpha}_{S,F} |h_{S,N}|^2 }{  \hat{\alpha}_{S,N} |h_{S,N}|^2 + 2 \sigma_{S,N}^2 }$ for  $\bm{s}_{F}$ demapping by $g_N^{\prime}$, and $\frac{ \hat{\alpha}_{S,F} |h_{S,F}|^2 }{  \hat{\alpha}_{S,N} |h_{S,F}|^2 + 2 \sigma_{S,F}^2 }$ for  $\bm{s}_{F}$ demapping by $g_{F,7}^{\prime} \subset g_F^{\prime}$. In this case, their new expressions are given by
\begin{align}
\hat{\bm{s} }_{N}  =  & g_{N}^{\prime} \bigg( \frac{1}{\omega_{N}} y_{S,N} \bigg) , \label{eq:pa-N} \\
\hat{\bm{s} }_{F}^{N}  =  & g_{N}^{\prime} \bigg( \frac{1}{\omega_{F}} y_{S,N} \bigg) , \label{eq:pa-NF} \\
\hat{\bm{s}}_{F} =  &  g_{F}^{\prime} \bigg( \frac{1}{\omega_{F}} y_{S,F}, \ y_{N,F} \bigg) , \label{eq:pa-F}
\end{align}
where the scaling factors  are defined as 
\begin{align}
\omega_{N} =  \sqrt{\frac{\hat{\alpha}_{S,N}}{{\alpha}_{S,N}}}, \ 
\omega_{F} =  \sqrt{\frac{\hat{\alpha}_{S,F}}{{\alpha}_{S,F}}}  .
\end{align}
Note that in \eqref{eq:pa-N} and \eqref{eq:pa-NF}, given two different inputs, $g_{N}^{\prime} (\cdot)$ is used twice to obtain $\hat{\bm{s} }_{N}$ and $\hat{\bm{s} }_{F}^{N}$, respectively. We prove in Appendix \ref{app:pa} that the SINR is exactly $\frac{ \hat{\alpha}_{S,N} |h_{S,N}|^2 }{  \hat{\alpha}_{S,F} |h_{S,N}|^2 + 2 \sigma_{S,N}^2 }$ for $\frac{1}{\omega_{N}} y_{S,N}$ in \eqref{eq:pa-N},  $\frac{ \hat{\alpha}_{S,F} |h_{S,N}|^2 }{  \hat{\alpha}_{S,N} |h_{S,N}|^2 + 2 \sigma_{S,N}^2 }$ for $ \frac{1}{\omega_{F}} y_{S,N}$ in \eqref{eq:pa-NF}, and $\frac{ \hat{\alpha}_{S,F} |h_{S,F}|^2 }{  \hat{\alpha}_{S,N} |h_{S,F}|^2 + 2 \sigma_{S,F}^2 }$ for $\frac{1}{\omega_{F}} y_{S,F}$ in \eqref{eq:pa-F}.

\subsection{Incorporation of Channel Coding}
\label{sec:c-coding}
Channel coding has been widely adopted to improve the communication  reliability \cite{book-clark2013error}. However, the conventional DNN based symbol-wise demapping \cite{ae-o2017introduction,deepNOMA-ye2020} cannot be directly connected to a soft channel decoder \cite{AE-bit-alberge2018deep,AE-bit-cammerer2020trainable}, such as the soft low-density parity-check code (LDPC) decoder \cite{LDPC-NOMA-pan2018sic} and polar code decoder \cite{pc-zheng2020threshold}. By contrast, our proposed scheme  in Section \ref{sec:proposed-CNOMA} outputs bit-wise soft information (c.f. \eqref{eq:det-1}, \eqref{eq:det-2}), enabling the straightforward 
cascade of a soft channel decoder.

Specifically, denote the  information bit blocks for UN and UF as $\bm{c}_{N}$ and $\bm{c}_{F}$, respectively. They are  encoded as binary codewords $ \langle \bm{s}_{N} \rangle = \mathcal{E} (\bm{c}_{N})$ and $ \langle \bm{s}_{F} \rangle  = \mathcal{E} ( \bm{c}_{F} )$ by channel encoder $\mathcal{E}( \cdot )$, and then split into multiple transmitted bit blocks (i.e., $\bm{s}_{N}$ and $\bm{s}_{F}$), which are sent into $f_{S}^{\prime}$.
At the receiver, the log-likelihood ratios (LLRs) of bits in $\bm{s}$ are calculated as 
\begin{align}
    \text{LLR}(\bm{s}(r)) = \log \Big( \frac{1-\hat{\bm{s}}(r)}{\hat{\bm{s}}(r)} \Big) ,  \ r \in \{1,2,\cdots,k\},
\end{align}
where we interpret $\hat{\bm{s}}(r)$ as the soft probability for bit $\bm{s}(r)$ with $\hat{\bm{s}}(r) = \Pr \{ \bm{s}(r)=1 | \hat{\bm{s}} \}$ \cite{intro-8849796}. The LLRs serve as the input of the soft channel decoder, denoted as $\mathcal{D} ( \cdot )$. 

At UN, we assume that it decodes its own information $\bm{c}_{N}$ as $\hat{\bm{c}}_{N} = \mathcal{D} ( \text{LLR} (\langle \hat{\bm{s}}_{N} \rangle ) ) $, but still  performs $ \hat{x}_{F}^{N} = f_{N}^{\prime} (\hat{\bm{s}}_{F}^{N})$ as in the uncoded case without decoding $\bm{c}_{F}$ (called demapping-and-forward). These two operations are separable because we use two parallel DNNs, i.e., $\textcircled{\scriptsize 4}$RxN-N and  $\textcircled{\scriptsize 5}$RxN-F, to obtain $\hat{\bm{s}}_{N}$ and $\hat{\bm{s}}_{F}^{N}$, respectively. Note that this parallel demapping can also reduce the error propagation compared to SIC. At UF, it decodes  $\bm{c}_{F}$ as $\hat{\bm{c}}_{F} = \mathcal{D} ( \text{LLR} (\langle \hat{\bm{s}}_{F} \rangle ) ) $. By contrast, the conventional SIC and JML decoding schemes need to decode $\hat{\bm{s}}_{N}$ and $\hat{\bm{s}}_{F}^{N}$ jointly.


\section{Simulation Results}
\label{sec:simu}

In this section, we perform simulation to verify the superiority of the proposed deep cooperative NOMA scheme, and compare it with OMA and the conventional cooperative NOMA scheme. In OMA, the BS transmits $x_{N}$ and $x_{F}$ to UN and UF, respectively, in two consecutive time slots, and there is no cooperation between UN and UF. Default parameters for simulation are: $k=2$ ($M_{N}=M_{F}=4$) and 
$\sigma_{S,F} = \sigma_{S,N} = \sigma_{N,F} = \sigma$, $\lambda_{S,F}=1$, $\lambda_{S,N}=\lambda_{N,F}$ for the three links.
We consider six scenarios (S1-S6), and their parameters are summarized in Table \ref{table:scenario-setup}, where ``cooperative link" refers to the BS to UN to UF link. 
Note that for S1-S4, we have $\big( \hat{\alpha}_{S,N},  \hat{\alpha}_{S,F} \big) =  ( {\alpha}_{S,N},  {\alpha}_{S,F}  ) $.
	\begin{table}[!ht] 
	\centering
	\captionsetup{font={small}}
	\caption{ 
		Parameters for scenarios S1-S6}
	\label{table:scenario-setup}
	\centering
	\scalebox{1}{
		\begin{tabular}{llll}
			\toprule
			Scenario  &	 $\lambda_{S,N}$ &		$( {\alpha}_{S,N},  {\alpha}_{S,F} )$ 
			&  Explanation 
			\\ 
			\midrule
			S1 &  10  & $( 0.4, 0.6 ) $ & Balanced PA \\
			S2 & 10	 & $( 0.25, 0.75 )$  & Optimized PA \\
			S3 & 6	& $( 0.25, 0.75 )$   & Weaker cooperative link  \\
			S4 & 6	& $( 0.1, 0.9 )$ & Unbalanced  PA \\
			S5 & 10 & $( 0.25, 0.75 )$   & \tabincell{l}{	PA mismatch:  $\big( \hat{\alpha}_{S,N},  \hat{\alpha}_{S,F} \big) = ( 0.3, 0.7 ) $  }		\\
			S6 & 10	& $( 0.25, 0.75 )$ & \tabincell{l}{	PA mismatch:  $\big( \hat{\alpha}_{S,N},  \hat{\alpha}_{S,F} \big) = ( 0.2, 0.8 ) $}  \\
			\bottomrule
	\end{tabular} 
	}
\end{table}

For the specific layer structure of each DNN module in Fig.~\ref{fig:AE-NOMA}, all three transmitters  ($\textcircled{\scriptsize 1}$, $\textcircled{\scriptsize 2}$ and $\textcircled{\scriptsize 6}$) have the same layer structure, with an input layer (dimension of $k_{N}$ or $k_{F}$) followed by $4$ hidden layers with $16$, $8$, $4$, and $2$ neurons, respectively. 
Modules $\textcircled{\scriptsize 3}$, $\textcircled{\scriptsize 7}$ and $\textcircled{\scriptsize 8}$  also have the same layer structure. There are three hidden layers of dimensions $64$, $32$ and $2$, respectively. Modules  $\textcircled{\scriptsize 4}$, $\textcircled{\scriptsize 5}$, and $\textcircled{\scriptsize 9}$ have three  hidden layers of dimensions $128$, $64$ and $32$, respectively, with output of dimension $k_N$ or $k_F$. We adopt tanh as the activation function for the hidden layers \cite{intro-8755300}.

We use Keras with TensorFlow backend to implement the proposed DNN architecture, which is first trained under AWGN channels at SNR$=5$~dB, and then $\{ g_{N}^{\prime},  g_{F}^{\prime} \} $ are fine-tuned under Rayleigh  fading channels (c.f. Section \ref{sec:self-train}) at a list of SNR values in $[ 15, 5, 6, 7, 30 ]$~dB to achieve a favorable error   performance in both low and high SNR regions. We have the learning rate $\tau = 0.001$ and $0.01$ for AWGN and Rayleigh fading channels, respectively.  After training, we test the DNN scheme for various SNRs, including those beyond the trained SNRs. In the uncoded case, the demapping rule for bit $\bm{s}(r)$ is $\text{LLR}(\bm{s}(r)) = \log \Big( \frac{1-\hat{\bm{s}}(r)}{\hat{\bm{s}}(r)} \Big) \ \overset{\bm{s}(r)=0}{\underset{\bm{s}(r)=1}{\gtrless}}  \ 0$.
\subsection{Network Losses $L_1$, $L_2$, and $L_3$ during Testing}
\begin{figure}[!t]
	\centering
	\subfigure[For S1]{\label{fig:loss-S1}
	\includegraphics[width=0.47\textwidth]{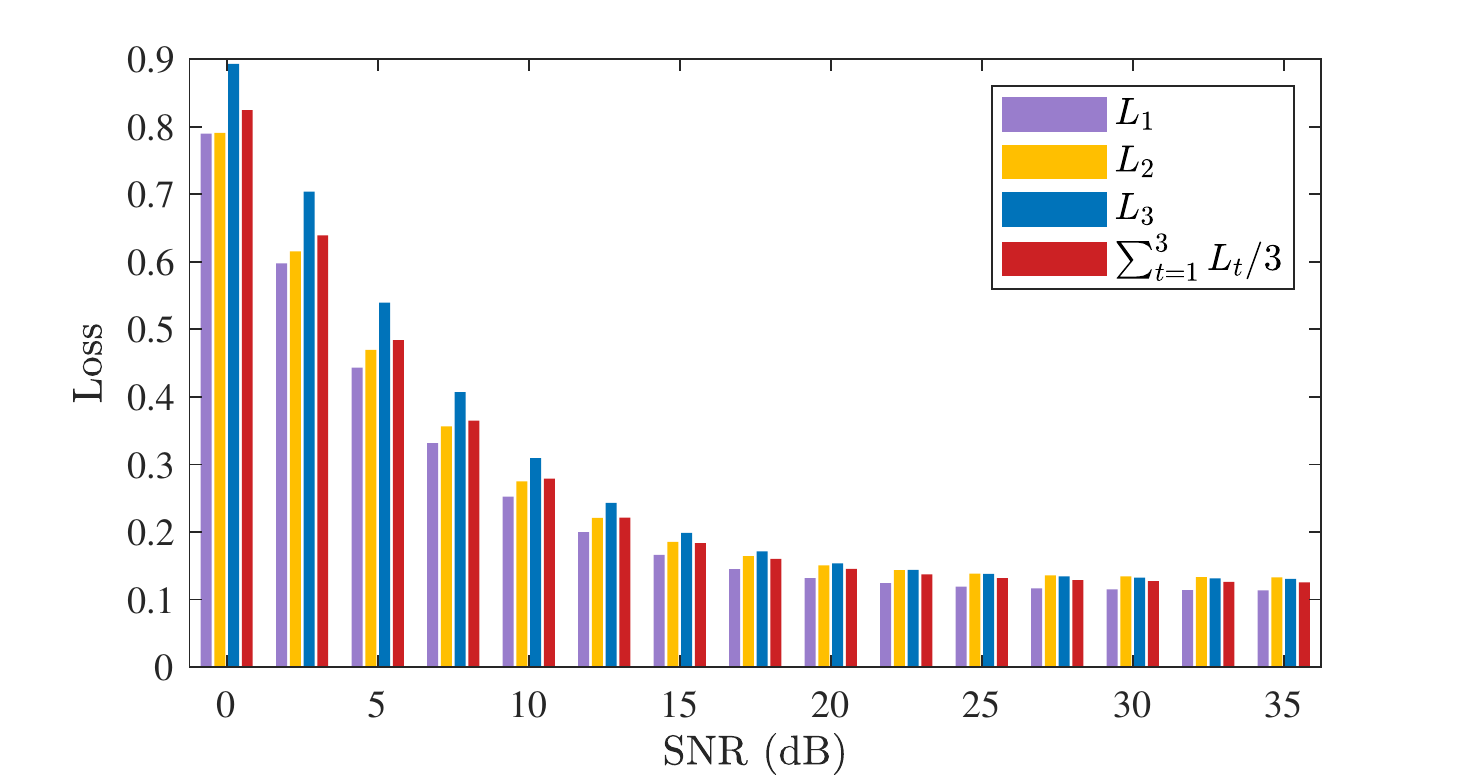} }
	\subfigure[For S2]{\label{fig:loss-S2}
	\includegraphics[width=0.47\textwidth]{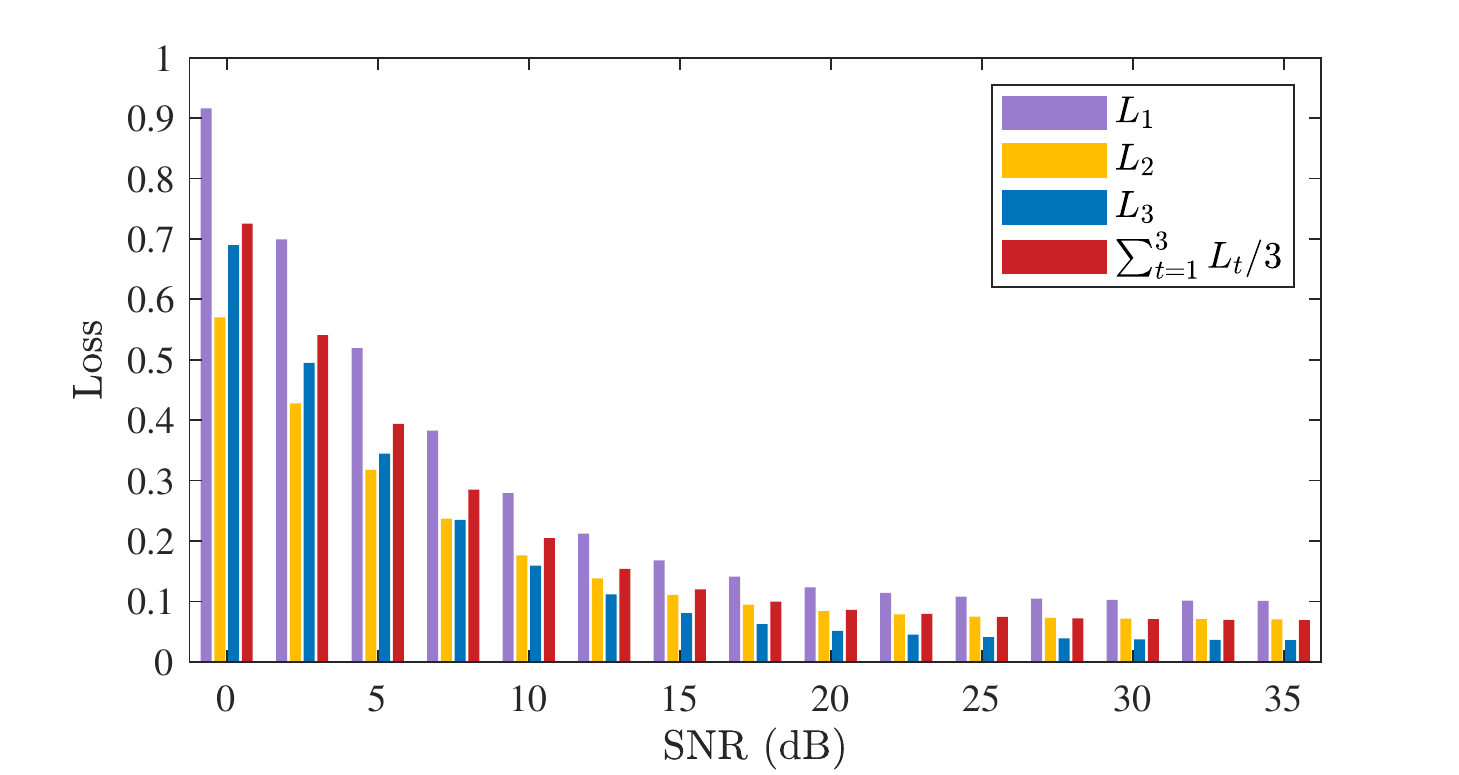} }
	\subfigure[For S3]{\label{fig:loss-S3}
	\includegraphics[width=0.47\textwidth]{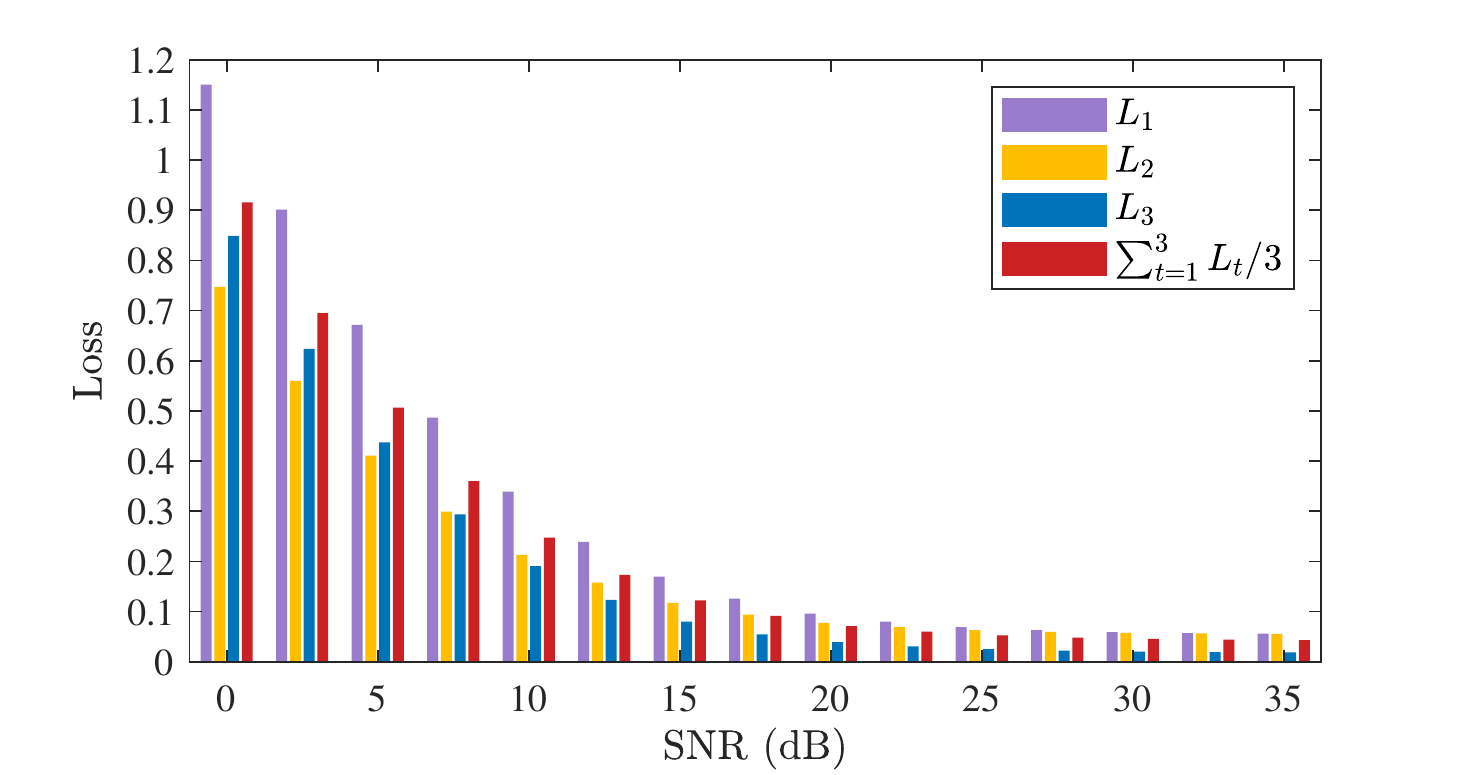} }
	\subfigure[For S4]{\label{fig:loss-S4}
	\includegraphics[width=0.47\textwidth]{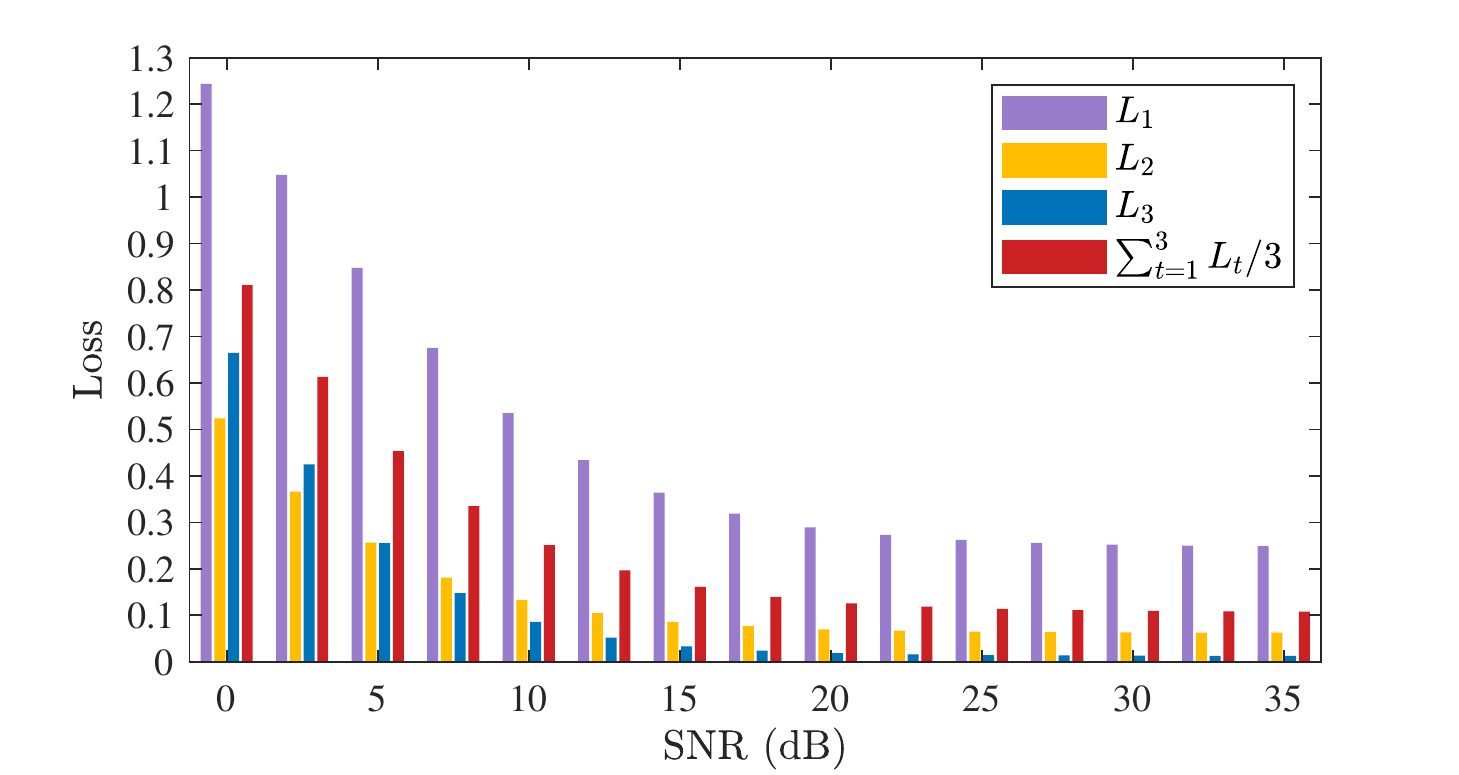} }
	\caption{Network losses $L_1$, $L_2$, $L_3$, and the average loss $\sum_{t=1}^3 L_t/3$ for different channel scenarios.}
	\label{fig:loss}
\end{figure} 
Upon obtaining the proposed DNN through training, in Fig.~\ref{fig:loss}, we check whether all the losses $L_1$, $L_2$, and $L_3$ can be significantly reduced by our proposed two-stage training method in Section \ref{sec:2stage}. 
For each SNR value, $8\times 10^5$ data bits are randomly generated for each user, divided into $B_t = 4\times 10^5$ data blocks with $k=2$ bits per block, and then sent into the DNN. We calculate $L_1$, $L_2$, and $L_3$  according to \eqref{eq:L-sample}, as well as the average loss $\sum_{t=1}^3 L_t/3$.

We can see that for all scenarios in Fig.~\ref{fig:loss}, as SNR increases, $L_1$, $L_2$ and $L_3$ each asymptotically decreases to  a small value, e.g.,  $0.13$ for $L_2$ in Fig.~\ref{fig:loss-S1}. The only exception is that $L_1$ in S4 (Fig.~\ref{fig:loss-S4}) asymptotically decreases to $0.25$, because of the relatively small PA coefficient ${\alpha}_{S,N} = 0.1$. Besides, $L_1$, $L_2$, and $L_3$ are all close to the average loss $\sum_{t=1}^3 L_t/3$ within $0.14$. These results indicate that the proposed two-stage training can significantly reduce $L_1$, $L_2$, and $L_3$, and provide a solution to the original MTL problem ({\bf P2}). 

\subsection{Learned Mappings by DNN Mapping Modules}
\begin{figure} [!t] 
	\centering
	\subfigure[For $\bm{s}_{N} \in \mathcal{M}_{N}$,  $\bm{s}_{F} \in \mathcal{M}_{F}$, and $\hat{\bm{s}}_{F}^{N} \in \mathcal{M}_{F}^{N}$, respectively]{
		\label{fig:cons-AE-x}
		\includegraphics[width=0.48\textwidth]{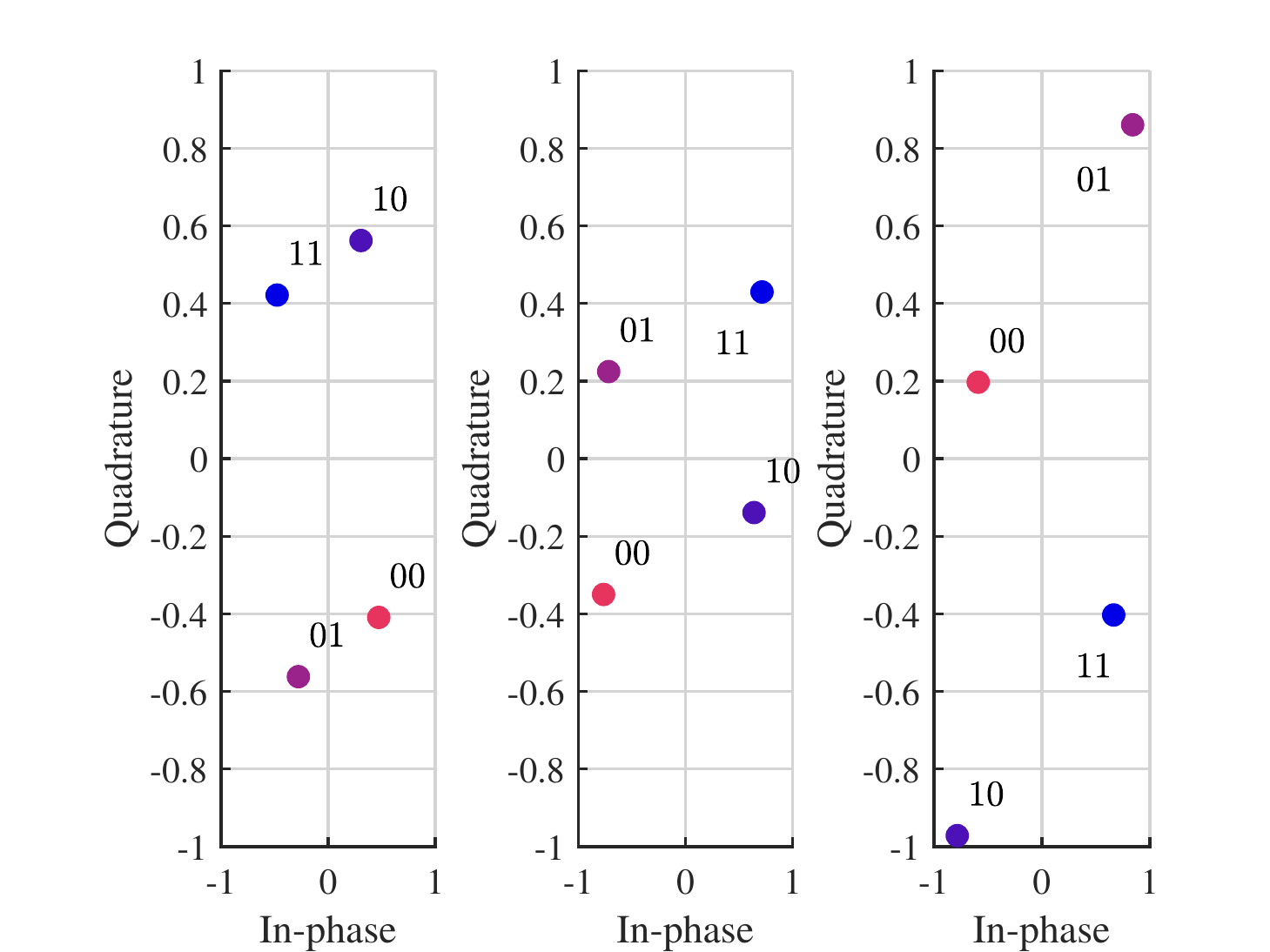}}
	\subfigure[For $\bm{s}_{S} \in \mathcal{M}_{S}$ (composite constellation)]{
		\label{fig:cons-AE-xS}
		\includegraphics[width=0.48\textwidth]{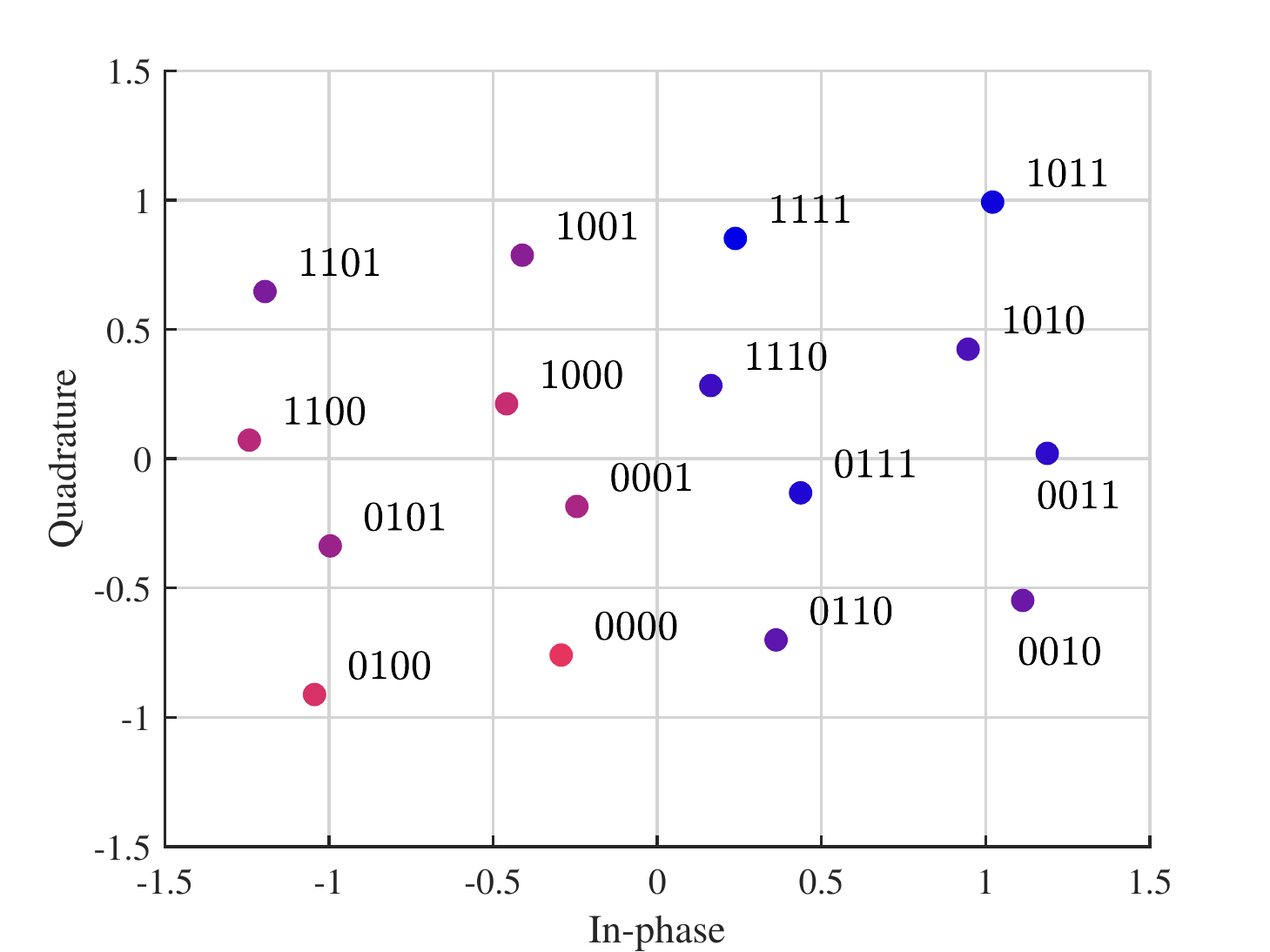}}	
	\caption{Learned constellations  by $f_{S}^{\prime}$ and $f_{N}^{\prime}$ with bit mapping for $( {\alpha}_{S,N}, {\alpha}_{S,F} )  = ( 0.4, 0.6 )$.}
	\label{fig:cons-AE-M4-s2}
\end{figure}
	\begin{figure}[!t] 
	\centering
	\subfigure[For $\bm{s}_{N}  \in \mathcal{M}_{N}$]{
		\label{fig:cons-AE-M4-xNb}
		\includegraphics[width=0.6\textwidth]{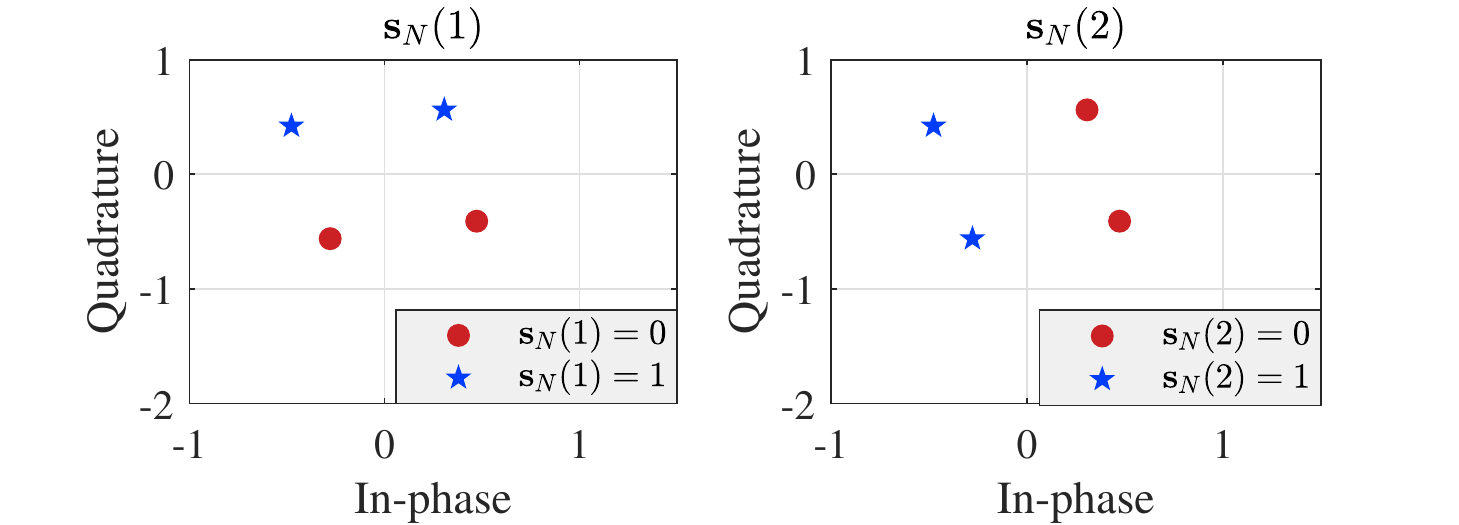}}
	\subfigure[For $\bm{s}_{S} \in \mathcal{M}_{S}$ (composite constellation)]{
		\label{fig:cons-AE-M4-cp}
		\includegraphics[width=0.6\textwidth]{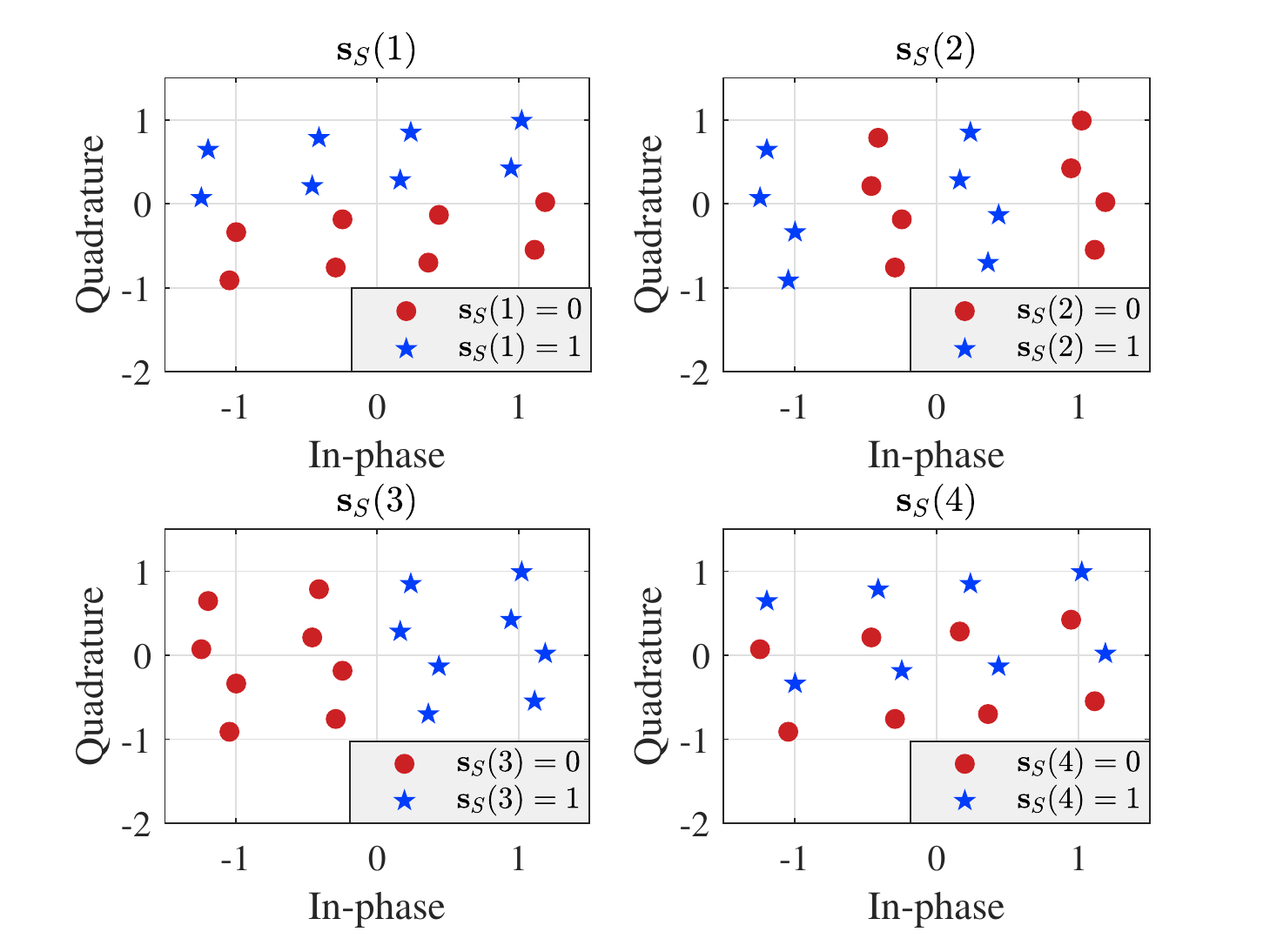}}	
	\caption{Learned constellations  by $f_{S}^{\prime}$ for the individual bit positions, where $( {\alpha}_{S,N}, {\alpha}_{S,F} )  = ( 0.4, 0.6 )$,  and the red and blue markers denote bit $0$ and $1$, respectively. } \label{fig:cons-AE-M4}
\end{figure}

As discussed in Section \ref{sec:AE-CoopNOMA}, the proposed DNN can learn mappings $\ (\{ 0,1 \}^{k_{N}}, \{ 0,1 \}^{k_{F}}) \to \mathcal{M}_{S}$ and $\hat{\bm{s}}_{F}^{N} \to  \mathcal{M}_{F}^{N} $ automatically, resulting in a new constellation and bit mapping. Fig.~\ref{fig:cons-AE-M4-s2} presents the learned constellations by $f_{S}^{\prime}$ and $f_{N}^{\prime}$ with bit mapping  for $( {\alpha}_{S,N}, {\alpha}_{S,F} )  = ( 0.4, 0.6 )$. Fig.~\ref{fig:cons-AE-x} shows the individual constellations $\bm{s}_{N} \in \mathcal{M}_{N}$,  $\bm{s}_{F} \in \mathcal{M}_{F}$, and $\hat{\bm{s}}_{F}^{N} \in \mathcal{M}_{F}^{N}$,  and it can be seen that   $\mathcal{M}_{N}$,  $\mathcal{M}_{F}$, and $\mathcal{M}_{F}^{N}$ all have learned parallelogram-like shapes with different orientations and aspect ratios. Fig.~\ref{fig:cons-AE-xS} shows the composite constellation $\mathcal{M}_{S}$, where the minimum Euclidean distance is improved significantly compared with that in Fig.~\ref{fig:cons-xS}, i.e., from $0.2$ to $0.36$.


In Section \ref{sec:proposed-CNOMA}, we use the bit-wise binary classification method to achieve the demappings $g_{N}^{\prime}$ and $g_{F}^{\prime}$. In Fig.~\ref{fig:cons-AE-M4}, we demonstrate that the two classes (bit $0$ and $1$) are separable by presenting the location of each individual bit. Specifically, the constellations 
$\bm{s}_{N} \in \mathcal{M}_{N}$ and $\bm{s}_{S} \in \mathcal{M}_{S}$ in Fig. 6 are presented here in a different form in Figs.~\ref{fig:cons-AE-M4-xNb} and \ref{fig:cons-AE-M4-cp}, respectively.
It is clearly shown that these two classes (bit $0$ and $1$) are easily separable for all bit positions. This indicates that the demapping  can be achieved. 

\subsection{Learned Distributions by DNN Demapping Modules}
\label{sec:simu-g}
	\begin{figure} [!t] 
	\centering
	\subfigure[$\textcircled{\scriptsize 3}$RxPreSN: $\hat{p}_{g_{N,3}^{\prime}} (  \hat{y}_{S,N} | y_{S,N} ) $ ]{
		\label{fig:cons-RxPreSN}
		\includegraphics[width=0.3\textwidth]{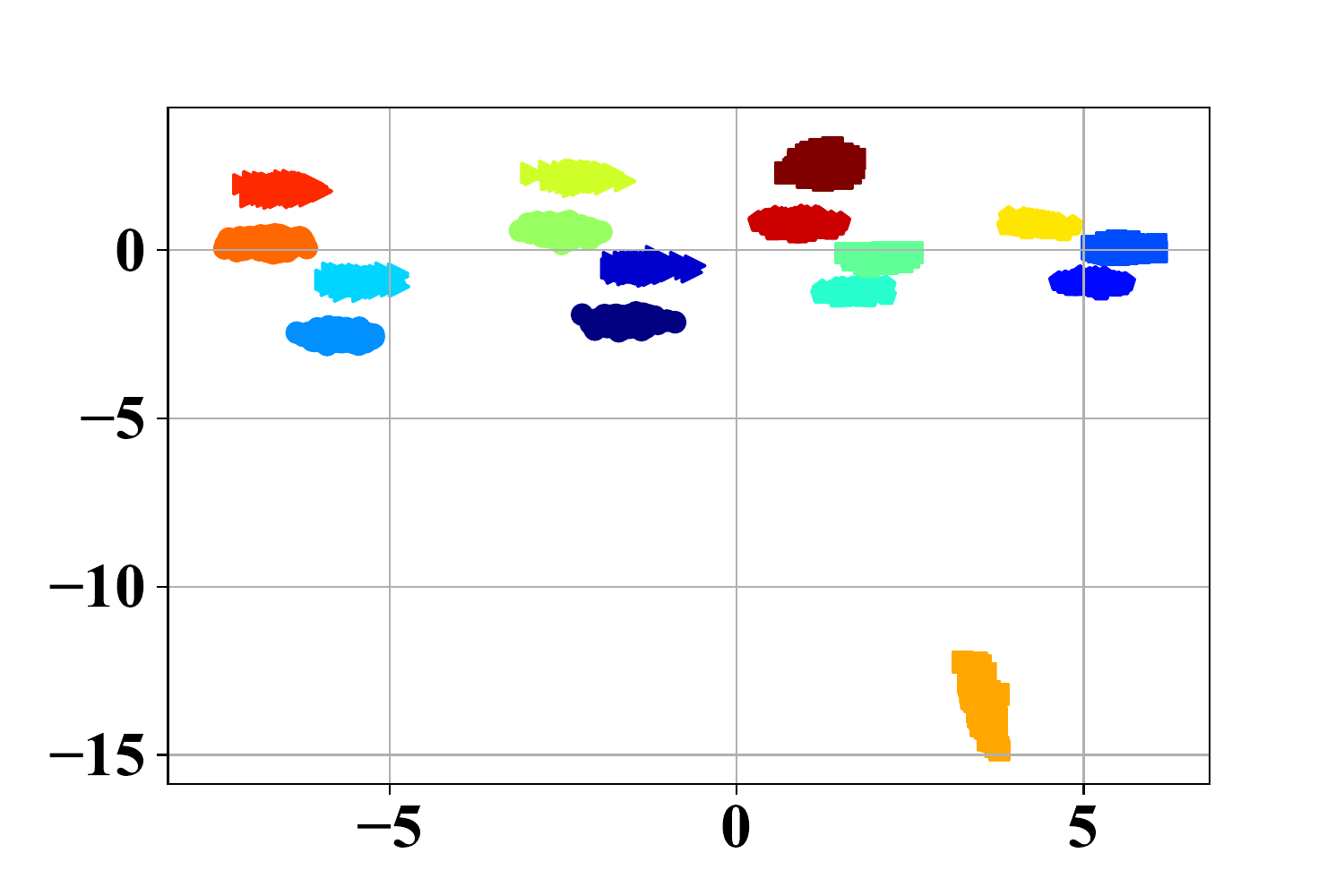}}
	\subfigure[$\textcircled{\scriptsize 7}$RxPreSF: $\hat{p}_{g_{F,7}^{\prime}} (  \hat{y}_{S,F}| y_{S,F} )$ ]{
		\label{fig:cons-RxPreSF}
	\includegraphics[width=0.3\textwidth]{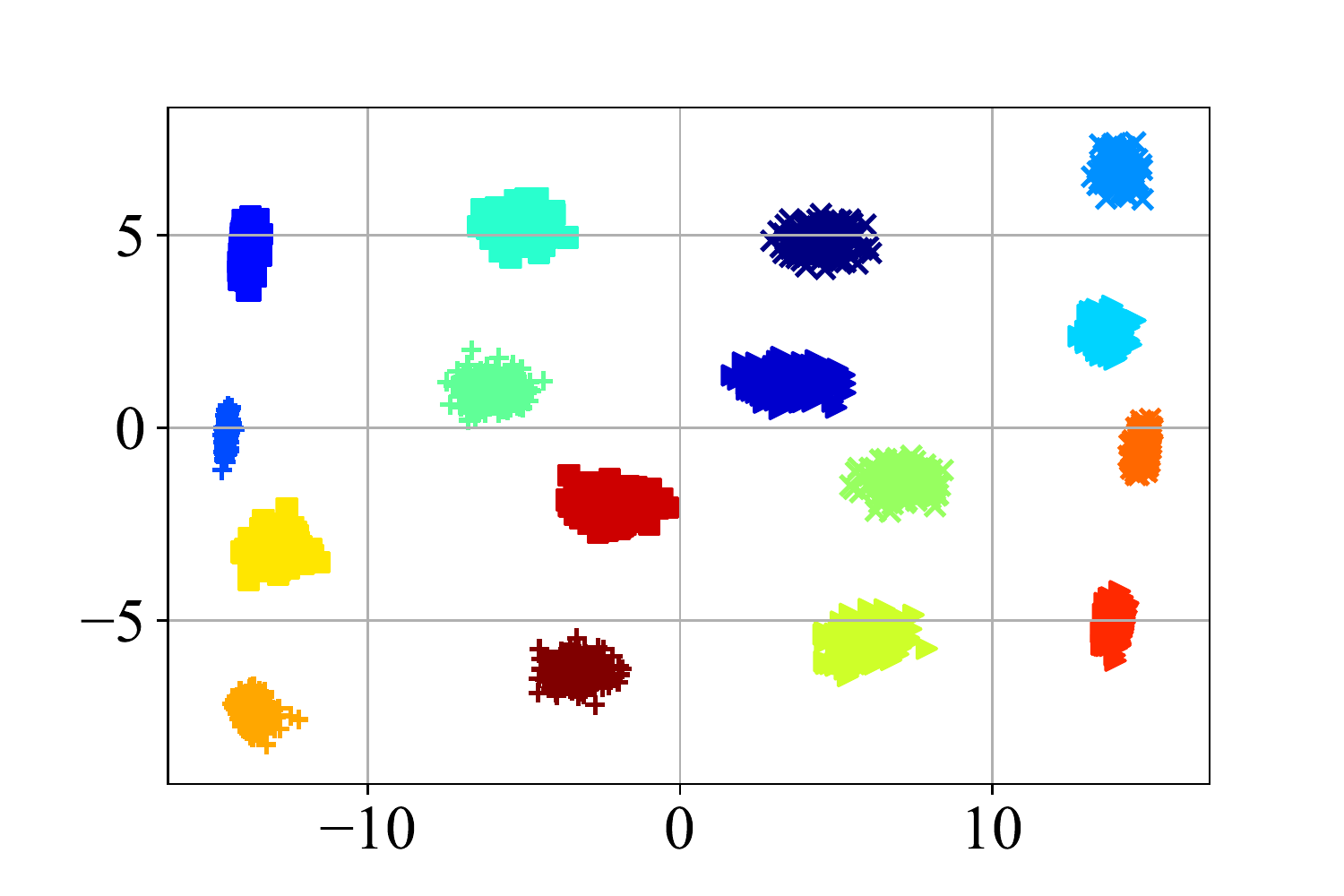}}
	\subfigure[$\textcircled{\scriptsize 8}$RxPreNF: $ \hat{p}_{g_{F,8}^{\prime}} (  \hat{y}_{N,F}| y_{N,F} )$]{
		\label{fig:cons-RxPreNF}
	\includegraphics[width=0.3\textwidth]{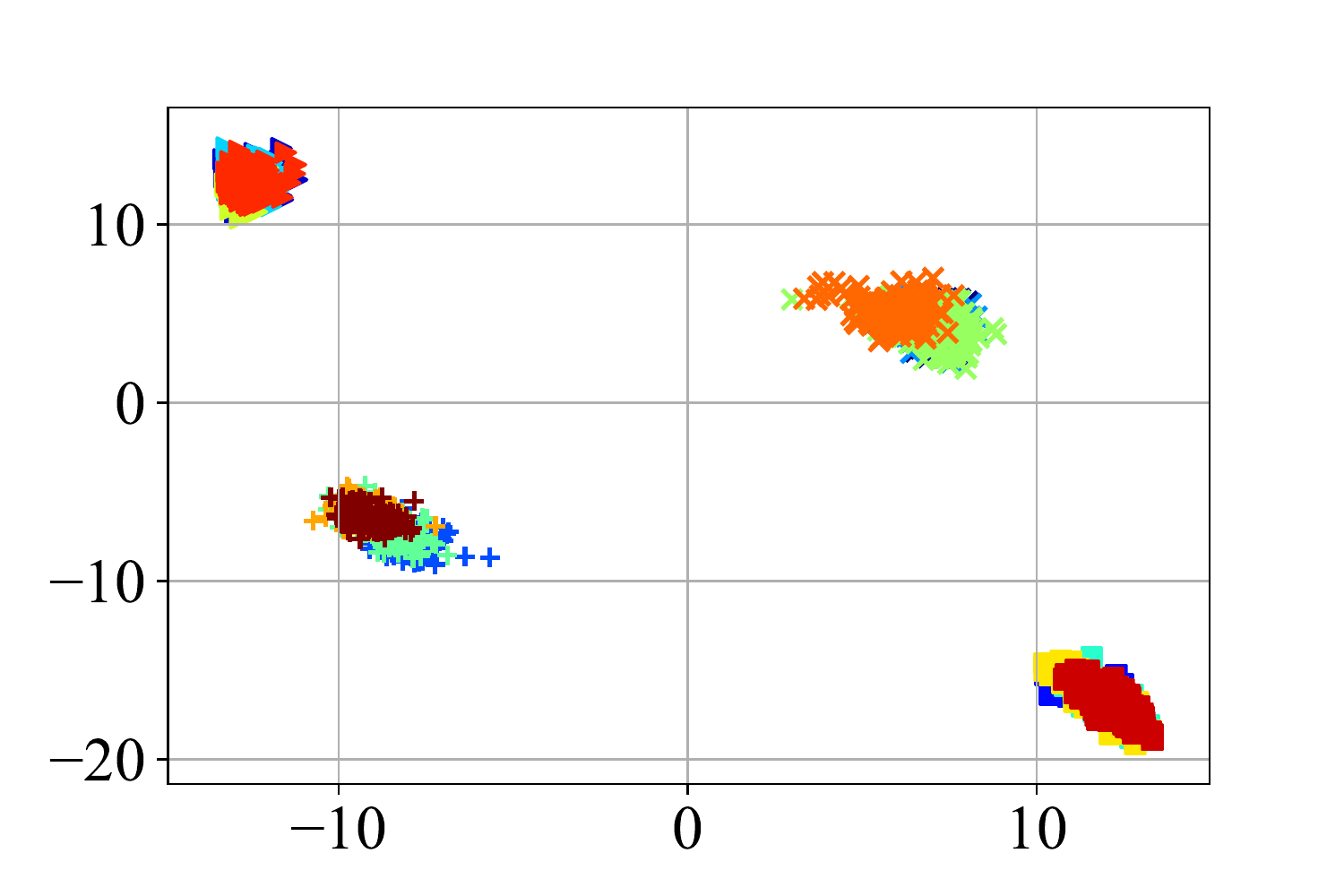}}	
	\subfigure[$p(  x_{S}| y_{S,N} ) \propto   p(y_{S,N}| x_{S} ) $]{
		\label{fig:cons-ySN}
	\includegraphics[width=0.3\textwidth]{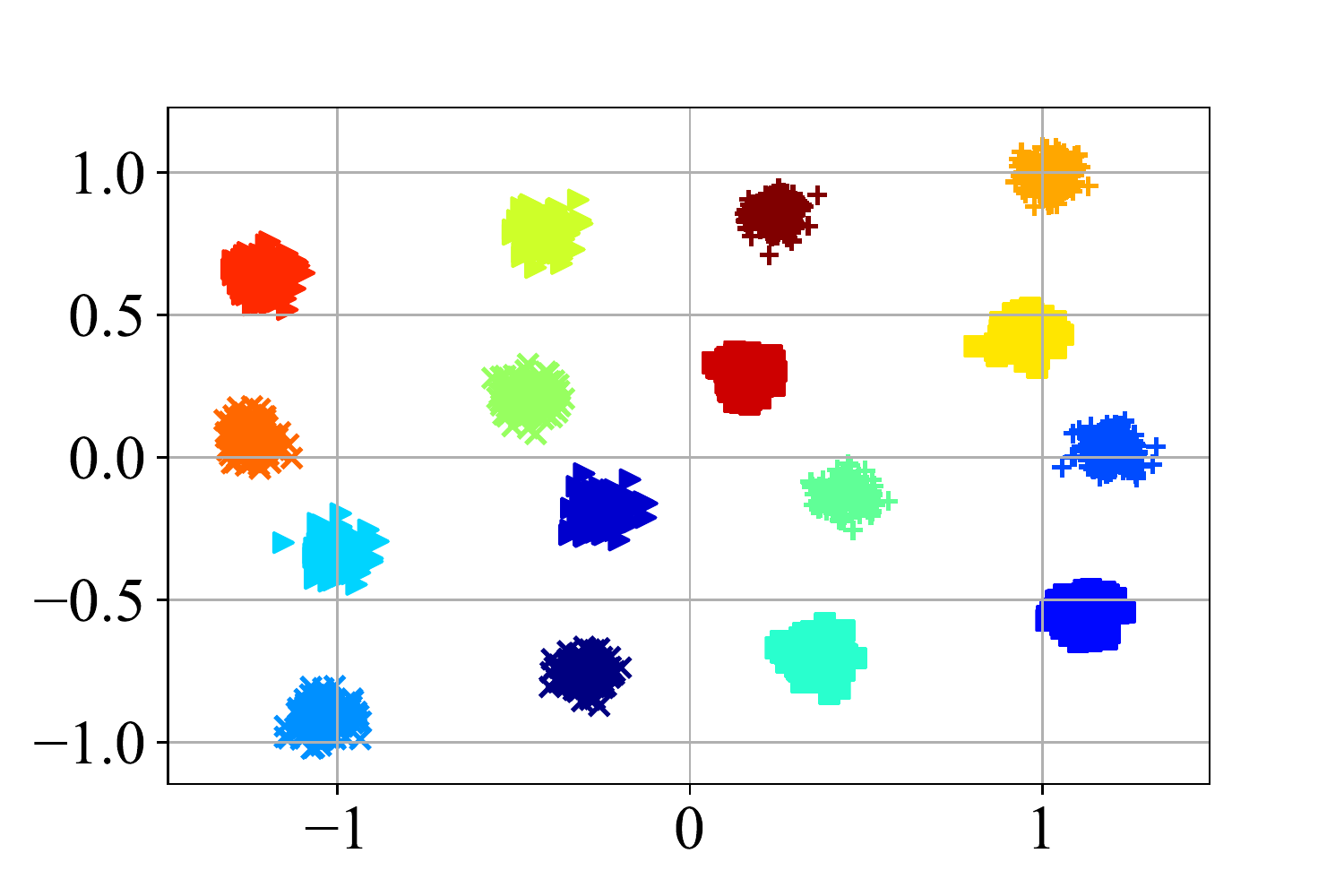}}
	\subfigure[$p(  x_{S}| y_{S,F} ) \propto  p(y_{S,F}| x_{S} ) $ ]{
		\label{fig:cons-ySF}
	\includegraphics[width=0.3\textwidth]{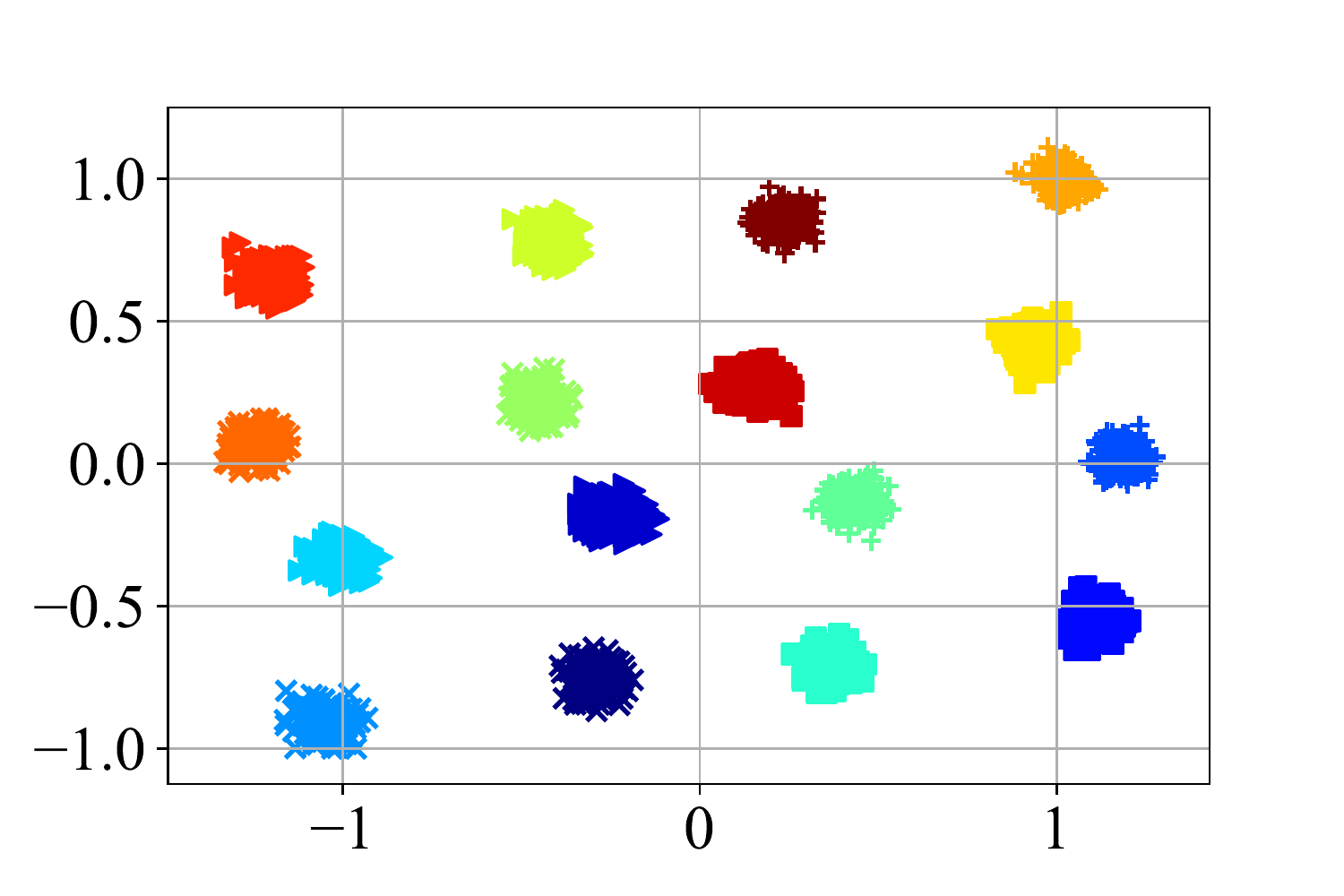}}
	\subfigure[$ p_{ f_{N}^{\prime} } (  \hat{\bm{s}}_{F}^{N} | y_{N,F} )  \propto   p_{ f_{N}^{\prime} } ( y_{N,F}  |  \hat{\bm{s}}_{F}^{N} ) $]{
	\label{fig:cons-yNF}
	\includegraphics[width=0.3\textwidth]{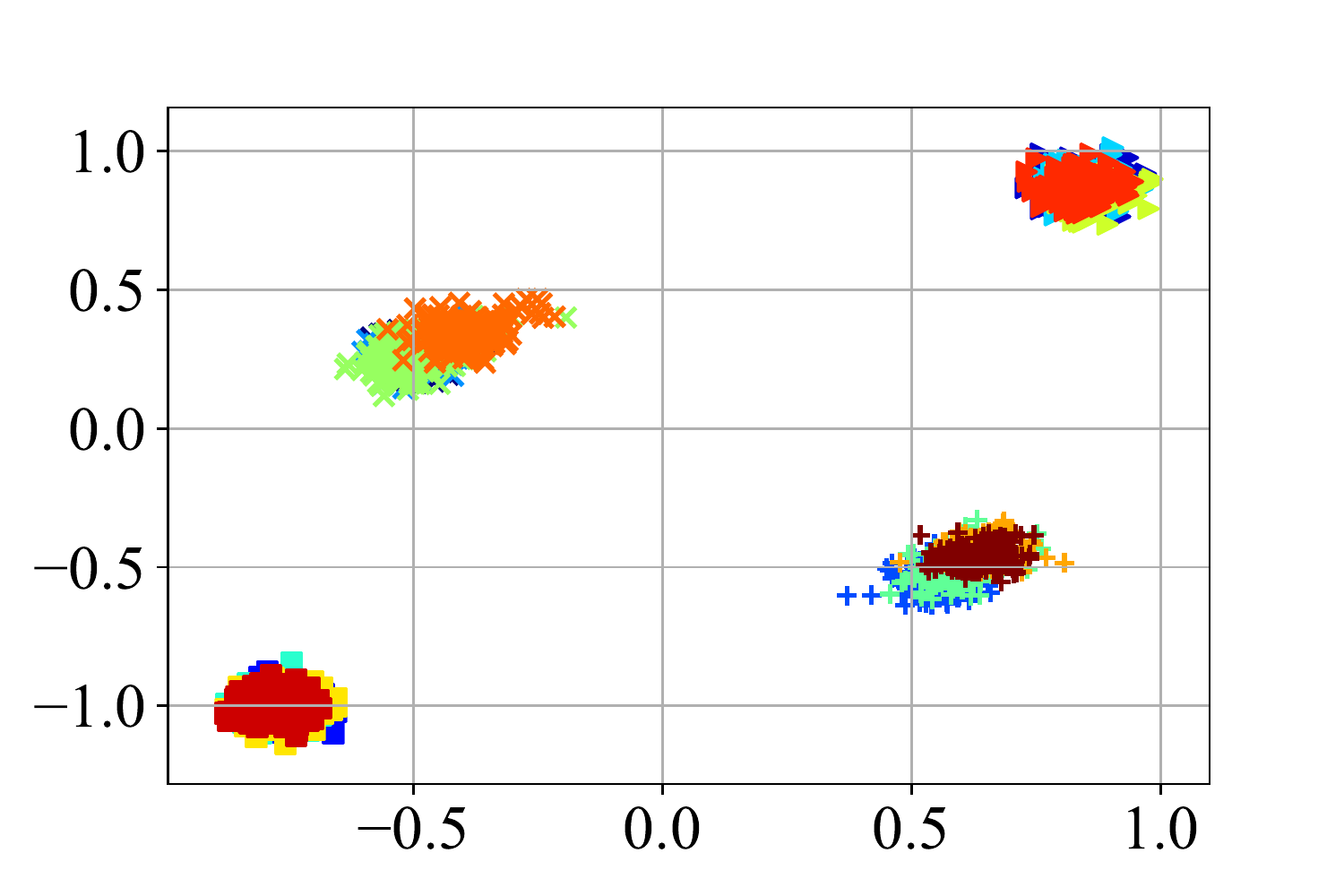}} %
	\caption{Signal clusters corresponding to the learned distributions for $\textcircled{\scriptsize 3}$RxPreSN, $\textcircled{\scriptsize 7}$RxPreSF, and $\textcircled{\scriptsize 8}$RxPreNF (top), and the respective true ones, where $ ( {\alpha}_{S,N}, {\alpha}_{S,F} )  = ( 0.4, 0.6 )$, $\bm{h} = [1,1,1]^T$, and SNR$=25$ dB. The x-axis and y-axis denote the in-phase and quadrature parts, respectively.} 
	\label{fig:cons-all} 
\end{figure}
	The learned distributions of $\textcircled{\scriptsize 3}$RxPreSN, $\textcircled{\scriptsize 7}$RxPreSF, and $\textcircled{\scriptsize 8}$RxPreNF for demapping and the corresponding true ones are shown in Table \ref{table:function-NN}. Here, to verify that $\textcircled{\scriptsize 3}$, $\textcircled{\scriptsize 7}$, and $\textcircled{\scriptsize 8}$ have successfully learned their  respective true distributions, we visualize these distributions in Fig.~\ref{fig:cons-all} by sampling, where each colored cluster consists of $200$ signal points. The results for 
	$\textcircled{\scriptsize 3}$, $\textcircled{\scriptsize 7}$, and $\textcircled{\scriptsize 8}$ are shown in Figs.~\ref{fig:cons-RxPreSN}, \ref{fig:cons-RxPreSF}, and \ref{fig:cons-RxPreNF}, respectively, while the corresponding true distributions in Figs.~\ref{fig:cons-ySN}, \ref{fig:cons-ySF}, and \ref{fig:cons-yNF}, respectively.

	It is shown that the two figures in the same column have  similar cluster shapes, indicating that $\textcircled{\scriptsize 3}$, $\textcircled{\scriptsize 7}$, and $\textcircled{\scriptsize 8}$ have successfully learned the true distributions.
	Besides, it can be seen that various forms of signal transformations have been learned. For example, Fig.~\ref{fig:cons-RxPreSN} can be regarded as a non-uniformly scaled version of Fig.~\ref{fig:cons-ySN}, Fig.~\ref{fig:cons-RxPreSF} can be regarded as a rotated and scaled version of Fig.~\ref{fig:cons-ySF}, while Fig.~\ref{fig:cons-RxPreNF} can be regarded as a mirrored and scaled version of  Fig.~\ref{fig:cons-yNF}. These transformations keep the original signal structure, and meanwhile can introduce more degrees of freedom to facilitate demapping. Similar observations are made in other scenarios.


\subsection{Uncoded BER Performance Comparison for  S1-S4}
\begin{figure}[!t]
	\centering
	\subfigure[For S1]{\label{fig:BER-comp-M4-S1}
	\includegraphics[width=0.47\textwidth]{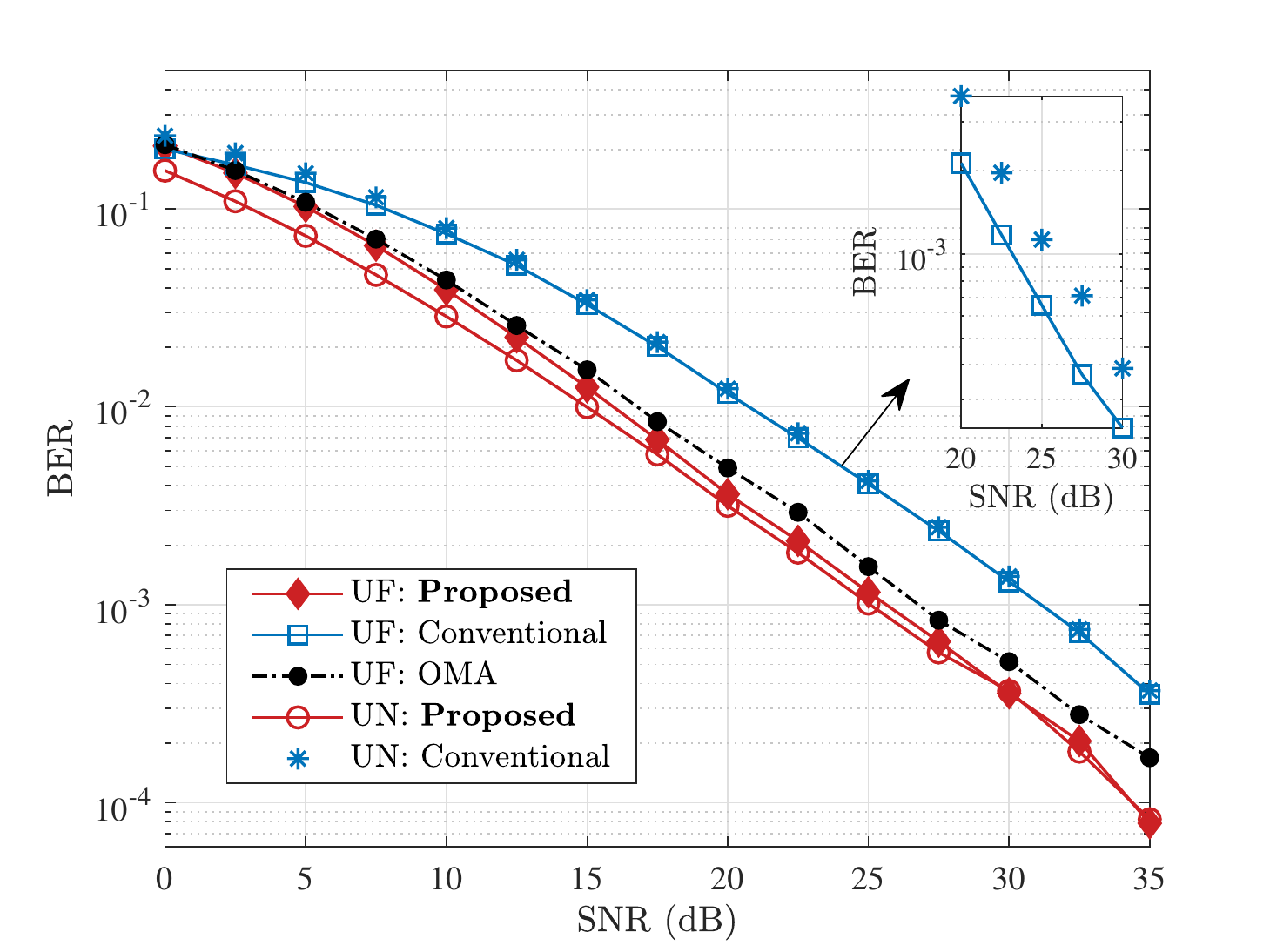} }
	\subfigure[For S2]{\label{fig:BER-comp-M4-S2}
	\includegraphics[width=0.47\textwidth]{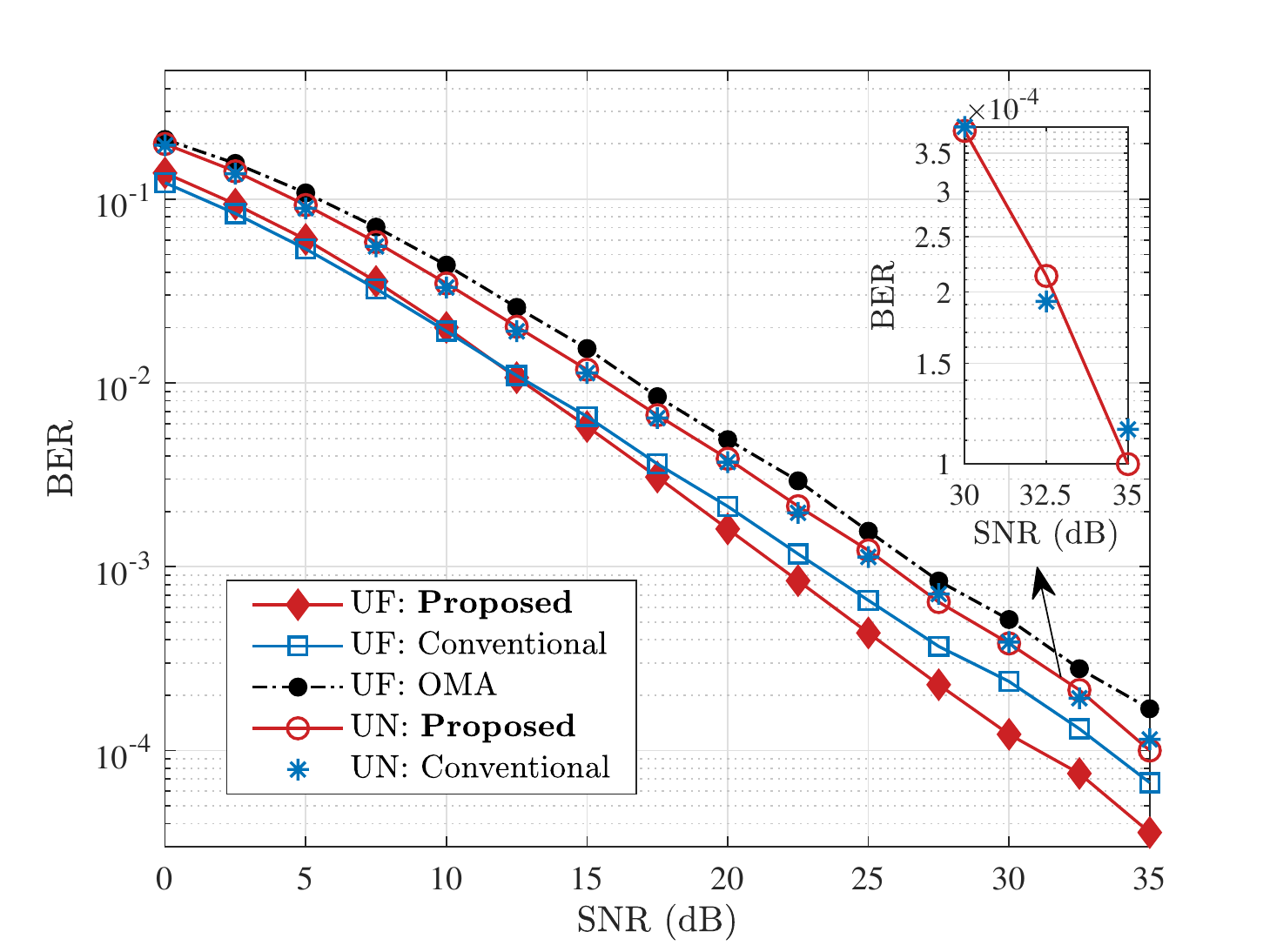} }
	\subfigure[For S3]{\label{fig:BER-comp-M4-S3}
	\includegraphics[width=0.47\textwidth]{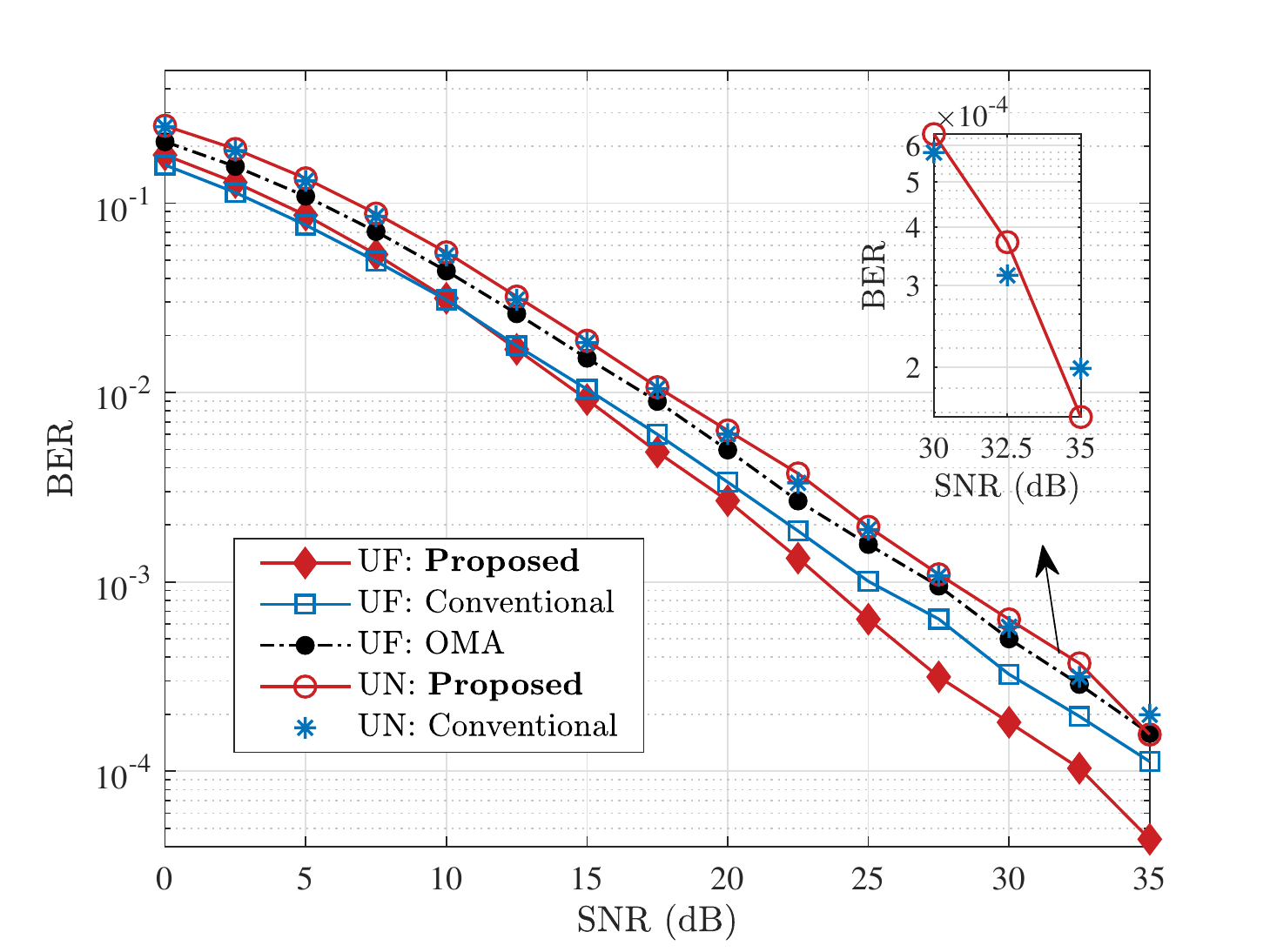} }
	\subfigure[For S4]{\label{fig:BER-comp-M4-S4}
	\includegraphics[width=0.47\textwidth]{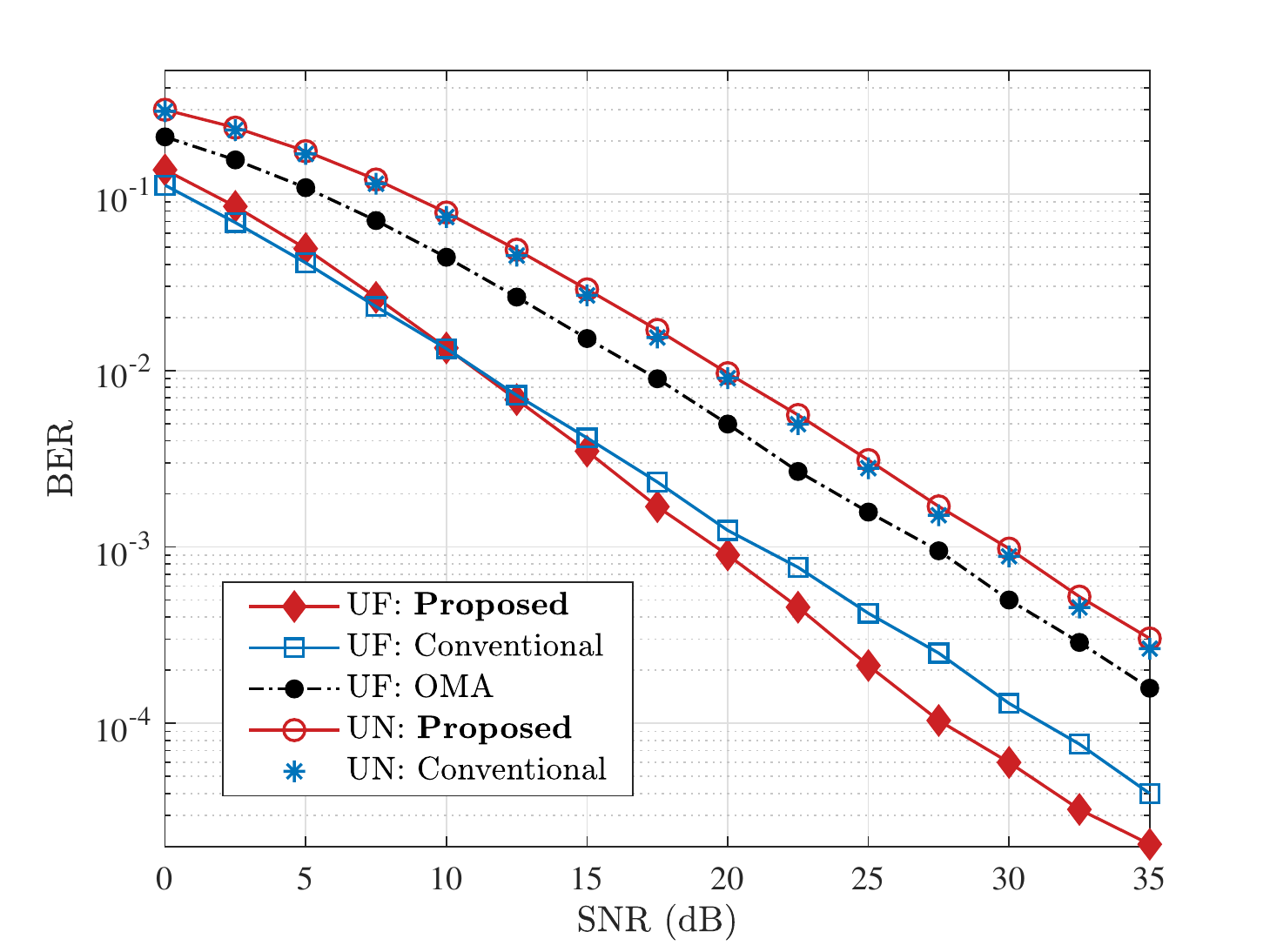} }
	\caption{BER performance comparison of the proposed deep cooperative NOMA scheme, OMA, and the conventional NOMA scheme for different channel scenarios.}
	\label{fig:BER-uncoded}
\end{figure}

Fig. \ref{fig:BER-uncoded} compares the uncoded BER performance of the proposed deep cooperative NOMA, OMA,  and the conventional NOMA for $( {\alpha}_{S,N},  {\alpha}_{S,F} ) = ( \hat{\alpha}_{S,N},  \hat{\alpha}_{S,F} )$, i.e., the PA coefficients for training and inference are the same.

We first consider the scenario S1 in Fig.~\ref{fig:BER-comp-M4-S1}. It is clearly shown that the proposed scheme significantly outperforms the conventional one by $6.25$~dB for both UN and UF, while outperforming the OMA by $1.25$~dB at BER=$10^{-3}$. It can also be seen that the conventional scheme is worse than the OMA scheme in S1, due to the lack of an appropriate PA. 

We then compare the BER with optimized PA coefficients $( {\alpha}_{S,N},  {\alpha}_{S,F}  ) $, as shown in  Fig.~\ref{fig:BER-comp-M4-S2} for S2.  We can see that for UF, the proposed scheme  outperforms the conventional one when SNR$\geq 12.5$~dB, while outperforming the OMA across the whole SNR range. For example, the performance gap between the proposed scheme and the conventional one (resp. OMA) is around $2.5$~dB (resp. $5$~dB) at BER=$10^{-4}$ (resp. $10^{-3}$). For UN, the proposed scheme has a similar BER performance with the  conventional one. Together with Fig.~\ref{fig:BER-comp-M4-S1}, we can see that the proposed scheme is robust to the PA.

In Fig.~\ref{fig:BER-comp-M4-S3}, we compare the BER in S3  with  channel conditions different from S1 and S2. Likewise, for UF, the proposed scheme outperforms the conventional one for SNR$> 12.5$~dB, e.g., by $3$~dB at BER=$10^{-4}$.  It outperforms the OMA across the whole SNR range, e.g., by $3$~dB at BER=$10^{-4}$.
Fig.~\ref{fig:BER-comp-M4-S4} compares the BER in S4 with an unbalanced PA, i.e., $( {\alpha}_{S,N},  {\alpha}_{S,F} ) = ( 0.1, 0.9 )$. Similar observations to Fig. \ref{fig:BER-comp-M4-S3} can be made, and the proposed scheme  outperforms both OMA and the conventional one. 
Moreover, we can see from Figs.~\ref{fig:BER-comp-M4-S2}-\ref{fig:BER-comp-M4-S4} that the proposed scheme shows a larger decay rate for UF BER for large SNRs, revealing that the demapping errors at UN are successfully learned and compensated at UF, achieving higher diversity orders.  
 
\subsection{Adaptation to Power Allocation for S5 and S6} 
\begin{figure}[!t]
	\centering
	\subfigure[For S5]{\label{fig:BER-comp-M4-S5}
	\includegraphics[width=0.47\textwidth]{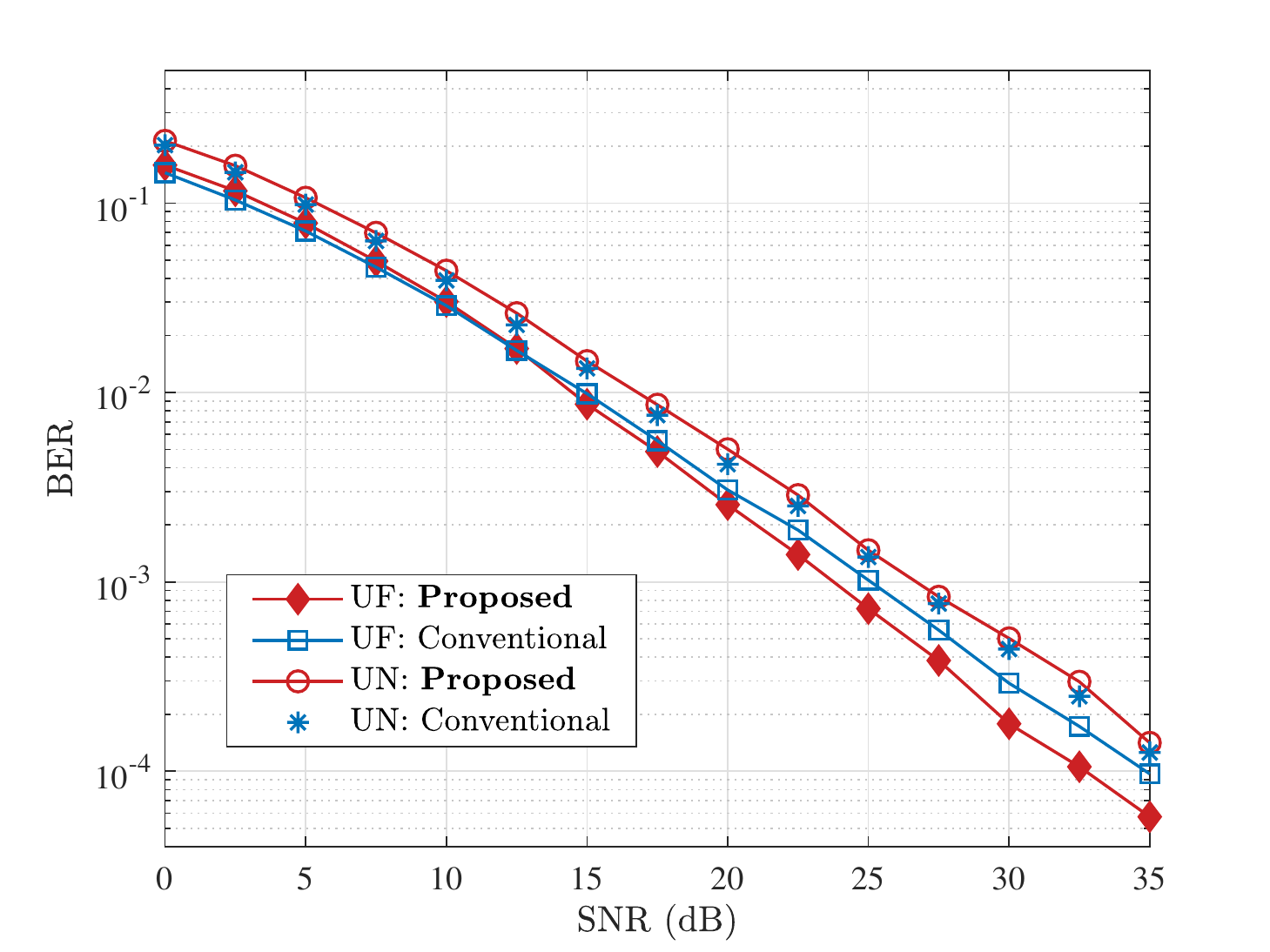} }
	\subfigure[For S6]{\label{fig:BER-comp-M4-S6}
	\includegraphics[width=0.47\textwidth]{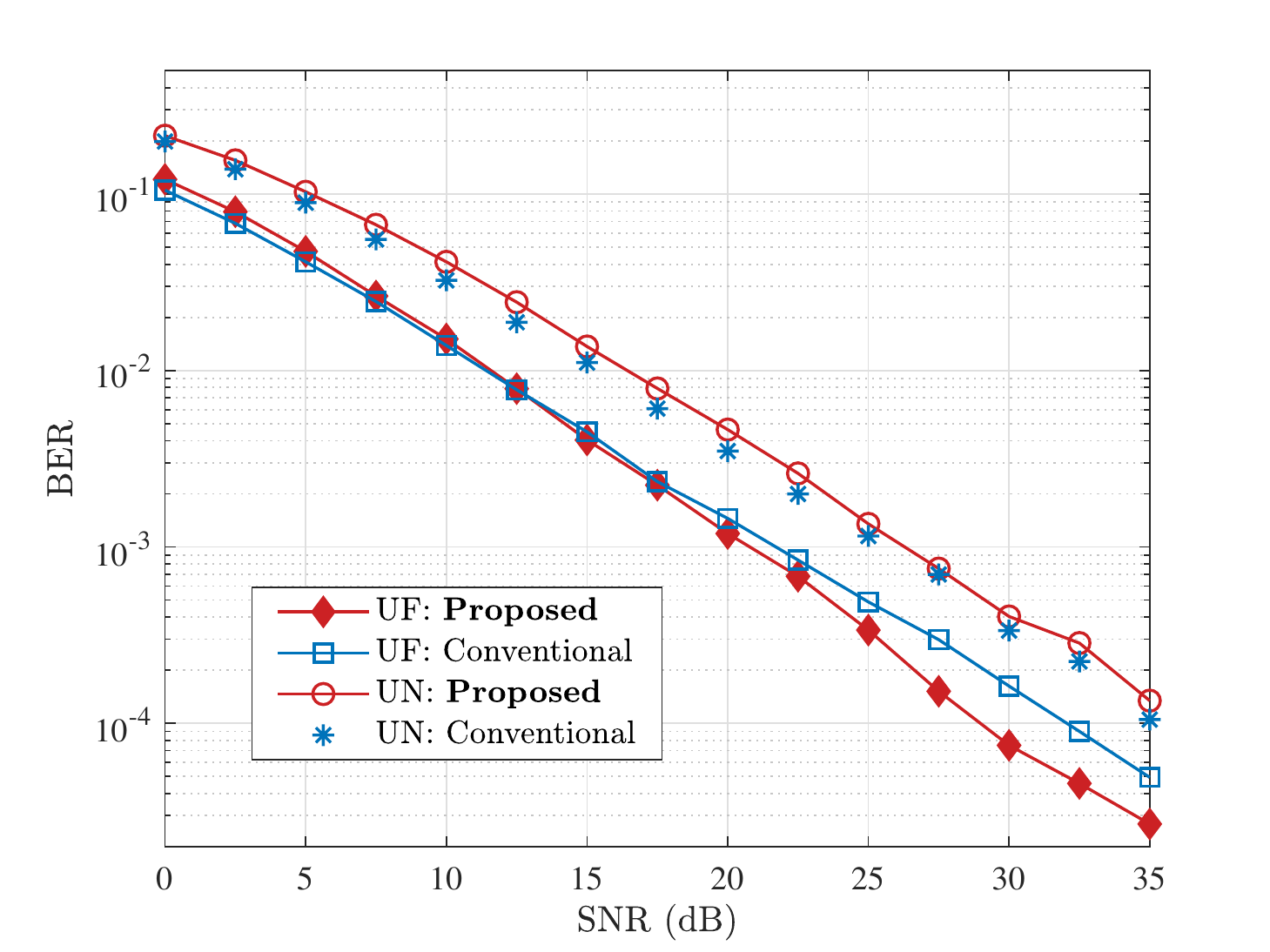} }
	\caption{BER performance comparison of the proposed deep cooperative NOMA and the conventional NOMA schemes with PA mismatch between training and inference.}
\end{figure} 
To demonstrate its adaptation to the mismatch between the training and inference PA discussed in Section \ref{sec:pa}, we validate the proposed scheme in S5 ($\hat{\alpha}_{S,F} < {\alpha}_{S,F} $) and S6 ($\hat{\alpha}_{S,F} > {\alpha}_{S,F} $) in Figs.~\ref{fig:BER-comp-M4-S5} and \ref{fig:BER-comp-M4-S6}, respectively. It can be seen that for UF, the proposed scheme outperforms the conventional one at SNR$>15$~dB. It can also be seen that the proposed scheme still achieves larger BER decay rates in both S5 and S6. These results clearly verify that, without carrying out a new training process, the proposed scheme 
can handle  the PA mismatch.  
\subsection{BER Performance Comparison with Channel Coding}
\begin{figure}[!t]
	\centering
	\subfigure[For S2]{\label{fig:BER-comp-coded-S2-NaF}
	\includegraphics[width=0.47\textwidth]{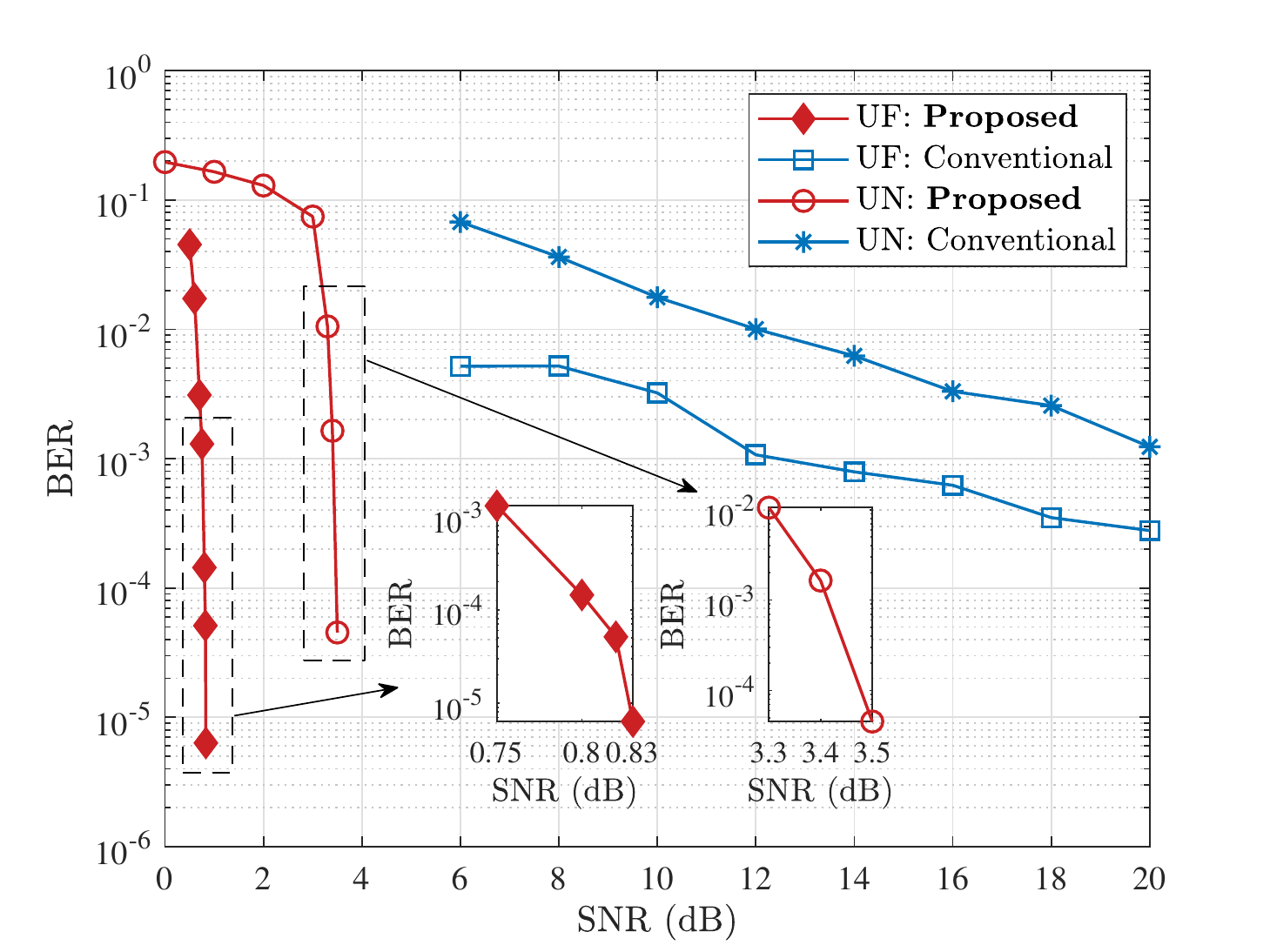} }
	\subfigure[For S4]{\label{fig:BER-comp-coded-S4-NaF}
	\includegraphics[width=0.47\textwidth]{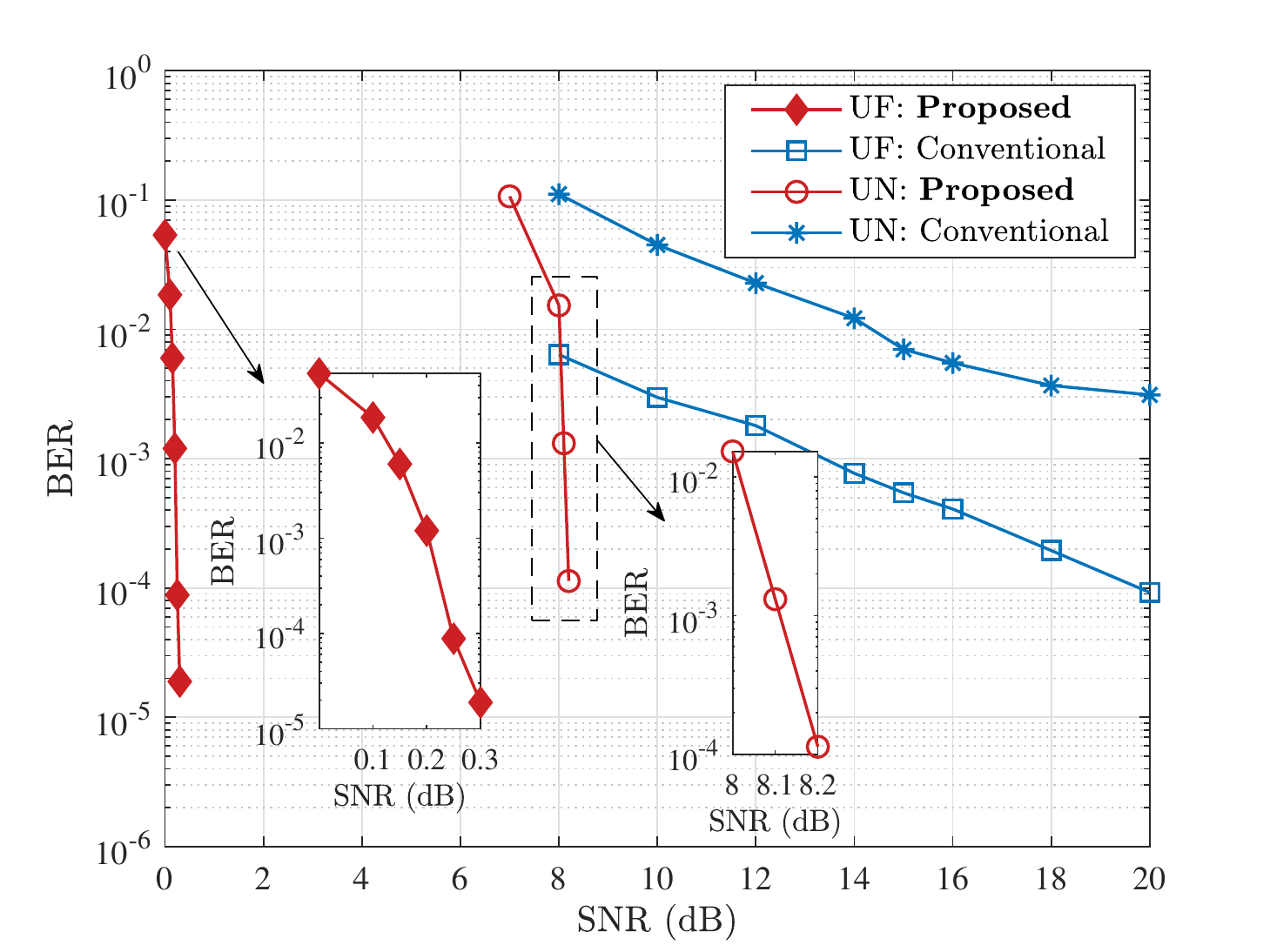} }
	\caption{BER performance comparison of the proposed deep cooperative NOMA and the conventional NOMA schemes with the LDPC code.}
	\label{fig:BER-LDPC}
\end{figure} 

In Fig.~\ref{fig:BER-LDPC}, we evaluate the coded BER performance with the LDPC code in S2 and S4. The code parity-check matrix comes from the DVB-S2 standard \cite{DVB-S2} with the rate $1/2$ and size of $32400 \times 64800$. Therefore, $\bm{c}_{N}$ and $\bm{c}_{F}$ have the length of  $32400$ bits, while the $\mathcal{E}( \cdot )$ encoded $ \langle \bm{s}_{N} \rangle $ and $ \langle \bm{s}_{F} \rangle$ have the length of  $64800$ bits. The LDPC decoder $\mathcal{D} ( \cdot ) $ is based on the classic belief propagation algorithm with soft LLR as input.  The coded BER is defined as $\Pr \{ \bm{c}_{J} \neq \hat{\bm{c}}_{J} \}$, $J \in \{ N, F\}$. For the conventional scheme,  UN adopts SIC due to its low computational complexity. Specifically, it first decodes  $\bm{c}_{F}$ as   $\hat{\bm{c}}_{F}^{N}   = \mathcal{D} ( \cdot )$, cancels the interference after re-encoding and re-modulating $\hat{\bm{c}}_{F}^{N}$, and then decodes $\hat{\bm{c}}_{N}$. Then, UN forwards the re-modulated signal to UF. Note that the  decoding is terminated on reaching  the maximum number of decoding iterations ($50$ here) or when all parity checks are satisfied.  

In both scenarios,  we observe a significant increasing decoding performance gap between the proposed and  conventional schemes.  
For example, in Fig.~\ref{fig:BER-comp-coded-S4-NaF}, to achieve BER=$10^{-4}$ for UF, the SNRs for the proposed and the conventional\footnote{The performance of the conventional scheme can also be found in \cite{LDPC-NOMA-pan2018sic}.} schemes are $0.25$ and $20$~dB, respectively, which shows a  gap more than $19$~dB. Similar observations can be made from Fig.~\ref{fig:BER-comp-coded-S2-NaF}. 
The performance superiority of the proposed scheme mainly originates from its utilizations of soft information and the parallel demapping at UN attributing to the error performance optimization. In the meantime, the performance of the conventional scheme is limited to the interference and error propagation \cite{LDPC-NOMA-pan2018sic}.  


\subsection{Computational Complexity Comparisons}

As discussed before, we adopt the offline-training and online-deploying mode for the proposed scheme. Therefore, we only need to consider the computational complexity in the online-deploying phase. Specifically, in the uncoded case, the complexity for signal detection is $\mathcal{O}(2^{k})$ for the conventional scheme. By contrast, the mapping-demapping complexity is $\mathcal{O}(k)$ for the proposed scheme, which is only linear in $k$. In the coded case, the conventional scheme includes two decoding processes to  jointly decode $\hat{\bm{s}}_{N}$ and $\hat{\bm{s}}_{F}^{N}$  at UN,  resulting in a high decoding complexity. The proposed scheme only involves a single decoding process to separately decode its own information $\hat{\bm{s}}_{N}$, so that a low-complexity demapping-and-forward scheme can be used for the UF signal.

\section{Conclusion}
In this paper, we proposed a novel deep cooperative NOMA scheme to optimize the BER performance. We developed a new hybrid-cascaded DNN architecture to represent the cooperative NOMA system, which can then be optimized in a holistic manner.  Multiple loss functions were constructed to quantify the BER performance, and a novel multi-task oriented two-stage training method was proposed to solve the end-to-end training problem in a self-supervised manner.  Theoretical perspective was then  established to reveal the learning mechanism of each DNN module.  Simulation results demonstrate the merits of our scheme over OMA and the conventional  NOMA scheme in various channel environments. As a main advantage, the proposed scheme can adapt to PA mismatch between training and inference, and can be incorporated with channel coding to combat signal deterioration.
In our future work, we will consider the system designs for high-order constellations, transmission rate adaptation\cite{concl-makki2020error}, and  grant-free access \cite{concl-8454392}, and to include more cooperative users \cite{concl-8989311,concl-8726376}.



%
%

\appendix
\subsection{Relationship among $\{ L_1, L_2 , L_3\}$} \label{app:P3}
Demapping at UN is described as in \eqref{eq:det-1}, and $\{ L_1, L_2 \}$ are the associated loss functions. The ultimate end-to-end  demapping at UF is described in \eqref{eq:det-2}, and $L_3$ is the associated end-to-end loss for the entire network. Let us observe \eqref{eq:det-2}. There are in total three random processes $\{\mathcal{C}_{S,F}, \mathcal{C}_{N,F}, \mathcal{C}_{S,N} \}$ (due to noise), while the remaining $\big\{ f_{S}^{\prime},  f_{N}^{\prime},  g_{N}^{\prime}, g_{F}^{\prime} \big\}$ are trainable modules.
$\big\{ f_{S}^{\prime},  f_{N}^{\prime},  g_{N}^{\prime}, g_{F}^{\prime} \big\}$ are determined through training to combat the randomness from $\{\mathcal{C}_{S,F}, \mathcal{C}_{N,F}, \mathcal{C}_{S,N} \}$. We can see that to solve \eqref{eq:det-2} exactly, $\{ h_{S,F}, h_{N,F}, h_{S,N} \}$ are all needed to describe $\{\mathcal{C}_{S,F}, \mathcal{C}_{N,F}, \mathcal{C}_{S,N} \}$ correspondingly. However, $h_{S,N}$ is practically not available at UF (recalling Section \ref{sec:sys-model}), meaning that $\mathcal{C}_{S,N}$ lacks description. This unavailable knowledge may potentially lead to a poor demapping performance.
We further observe that conditioned on the case of $\hat{\bm{s}}_{F}^{N} =\bm{s}_{F}$, \eqref{eq:det-2} can be described as
\begin{align}
\hat{\bm{s}}_{F} =   g_{F}^{\prime}  ( & \mathcal{C}_{S,F} \circ f_{S}^{\prime}  (\bm{s}_{N}, \bm{s}_{F}) ,  \mathcal{C}_{N,F} \circ \underbrace{f_{N}^{\prime}}_{\text{Input: }\hat{\bm{s}}_{F}^{N}}  \circ \underbrace{g_{N}^{\prime} \circ \mathcal{C}_{S,N} \circ f_{S}^{\prime}  (\bm{s}_{N}, \bm{s}_{F}) }_{\text{ Output:  } \big(\hat{\bm{s}}_{N}, \hat{\bm{s}}_{F}^{N} \big)} \vert \hat{\bm{s}}_{F}^{N} =\bm{s}_{F} ) \notag \\
= g_{F}^{\prime}   ( &   \mathcal{C}_{S,F} \circ f_{S}^{\prime}  (\bm{s}_{N}, \bm{s}_{F})  , \mathcal{C}_{N,F} \circ  f_{N}^{\prime} ( \bm{s}_{F}) \vert \hat{\bm{s}}_{F}^{N} =\bm{s}_{F}  ), \label{eq:det-2-ax}
\end{align}	
where the description of $\mathcal{C}_{S,N}$ can be avoided. This observation inspires us to maximize $\Pr \big\{ \hat{\bm{s}}_{F}^{N} =\bm{s}_{F} \big\}$ to achieve \eqref{eq:det-2-ax} and solve \eqref{eq:det-2} exactly with a high probability.
It needs to be pointed out that the case $\hat{\bm{s}}_{F}^{N} =\bm{s}_{F}$ means that the demapping \eqref{eq:det-1} at UN succeeds for $\bm{s}_{F}$ ($L_2$   is sufficiently small). The above analysis reveals the causal structure between \eqref{eq:det-1} and \eqref{eq:det-2}, which motivates us to perform optimization first  for \eqref{eq:det-1} and then for \eqref{eq:det-2}. Note that $\Pr \big\{ \hat{\bm{s}}_{F}^{N} =\bm{s}_{F} \big \}$ can be maximized (or equivalently, $L_2$ can be minimized) through training stage I to achieve \eqref{eq:det-2-ax}. 
Besides, considering the causal structure, in stage II, the modules learned from stage I, i.e., $f_{S}^{\prime}$ and $ g_{N}^{\prime}$, are fixed. 
\subsection{Derivations of \eqref{eq:LossNN-1} and \eqref{eq:LossNN}}
\label{app:A}
First, \eqref{eq:LossNN-1} can be derived according to \eqref{eq:BMI-deri-iid} by averaging over the channel output $y_{S,N}$. Then, by applying 
$I_{f_{S}^{\prime} }( {\bf  S}_{N}(r) , {\bf  S}_{F}(r) ; Y_{S,N} ) = I_{f_{S}^{\prime} }( {\bf  S}_{N}(r)  ; Y_{S,N} ) + I_{f_{S,2}^{\prime} }( {\bf  S}_{F}(r) ; Y_{S,N} | {\bf  S}_{N}(r) )$ from information theory and $p_{ f_{S}^{\prime}} (  \bm{s}_{N}(r) | y_{S,N} ) = \int_{x_{S}} p_{ f_{S}^{\prime}}  (  \bm{s}_{N}(r) | x_{S} )
p(  x_{S}| y_{S,N} ) d x_{S}$ from probability theory (to include the composite signal $x_{S}$), \eqref{eq:LossNN} can be derived from \eqref{eq:LossNN-1}.
\subsection{Proof of SINR Values for \eqref{eq:pa-N}, \eqref{eq:pa-NF}, and \eqref{eq:pa-F}} \label{app:pa}	
Taking a closer look at  \eqref{eq:pa-N} and \eqref{eq:pa-NF}, the respective inputs of  $g_{N}^{\prime}(\cdot)$ can be written as
\begin{align}
    \frac{1}{\omega_{N}} y_{S,N}  = &  h_{S,N} \Big( \sqrt{{\alpha}_{S,N}} x_{N} + \frac{1}{\omega_{N}} \sqrt{\hat{\alpha}_{S,F} } x_{F} \Big) + \frac{1}{\omega_{N}} n_{S,N}, \label{eq:pa-app-N} \\
    \frac{1}{\omega_{F}}y_{S,N}  =  &  h_{S,N} \Big( \sqrt{{\alpha}_{S,F}} x_{F} + \frac{1}{\omega_{F}} \sqrt{\hat{\alpha}_{S,N}} x_{N}\Big) + \frac{1}{\omega_{F}}n_{S,N}. \label{eq:pa-app-NF}
\end{align}
The SINRs can be calculated as $\frac{ \hat{\alpha}_{S,N} |h_{S,N}|^2 }{  \hat{\alpha}_{S,F} |h_{S,N}|^2 + 2 \sigma_{S,N}^2 }$ from \eqref{eq:pa-app-N} and $\frac{ \hat{\alpha}_{S,F} |h_{S,N}|^2 }{  \hat{\alpha}_{S,N} |h_{S,N}|^2 + 2 \sigma_{S,N}^2 }$ from \eqref{eq:pa-app-NF}. Similar proof can be given for \eqref{eq:pa-F}.

\bibliographystyle{IEEEtran} 
\bibliography{ref_coopNOMA}

\begin{thebibliography}{10}
\providecommand{\url}[1]{#1}
\csname url@samestyle\endcsname
\providecommand{\newblock}{\relax}
\providecommand{\bibinfo}[2]{#2}
\providecommand{\BIBentrySTDinterwordspacing}{\spaceskip=0pt\relax}
\providecommand{\BIBentryALTinterwordstretchfactor}{4}
\providecommand{\BIBentryALTinterwordspacing}{\spaceskip=\fontdimen2\font plus
\BIBentryALTinterwordstretchfactor\fontdimen3\font minus
  \fontdimen4\font\relax}
\providecommand{\BIBforeignlanguage}[2]{{%
\expandafter\ifx\csname l@#1\endcsname\relax
\typeout{** WARNING: IEEEtran.bst: No hyphenation pattern has been}%
\typeout{** loaded for the language `#1'. Using the pattern for}%
\typeout{** the default language instead.}%
\else
\language=\csname l@#1\endcsname
\fi
#2}}
\providecommand{\BIBdecl}{\relax}
\BIBdecl

\bibitem{Intro-NOMA-book-vaezi2019multiple}
M.~Vaezi, Z.~Ding, and H.~V. Poor, \emph{Multiple access techniques for 5{G}
  wireless networks and beyond}.\hskip 1em plus 0.5em minus 0.4em\relax Berlin,
  German: Springer, 2019.

\bibitem{NOMA-liu2020non}
Y.~Liu, Z.~Qin, and Z.~Ding, \emph{Non-Orthogonal Multiple Access for Massive
  Connectivity}.\hskip 1em plus 0.5em minus 0.4em\relax Switzerland: Springer,
  2020.

\bibitem{Intro-NOMA-ding2017}
Z.~{Ding}, Y.~{Liu}, J.~{Choi}, Q.~{Sun}, M.~{Elkashlan}, I.~{Chih-Lin}, and
  H.~V. {Poor}, ``Application of non-orthogonal multiple access in {LTE} and
  5{G} networks,'' \emph{IEEE Commun. Mag.}, vol.~55, no.~2, pp. 185--191, Feb.
  2017.

\bibitem{intro-NOMA-dai2018asur}
L.~{Dai}, B.~{Wang}, Z.~{Ding}, Z.~{Wang}, S.~{Chen}, and L.~{Hanzo}, ``A
  survey of non-orthogonal multiple access for 5{G},'' \emph{IEEE Commun.
  Surveys Tuts.}, vol.~20, no.~3, pp. 2294--2323, May 2018.

\bibitem{coNOMA-ding2015}
Z.~{Ding}, M.~{Peng}, and H.~V. {Poor}, ``Cooperative non-orthogonal multiple
  access in 5{G} systems,'' \emph{IEEE Commun. Lett.}, vol.~19, no.~8, pp.
  1462--1465, Aug. 2015.

\bibitem{intro-CRS-NOMA-kim2015cap}
J.~{Kim} and I.~{Lee}, ``Capacity analysis of cooperative relaying systems
  using non-orthogonal multiple access,'' \emph{IEEE Commun. Lett.}, vol.~19,
  no.~11, pp. 1949--1952, Nov. 2015.

\bibitem{coopNOMA2017-8108407}
Z.~{Wei}, L.~{Dai}, D.~W.~K. {Ng}, and J.~{Yuan}, ``Performance analysis of a
  hybrid downlink-uplink cooperative {NOMA} scheme,'' in \emph{2017 IEEE 85th
  Vehicular Technology Conference (VTC Spring)}, Sydney, Australia, June 2017,
  pp. 1--7.

\bibitem{coNOMA-AF-eb2019}
O.~{Abbasi}, A.~{Ebrahimi}, and N.~{Mokari}, ``{NOMA} inspired cooperative
  relaying system using an {AF} relay,'' \emph{IEEE Wireless Commun. Lett.},
  vol.~8, no.~1, pp. 261--264, Sep. 2019.

\bibitem{copNOMA-zhou2018dynamic}
Y.~Zhou, V.~W. Wong, and R.~Schober, ``Dynamic decode-and-forward based
  cooperative {NOMA} with spatially random users,'' \emph{IEEE Trans. Wireless
  Commun.}, vol.~17, no.~5, pp. 3340--3356, Mar. 2018.

\bibitem{copNOMA-liu2016cooperative}
Y.~Liu, Z.~Ding, M.~Elkashlan, and H.~V. Poor, ``Cooperative non-orthogonal
  multiple access with simultaneous wireless information and power transfer,''
  \emph{IEEE J. Sel. Areas Commun.}, vol.~34, no.~4, pp. 938--953, Mar. 2016.

\bibitem{BER-coNOMA-kara2019}
F.~{Kara} and H.~{Kaya}, ``On the error performance of cooperative-{NOMA} with
  statistical {CSIT},'' \emph{IEEE Commun. Lett.}, vol.~23, no.~1, pp.
  128--131, Jan. 2019.

\bibitem{BER-coNOMA-kara2020}
------, ``Error probability analysis of {NOMA}-based diamond relaying
  network,'' \emph{IEEE Trans. Veh. Technol.}, vol.~69, no.~2, pp. 2280--2285,
  Feb. 2020.

\bibitem{NOMA-ber-li2019spatial}
Q.~Li, M.~Wen, E.~Basar, H.~V. Poor, and F.~Chen, ``Spatial modulation-aided
  cooperative {NOMA}: Performance analysis and comparative study,'' \emph{IEEE
  J. Sel. Topics Signal Process.}, vol.~13, no.~3, pp. 715--728, Feb. 2019.

\bibitem{book-goodfellow2016deep}
I.~Goodfellow, Y.~Bengio, and A.~Courville, \emph{Deep learning}.\hskip 1em
  plus 0.5em minus 0.4em\relax MIT press, 2016.

\bibitem{review-DL-Pqin2019deep}
Z.~Qin, H.~Ye, G.~Y. Li, and B.-H.~F. Juang, ``Deep learning in physical layer
  communications,'' \emph{IEEE Wireless Commun.}, vol.~26, no.~2, pp. 93--99,
  Mar. 2019.

\bibitem{intro-zhang2019deep}
C.~Zhang, P.~Patras, and H.~Haddadi, ``Deep learning in mobile and wireless
  networking: A survey,'' \emph{IEEE Commun. Surveys Tuts.}, vol.~21, no.~3,
  pp. 2224--2287, Mar. 2019.

\bibitem{onoffline-he2019model}
H.~He, S.~Jin, C.-K. Wen, F.~Gao, G.~Y. Li, and Z.~Xu, ``Model-driven deep
  learning for physical layer communications,'' \emph{IEEE Wireless Commun.},
  vol.~26, no.~5, pp. 77--83, May 2019.

\bibitem{6G-letaief2019roadmap}
K.~B. Letaief, W.~Chen, Y.~Shi, J.~Zhang, and Y.-J.~A. Zhang, ``The roadmap to
  6{G}: {AI} empowered wireless networks,'' \emph{IEEE Commun. Mag.}, vol.~57,
  no.~8, pp. 84--90, Aug. 2019.

\bibitem{ch-est-dong2019deep}
P.~Dong, H.~Zhang, G.~Y. Li, I.~S. Gaspar, and N.~NaderiAlizadeh, ``Deep
  {CNN}-based channel estimation for mm{W}ave massive {MIMO} systems,''
  \emph{IEEE J. Sel. Topics Signal Process.}, vol.~13, no.~5, pp. 989--1000,
  July 2019.

\bibitem{DL-MIMO-he2020model}
H.~He, C.-K. Wen, S.~Jin, and G.~Y. Li, ``Model-driven deep learning for {MIMO}
  detection,'' \emph{IEEE Trans. Signal Process.}, vol.~68, pp. 1702--1715,
  Feb. 2020.

\bibitem{ours-lu2020deep}
Y.~Lu, P.~Cheng, Z.~Chen, Y.~Li, W.~H. Mow, and B.~Vucetic, ``Deep autoencoder
  learning for relay-assisted cooperative communication systems,'' \emph{IEEE
  Trans. Commun.}, 2020, early access.

\bibitem{ae-o2017introduction}
T.~O'Shea and J.~Hoydis, ``An introduction to deep learning for the physical
  layer,'' \emph{IEEE Trans. Cogn. Commun. Netw.}, vol.~3, no.~4, pp. 563--575,
  Dec. 2017.

\bibitem{ae-jsac-aoudia2019model}
F.~A. {Aoudia} and J.~{Hoydis}, ``Model-free training of end-to-end
  communication systems,'' \emph{IEEE J. Sel. Areas Commun.}, vol.~37, no.~11,
  pp. 2503--2516, Nov. 2019.

\bibitem{deepNOMA-ye2020}
N.~{Ye}, X.~{Li}, H.~{Yu}, L.~{Zhao}, W.~{Liu}, and X.~{Hou}, ``Deep{NOMA}: A
  unified framework for {NOMA} using deep multi-task learning,'' \emph{IEEE
  Trans. Wireless Commun.}, pp. 1--1, Jan. 2020.

\bibitem{DL-inte-9061001}
N.~{Kato}, B.~{Mao}, F.~{Tang}, Y.~{Kawamoto}, and J.~{Liu}, ``Ten challenges
  in advancing machine learning technologies toward 6{G},'' \emph{IEEE Wireless
  Commun.}, vol.~27, no.~3, pp. 96--103, Apr. 2020.

\bibitem{NOMA-coop-xu2016novel}
M.~Xu, F.~Ji, M.~Wen, and W.~Duan, ``Novel receiver design for the cooperative
  relaying system with non-orthogonal multiple access,'' \emph{IEEE Commun.
  Lett.}, vol.~20, no.~8, pp. 1679--1682, June 2016.

\bibitem{NOMA-cons-ye2017constellation}
N.~Ye, A.~Wang, X.~Li, W.~Liu, X.~Hou, and H.~Yu, ``On constellation rotation
  of {NOMA} with {SIC} receiver,'' \emph{IEEE Commun. Lett.}, vol.~22, no.~3,
  pp. 514--517, Dec. 2017.

\bibitem{intro-AF-classic}
J.~N. {Laneman}, D.~N.~C. {Tse}, and G.~W. {Wornell}, ``Cooperative diversity
  in wireless networks: Efficient protocols and outage behavior,'' \emph{IEEE
  Trans. Inf. Theory}, vol.~50, no.~12, pp. 3062--3080, Dec. 2004.

\bibitem{book-deb2014}
K.~Deb and K.~Deb, \emph{Multi-objective optimization}.\hskip 1em plus 0.5em
  minus 0.4em\relax Boston, MA: Springer US, 2014, pp. 403--449.

\bibitem{MTL-ruder2017overview}
S.~Ruder, ``An overview of multi-task learning in deep neural networks,''
  \emph{arXiv preprint arXiv:1706.05098}, 2017.

\bibitem{Sync-dorner2018deep}
S.~{D{\"o}rner}, S.~{Cammerer}, J.~{Hoydis}, and S.~t.~{Brink}, ``Deep learning
  based communication over the air,'' \emph{IEEE J. Sel. Topics Signal
  Process.}, vol.~12, no.~1, pp. 132--143, Feb. 2018.

\bibitem{MI-book-cover2012elements}
T.~M. Cover and J.~A. Thomas, \emph{Elements of information theory}.\hskip 1em
  plus 0.5em minus 0.4em\relax John Wiley \& Sons, Nov. 2012.

\bibitem{PA-popovic2017amping}
Z.~Popovic, ``Amping up the {PA} for 5{G}: Efficient {G}a{N} power amplifiers
  with dynamic supplies,'' \emph{IEEE Microw. Mag.}, vol.~18, no.~3, pp.
  137--149, May 2017.

\bibitem{PA-sun2019behavioral}
J.~Sun, W.~Shi, Z.~Yang, J.~Yang, and G.~Gui, ``Behavioral modeling and
  linearization of wideband {RF} power amplifiers using {B}i{LSTM} networks for
  5{G} wireless systems,'' \emph{IEEE Trans. Veh. Technol.}, vol.~68, no.~11,
  pp. 10\,348--10\,356, June 2019.

\bibitem{book-clark2013error}
G.~C. Clark~Jr and J.~B. Cain, \emph{Error-correction coding for digital
  communications}.\hskip 1em plus 0.5em minus 0.4em\relax Springer Science \&
  Business Media, 2013.

\bibitem{AE-bit-alberge2018deep}
F.~Alberge, ``Deep learning constellation design for the {AWGN} channel with
  additive radar interference,'' \emph{IEEE Trans. Commun.}, vol.~67, no.~2,
  pp. 1413--1423, Oct. 2018.

\bibitem{AE-bit-cammerer2020trainable}
S.~Cammerer, F.~A. Aoudia, S.~D{\"o}rner, M.~Stark, J.~Hoydis, and
  S.~Ten~Brink, ``Trainable communication systems: Concepts and prototype,''
  \emph{IEEE Trans. Commun.}, June 2020.

\bibitem{LDPC-NOMA-pan2018sic}
L.~{Yuan}, J.~{Pan}, N.~{Yang}, Z.~{Ding}, and J.~{Yuan}, ``Successive
  interference cancellation for {LDPC} coded nonorthogonal multiple access
  systems,'' \emph{IEEE Trans. Veh. Technol.}, vol.~67, no.~6, pp. 5460--5464,
  Apr. 2018.

\bibitem{pc-zheng2020threshold}
H.~Zheng, S.~A. Hashemi, A.~Balatsoukas-Stimming, Z.~Cao, T.~Koonen, J.~Cioffi,
  and A.~Goldsmith, ``Threshold-based fast successive-cancellation decoding of
  polar codes,'' \emph{arXiv preprint arXiv:2005.04394}, 2020.

\bibitem{intro-8849796}
T.~{Sypherd}, M.~{Diaz}, L.~{Sankar}, and P.~{Kairouz}, ``A tunable loss
  function for binary classification,'' in \emph{2019 IEEE International
  Symposium on Information Theory (ISIT)}, Paris, France, July 2019, pp.
  2479--2483.

\bibitem{intro-8755300}
M.~{Chen}, U.~{Challita}, W.~{Saad}, C.~{Yin}, and M.~{Debbah}, ``Artificial
  neural networks-based machine learning for wireless networks: A tutorial,''
  \emph{IEEE Commun. Surveys Tuts.}, vol.~21, no.~4, pp. 3039--3071, July 2019.

\bibitem{DVB-S2}
S.~S.~U. Ghouri, S.~Saleem, and S.~S.~H. Zaidi, ``Enactment of {LDPC} code over
  {DVB-S2} link system for {BER} analysis using {MATLAB},'' in \emph{Advances
  in Computer Communication and Computational Sciences}, S.~K. Bhatia,
  S.~Tiwari, K.~K. Mishra, and M.~C. Trivedi, Eds.\hskip 1em plus 0.5em minus
  0.4em\relax Singapore: Springer, 2019, pp. 743--750.

\bibitem{concl-makki2020error}
B.~Makki, T.~Svensson, and M.~Zorzi, ``An error-limited {NOMA}-{HARQ} approach
  using short packets,'' \emph{arXiv preprint arXiv:2006.14315}, 2020.

\bibitem{concl-8454392}
L.~{Liu}, E.~G. {Larsson}, W.~{Yu}, P.~{Popovski}, C.~{Stefanovic}, and E.~{de
  Carvalho}, ``Sparse signal processing for grant-free massive connectivity: A
  future paradigm for random access protocols in the {I}nternet of {T}hings,''
  \emph{IEEE Signal Process. Mag.}, vol.~35, no.~5, pp. 88--99, Sep. 2018.

\bibitem{concl-8989311}
X.~{Wu}, A.~{Ozgur}, M.~{Peleg}, and S.~S. {Shitz}, ``New upper bounds on the
  capacity of primitive diamond relay channels,'' in \emph{2019 IEEE
  Information Theory Workshop (ITW)}, Visby, Sweden, Aug. 2019, pp. 1--5.

\bibitem{concl-8726376}
N.~{Wu}, X.~{Zhou}, and M.~{Sun}, ``Incentive mechanisms and impacts of
  negotiation power and information availability in multi-relay cooperative
  wireless networks,'' \emph{IEEE Trans. Wireless Commun.}, vol.~18, no.~7, pp.
  3752--3765, July 2019.

\end{thebibliography}
\ifCLASSOPTIONcaptionsoff
\newpage
\fi





%

\end{document}